\documentclass[pra,aps,amsmath,amssymb,superscriptaddress,twocolumn,longbibliography]{revtex4-1}
\usepackage{graphicx,multirow}
\usepackage{color,outlines}
\usepackage[english]{babel}
\usepackage{amsthm}
\usepackage[sc,osf]{mathpazo}
\usepackage{hyperref}
\usepackage{physics}
\usepackage{outlines}

\usepackage{tabularx} 
\usepackage{array}
\usepackage{subcaption}
\usepackage{float}
\usepackage{tikz}
\usepackage{adjustbox}
\newcolumntype{C}{>{\centering\arraybackslash}X}

\usepackage[sc,osf]{mathpazo}
\usetikzlibrary {positioning}

\newcommand{\bla}{\color{black}}

\newtheorem{defn}{Definition}

\definecolor{green2}{RGB}{0,100,0}

\begin{document}

\title{Analogs of absolutely maximally entangled states in nonlocal correlations via the sheaf-theoretic framework and its applications}

\author{Nripendra Majumdar}
\email{nripendrajoin123@gmail.com}
\author{S. Aravinda}
\email{aravinda@iittp.ac.in}
\affiliation{Department of Physics, Indian Institute of Technology Tirupati, Tirupati, India~517619}  
\begin{abstract} 
   The foundational work by Bell led to an interest in understanding non-local correlations that arise from entangled states shared between distinct, spacelike-separated parties, which formed a foundation for the theory of quantum information processing. We investigate the question of maximal correlations analogous to the maximally entangled states defined in the entanglement theory of multipartite systems. In this work, we define the maximality of nonlocal correlation as being analogous to the absolutely maximally entangled state. 
   To formalize this, we employ the sheaf-theoretic framework for contextuality, which generalizes non-locality. This provides a metric for correlations called contextual fraction (CF), which ranges from $0$ (non-contextual) to $1$ (maximally contextual). Using this, we have defined the absolutely maximal contextual correlations (AMCC), which are maximally contextual and have maximal marginals. The Popescu-Rohrlich (PR) box serves as the bipartite example, and we construct various extensions of such correlations in the tripartite case. An infinite family of various forms of AMCC is constructed using the parity check and the constraint satisfiability problem (CSP) construction. We also demonstrate the existence of maximally contextual correlations, which do not exhibit maximal marginals, and refer to them as non-AMCC. Furthermore, we showed that GHZ correlations in the $(n,2,2)$ setting give rise to AMCCs for the particular choice of measurement settings. The results are further applied to secret sharing and randomness extraction using AMCCs.
   
\end{abstract}

\maketitle

\section{Introduction}
 
The remarkable analysis by Einstein, Podolsky, and Rosen (EPR) \cite{PhysRev.47.777} revealed the peculiar behavior of nature and questioned the completeness of quantum mechanics. The quantum mechanical state used in their argument exhibits nonclassical correlations, which Schrödinger later termed entanglement.  Attempts to address the questions raised by EPR culminated in the seminal work of Bell \cite{bell1964einstein}, who developed a theoretical framework to test the foundations of quantum theory. Bell’s theorem demonstrates the incompatibility of quantum mechanical predictions with any theory that combines the assumptions of determinism, locality, and measurement independence.

The exploration of entanglement has led to profound insights and practical advantages, underpinning the entire field of quantum information theory and quantum computation. A foundational framework for this study treats entanglement as a resource that can be manipulated through local operations and classical communication (LOCC) \cite{RevModPhys.81.865}. For bipartite systems, asymptotic entanglement transformation via LOCC has been extensively studied, along with measures of pure-state bipartite entanglement \cite{PhysRevA.56.R3319, PhysRevLett.83.436}. From these works, the concept of a maximally entangled state in bipartite systems is well established \cite{PhysRevLett.83.436, PhysRevA.56.R3319, PhysRevA.62.062314}. The structure becomes richer for multipartite systems. Even for three qubits, two inequivalent classes exist under stochastic LOCC (SLOCC) \cite{PhysRevA.62.062314}, requiring new measures to quantify genuine multipartite entanglement. In general, the concepts of entanglement transformation, quantification, and maximal entanglement become intractable for multipartite systems due to the exponential complexity in the number of parameters. Consequently, despite extensive research \cite{verstraete2002four, kraus2010local, de2013maximally, miyake2003classification, gour2011necessary, spee2017entangled, li2024identifying, gunn2023approximate, ma2011measure, xie2021triangle}, fundamental questions remain open: Does a unambiguous concept of a ``maximally entangled state" exist for multipartite systems, and if so, how should it be defined?

 The correlations obtained by measuring an entangled state constitute another important form of nonclassicality, both from a resource-theoretic and an applied perspective. In a Bell experiment, these correlations can be viewed abstractly as a ``black box" characterized by inputs and outputs, with its internal physical structure being irrelevant. This abstraction makes them a powerful resource for device-independent protocols \cite{randomnessextraction, DImoreno2020device}. Correlations that violate a Bell inequality are termed nonlocal and belong to a strictly larger set than those admitting a local hidden-variable model \cite{AshutoshRaiGeometryofquantumset2019}. Cirel'son established that the quantum maximum for the paradigmatic Clauser-Horne-Shimony-Holt (CHSH) expression is $2\sqrt{2}$ \cite{cirel1980quantum}. More generally, postquantum nonlocal correlations—supersets of the quantum set—have also been studied, as exemplified by the Popescu-Rohrlich box \cite{popescu1994quantum}.

The theories of entanglement and nonlocal correlations are deeply intertwined, presenting significant conceptual intricacies~\cite{schmid2023understanding}. This interplay is clarified by the framework of local operations and shared randomness (LOSR) \cite{LOSRSchmid2020typeindependent, LOSRTRosset2020}, which serves as the correlational analog to LOCC in entanglement theory. To understand genuine multipartite correlations, a more general operational approach employs wiring and classical communication prior to inputs (WCCPI) \cite{gallego2012operational, Amit2020Operationalcharacterization}, demonstrating that earlier definitions were incomplete \cite{bancal2011definition}. A fundamental difficulty arises from a key divergence: while the classification of pure entangled states into equivalence classes under LOCC and SLOCC is well-established, no analogous structure exists for nonlocal correlations. In the bipartite case, LOCC yields a single equivalence class of maximally entangled states—a definitive ``golden unit." However, no corresponding unique unit has been identified for correlations. Concepts such as single-copy or asymptotic distillation do not translate directly from entanglement theory, preventing a straightforward mapping from maximal entanglement to maximal correlation. Consequently, many correlation sets are incomparable in the asymptotic limit \cite{ghosh2024quantum}. Thus, a central challenge is to formulate rigorous definitions for maximal entanglement and maximal correlations in genuine multipartite systems, an area that remains actively unresolved.

 Correlations from an experimental setup can be viewed more generally using a sheaf-theoretic approach \cite{abramsky2011sheaf}. Any dataset described by $(n, m, o)$, where $n$ is the number of parties, $m$ is the number of measurements per party, and $o$ is the set of possible outcomes, can be formalized as a measurement scenario. In this framework, compatible sets of measurements (called contexts) and their possible outcomes are organized so that probability distributions over outcomes define the empirical model. A no-signaling empirical model is one in which outcome probabilities do not depend on which other measurements are performed, generalized by the compatibility condition. If these probabilities can be derived from a single global assignment, the model is noncontextual; otherwise, it is contextual.
 
 Contextuality is a nonclassical feature indicating the impossibility of assignment to measurements independent of the measurement context. The Kochen-Specker (KS) theorem \cite{kS2011problem}, in this regard, provides a foundational understanding of such logical proofs. The Mermin square proofs \cite{mermin1990simple} with $9$ observables and the 18-ray vector \cite{18vectorcabello1996bell}, which contains the first minimal set of projectors after the 117-vector \cite{kS2011problem}, followed by the 13-ray vector \cite{13vector2012state} proof, show examples of state-independent KS theorem proofs. Fine's theorem \cite{fine1982hidden} is one of the pieces of evidence that leads us to the equivalence between the violation of the Bell inequality (CHSH) \cite{clauser1969proposed} and the impossibility of a global joint probability distribution. The Klyachko-Can-Binicioğlu-Shumovsky (KCBS) inequality \cite{kcbs2008simple} provides one of the simplest state-dependent proofs for a qutrit system, demonstrating that spatial separation is not required and that contextuality can be exhibited in a single system. In this respect, Bell nonlocality can be viewed as a special case of contextuality, which requires extra constraints such as specially separated parties and no-signaling conditions. The sheaf-theoretic approach makes this point more prominent and provides a general mathematical tool.
Within this framework, any empirical model can be represented as a linear system: if this system admits a non-negative solution, the model is noncontextual; otherwise, it is contextual. In the possibilistic case, a stronger criterion defines the notion of a strongly contextual model. Any empirical model can be expressed as a convex combination of a noncontextual model (with coefficient noncontextual fraction, $\mathrm{NCF} = p$) and a strongly contextual model (with coefficient contextual fraction $\mathrm{CF} = 1-p$). If $\mathrm{CF}=1$ ($p = 0$), the model is maximally contextual; if $\mathrm{CF}=0$ ($p = 1$), the model is noncontextual \cite{CF}. It has been shown that a model is strongly contextual if and only if it is maximally contextual \cite{abramsky2011sheaf}. 

These results and the framework have broader implications for our understanding of computational advantage. In particular, in quantum computation, contextuality enables transitions from limited computational power, characterized by the parity complexity class, to full quantum universality \cite{comppower}. This transition has been demonstrated using Greenberger-Horne-Zeilinger (GHZ)-type correlations and Mermin-type arguments, and the computational advantage of the measurement-based quantum computation (MBQC) model has been understood \cite{CMBQC, CF}. Contextuality also admits various mathematical characterizations, such as cohomological or topological obstruction \cite{ topologicalproofscontextualityquantum, cohomandnonlocalityandcontextuality, cohomologicalframeworkcontextualquantum, ContextualityasresourceinMBQC, walleghem2024refinedfrauchigerrennerparadoxbased, stringordermeasurementbased, homotopicalapproach, Peresconjectureforcontextuality, contextualityadvancesonewaycommunication}. Such a global obstruction can be geometrically visible in the bundle diagram  \cite{CinBundlediagram}. 

Although contextuality is an abstract concept, its experimental validation and implications have been extensively studied, including Mermin's square and the KCBS inequality \cite{Merminsquareexperiment2009, KCBSexperiment2017}. Quantifying and extending contextuality in quantum many-body systems is being investigated, which has motivated the translation of contextuality into a nonlocal game. Contextuality undermines the classical strategy \cite{scalabletestsquantumcontextuality, manybodycontextualityselftestingquantum}. This also implies the connection between the phase of the quantum many-body system and contextuality. Recently, the Google research team demonstrated the implications of quantum contextuality for classical strategies with limited resources on a real-time quantum processor. That also shows the scalability of contextuality and operational relevance in the multiparty system \cite{kumar2025qcsep}. Contextuality has also been shown to certify intrinsic randomness, and its scalability is understood from a graph-theoretic perspective, which is experimentally feasible \cite{contextualityandrandomness}. All the above instances involve parity conditions, which exhibit the importance of maximal contextuality.

It is evident that contextuality is a more general phenomenon than nonlocality and that correlations exhibiting maximal contextuality are termed maximal contextual correlations, as they do not contain any local fraction \cite{FNSandallvsnothingequivalece, AshutoshRaiGeometryofquantumset2019}. To extend the idea of genuine multipartite entanglement, a special class of entanglement is defined as absolutely maximally entangled (AME) states \cite{AME}. These pure states are maximally entangled, such that for any bipartite reduction, they yield maximally mixed states.  In this backdrop, we would like to ask the question of what are the analogs of AME states in the theory of correlations.  In this work, we extend this to genuine multipartite correlations and define a class of correlations as absolutely maximal contextual correlations (AMCC). These correlations represent the maximal contextual correlations $(CF=1)$, which are also maximally random marginals. In the $(2,2,2)$ scenario, there are exactly eight such AMCCs, which are Popescu-Rohrlich (PR) boxes.  However, in a multipartite scenario like $(3,2,2)$, there are infinitely many such correlations that qualify as AMCCs, which means they are face-no-signaling correlations having zero local fraction and maximal marginality. The mathematical construction of these correlations can be accomplished using the parity check or the constraint satisfaction problem (CSP) approach. Apart from AMCCs, there exists another class of correlations that are maximally contextual but not maximally random marginals. The AMCCs correspond to correlations that are analogous to absolutely maximally entangled (AME) states. This property is of major significance in device-independent randomness extraction due to the maximal marginal property, and such correlations can also be utilized in secure communication protocols.

\bla 

The paper is organized into the following sections. In Sec.~\ref{sec 00}, we will discuss the mathematical formalism of correlations and the corresponding convex polytope structure. It also incorporates the mathematical tool of sheaf theory and several useful definitions to facilitate an understanding of contextuality and its various forms. In Sec.~\ref{sec 01}, we define the AMCC and present some evidence based on it. Later in the section, we provide the construction of AMCCs. We propose two types of constructions here, based on the structure of correlations: the parity-check construction and the constraint-satisfaction problem (CSP) approach. In addition, we provide a proof that $(n, 2, 2)$ GHZ correlations are AMCCs. Lastly, in Sec.~\ref{02}, we have shown some remarkable applications, demonstrating how the AMCC will be used for randomness extraction and secret-sharing purposes.

\section{Formalism}
\label{sec 00}
We begin with a simple picture: two observers, Alice and Bob, share a joint state, and each can perform local measurements on their own systems. Alice can choose between two dichotomous measurements, \( \{X_1, X_1^\prime\} \), and Bob likewise chooses from \( \{X_2, X_2^\prime\} \). Each measurement yields one of two possible outcomes, labeled \( 0 \) or \( 1 \). The experimental statistics are fully described by the set of joint probabilities \( p(x_1, x_2 \mid X_1, X_2) \), giving the probability that Alice obtains outcome \( x_1 \) and Bob obtains outcome \( x_2 \) when they choose measurements \( X_1 \) and \( X_2 \), respectively. This setup is often described as a \emph{black-box scenario}: each observer holds a device to which they input a measurement choice (\( X_1 \) or \( X_1^\prime \) for Alice, \( X_2 \) or \( X_2' \) for Bob) and receives an outcome (\( x_1 \) or \( x_2 \)) in return. The internal workings of the devices are unknown; only the input-output statistics are accessible. This is known as the \((2,2,2)\) Bell scenario, where the triplet denotes two parties, each with two measurement choices and two outcomes per measurement.

Consider a general $n$ partite Bell scenario, where \(n\) denotes the number of parties indexed by \(i \in \{1,\dots,n\}\), for each party \(i\) there are \(m_i\) measurement settings denoted \(X_i \in \{1,\dots,m_i\}\), and for each measurement \(X_i=j\) of party \(i\) there are \(o_{ij}\) possible outcomes denoted \(x_i \in \{1,\dots,o_{ij}\}\). The statistics are described by the joint conditional probability distribution \(p(x_1,\dots,x_n \mid X_1,\dots,X_n)\), which must satisfy: normalization, \(\sum_{x_1,\dots,x_n} p(x_1,\dots,x_n \mid X_1,\dots,X_n) = 1\) for all \(X_1,\dots,X_n\); positivity, \(p(x_1,\dots,x_n \mid X_1,\dots,X_n) \geq 0\) for all outcomes and settings; and the no-signaling condition, which requires that for each party \(j\) the marginal distribution of the remaining parties is independent of party \(j\)'s measurement choice:
\begin{widetext}
    \begin{equation}
        \sum_{x_j}p(x_1,\cdots,x_j,\cdots,x_n\mid X_1,\cdots,X_j,\cdots,X_n)=
\sum_{x_j}p(x_1,\cdots,x_j,\cdots,x_n \mid X_1,\cdots,X_j',\cdots,X_n)
\label{eq: no-signaling condition}
    \end{equation}
\end{widetext}
for all \(j\), all outcomes \(x_1,\dots,x_{j-1},x_{j+1},\dots,x_n\), and all settings \(X_1,\dots,X_{j-1},X_j,X_j',X_{j+1},\dots,X_n\). The set of probability distributions satisfying these constraints forms a convex polytope known as the no-signaling polytope \cite{pironio2005lifting, AshutoshRaiGeometryofquantumset2019}, with dimension

\[
\dim P = \prod_{i=1}^n \left( \sum_{j=1}^{m_i} (o_{ij} - 1) + 1 \right) - 1.
\]

The correlations are called \emph{local} if they can be reproduced by a local hidden-variable model, i.e.,
\begin{widetext}
\begin{equation}
    p(x_1,\dots,x_n\mid X_1,\dots,X_n)= \int d\lambda\,\rho(\lambda)\,p(x_1\mid X_1, \lambda)\cdots p(x_n\mid X_n,\lambda),
\end{equation}
\end{widetext}
where $\rho(\lambda)\geq 0$, $\int d\lambda\,\rho(\lambda)=1$, and $p(x_i\mid X_i, \lambda)$ denotes the probability of obtaining outcome $x_i$ for measurement $X_i$ given the hidden variable $\lambda$. Here $\lambda$ represents a set of preassigned values for all possible measurement outcomes. The corresponding deterministic probability vectors are defined as
\begin{widetext}
   \begin{equation}
    p^{\lambda}(x_1,\dots, x_n\mid X_1,\dots,X_n)=\begin{cases}
1, & \text{if}\quad \forall i,\; \lambda_{iX_i}=x_i,\; i\in \{1,\dots, n\},\\
0, & \text{otherwise}.
\end{cases}
\end{equation} 
\end{widetext}
The set of local correlations $P$ is the convex hull of these deterministic points, forming the \emph{local polytope} \cite{pironio2005lifting, AshutoshRaiGeometryofquantumset2019}
\begin{equation*}
    P=\left\{p\in \mathbb{R}^r \;|\; p=\sum_{\lambda \in \Lambda}\rho(\lambda)p^{\lambda},\; \rho(\lambda)\geq 0,\; \sum_{\lambda \in \Lambda}\rho(\lambda)=1 \right\}.
\end{equation*}

For the Bell polytope $P \subseteq \mathbb{R}^r$, consider an inequality $\mathbf{a}\cdot\mathbf{p}\geq a_0$, where $(\mathbf{a},a_0)\in \mathbb{R}^{r+1}$. This inequality is called a \emph{valid Bell inequality} for $P$ if it holds for all $\mathbf{p}\in P$. To verify validity, it suffices to check that it is satisfied by all extreme points of $P$, i.e., by the deterministic vectors. Given such an inequality, the corresponding face of the polytope is defined by the equality condition $F=\{\mathbf{p}\in P \;|\; \mathbf{a}\cdot\mathbf{p}= a_0\}$. A face with dimension exactly one less than that of the full polytope is called a \emph{facet} \cite{pironio2005lifting, AshutoshRaiGeometryofquantumset2019}. Thus, for any proper face $F$ (with $F\neq \varnothing$ and $F\neq P$), the dimension satisfies $\dim F \leq \dim P - 1$.  These proper faces, which also satisfy the no-signaling and normalization conditions, are defined by zero constraints: 
\begin{equation}
F^{\text{NS}} = \{\mathbf{p} \in \mathbf{NS}\;|\; p = 0\},
\label{eq: properface}
\end{equation} 
where $p$ is an element of $\mathbf{p} = \{p(x_1, \cdots, x_i \mid X_1, \cdots, X_i)_{i \leq \dim P - 1}\}$, and $\mathbf{NS} \subset \mathbb{R}^r$ denotes the set of no-signaling correlations. By imposing these zero constraints, the correlations are brought to their maximally reduced form, corresponding to the lowest-dimensional face. The resulting face correlations yield a set of deterministic constraints that cannot be satisfied simultaneously by any global assignment. In other words, this construction leads directly to all-versus-nothing (AVN) proofs \cite{cohom_AVN_abramsky_et_al2015} (see Appendix~\ref{app: prope facet}). Hence, all the zeros define a face of the no-signaling polytope, and every deterministic local vertex $\mathbf{p}^L\in P$  violates at least one zero constraint. This implies that $F^{\text{NS}}\cap\mathbf{p}^L=\phi$; in other words, the correlations on the face $F^{\text{NS}}$ in Eq.~(\ref{eq: properface}) do not contain any local components \cite{FNSandallvsnothingequivalece, AshutoshRaiGeometryofquantumset2019}. These face correlations are referred to as maximal correlations, which are discussed in the next section.

\subsection{ Contextual correlations} 

The Bell-scenario $(n,m,o)$ is an empirical dataset obtained in an experiment. Such an empirical model is described using a measurement scenario, which is a triplet $\langle \mathcal{X}, \mathcal{M}, \mathcal{O} \rangle$, where $\mathcal{X}$ is a set of all observables, $\mathcal{M}$ is a measurement cover, and $\mathcal{O}$ is a set of all the outcomes obtained from each measurement. We will elaborate on these definitions in the following paragraphs.

\begin{defn}[Measurement Cover] 
    The measurement cover is a set $\mathcal{M}\subset \mathcal{P(X})$ such that
    \begin{enumerate}
        \item $\bigcup \mathcal{M}=\mathcal{X}$ and
        \item the following chain condition applies: if $\mathcal{C},\mathcal{C}^\prime\in \mathcal{M}$ and $\mathcal{C}\subseteq\mathcal{C}^\prime$, then $\mathcal{C}=\mathcal{C}^\prime$.
    \end{enumerate}
\end{defn} 

For example, let us take the $(2, 2, 2)$-Bell scenario, as introduced earlier. We can write the measurement set:
\begin{equation}
    \mathcal{X}=\{X_1, X_1^\prime, X_2, X_2^\prime\},
\end{equation}
the measurement cover,
\begin{equation}
    \mathcal{M}=\{\{X_1,X_2\},\{X_1,X_2^\prime\},\{X_1^\prime,X_2\},\{X_1^\prime,X_2^\prime\}\},
    \label{9}
\end{equation}
and the outcome set $\mathcal{O}=\{0,1\}$. The joint outcome corresponding to each context $\mathcal{C}\in \mathcal{M}$ is called a section $\mathrm{s}$.

\begin{defn}[section]
    A section $s$ is a mapping from $\mathcal{U}\subset\mathcal{X}$ to $\mathcal{O}$, i.e.,
    \begin{equation}
        s:\mathcal{U}\to\mathcal{O}.
    \end{equation}
\end{defn}

The set of all sections is defined as the \emph{sheaf of events} (or a presheaf) $\mathcal{E}: P(X)^{\text{op}} \to \mathsf{Set}$, which is a functor from the opposite category of the poset \(P(X)\) to the category of sets \cite{abramsky2011sheaf}. Considering the previous example, for $\mathcal{C}_1=\{X_1,Y_1\}\in\mathcal{M}$, the set of events (sections) is  
\begin{equation}
    \mathcal{E}(\mathcal{C}_1)=\{(0,0),(0,1),(1,0),(1,1)\}.
\end{equation}

Weight assignments to the sheaf of events are defined via a functor
\begin{equation}
    \mathcal{D}_{\mathcal{R}}\mathcal{E}:P(X)\to \mathsf{Set},
    \label{eq:distribution-functor}
\end{equation}
where $\mathcal{D}_{\mathcal{R}}:\mathsf{Set}\to \mathsf{Set}$ is the distribution functor that sends a set $\mathcal{E}$ to
\[
\mathcal{D}_{\mathcal{R}}(\mathcal{E}) = \{\mathcal{P} : \mathcal{E} \to \mathbb{R}_{\geq 0} \mid \sum_{x \in \mathcal{E}} \mathcal{P}(x) = 1\},
\]
and $\mathcal{R} = \{\mathbb{R}_{\geq 0}, +, \cdot, 0, 1\}$ is a semiring of non‑negative reals. Applying this functor to the sheaf $\mathcal{E}$ yields a presheaf of probability distributions.

If $\mathrm{e} \in \mathcal{D}_{\mathcal{R}}(\mathcal{E}(U^\prime))$ is a distribution over sections of $U^\prime\subseteq \mathcal{X}$, then there is a restriction map \(\text{res}_U^{U^\prime}\) for distributions such that
\begin{equation}
    \begin{split}
        \mathcal{D}_\mathcal{R}(\mathcal{E}(U^\prime)) &\xrightarrow{\text{res}_U^{U^\prime}} \mathcal{D}_{\mathcal{R}}(\mathcal{E}(U)),\\
        \mathrm{e} &\mapsto \mathrm{e}|_U.
    \end{split}
    \label{eq:restriction-map}
\end{equation}
Here $\mathrm{e}|_U$ is obtained by summing over all $\mathrm{s}^\prime \in \mathcal{E}(U^\prime)$ that restrict to a given $\mathrm{s} \in \mathcal{E}(U)$:
\begin{equation}
    \mathrm{e}|_U(\mathrm{s}) = \sum_{\substack{\mathrm{s}^\prime \in \mathcal{E}(U^\prime) \\ \mathrm{s}^\prime|_U = \mathrm{s}}} \mathrm{e}(\mathrm{s}^\prime).
    \label{eq:marginal-sum}
\end{equation}
Thus $\mathrm{e}|_U$ is the marginal distribution of $\mathrm{e}$ over the subset $U$.

\begin{defn}[Empirical model]
    Given a measurement scenario $\langle \mathcal{X}, \mathcal{M}, \mathcal{O} \rangle$, an \emph{empirical model} is a family of distributions $\{\mathrm{e}_{\mathcal{C}}\}_{\mathcal{C}\in\mathcal{M}}$, where each $\mathrm{e}_{\mathcal{C}}\in \mathcal{D}_{\mathcal{R}}(\mathcal{E}(\mathcal{C}))$.
\end{defn}

\begin{defn}[Generalized no‑signaling] 
    Let $\mathcal{C}_1, \mathcal{C}_2\in \mathcal{M}$ with $\mathrm{e}_{\mathcal{C}_1}\in \mathcal{D}_\mathcal{R}\mathcal{E}(\mathcal{C}_1)$ and $\mathrm{e}_{\mathcal{C}_2}\in \mathcal{D}_\mathcal{R}\mathcal{E}(\mathcal{C}_2)$. The generalized no‑signaling condition is
    \begin{equation}
        \mathrm{e}_{\mathcal{C}_1}|_{\mathcal{C}_1\cap \mathcal{C}_2}=\mathrm{e}_{\mathcal{C}_2}|_{\mathcal{C}_1\cap \mathcal{C}_2}.
        \label{eq:generalized-nosignal}
    \end{equation}
    An empirical model satisfying this compatibility condition is called a \emph{no‑signaling empirical model}.
\end{defn}

The set of all sections over $\mathcal{X}$, i.e. $\mathcal{E}(\mathcal{X}) = \mathcal{O}^{\mathcal{X}}$, is called \emph{global sections}. A global section is a map $g:\mathcal{X}\to \mathcal{O}$, and a distribution over these sections is called a \emph{global distribution} $\mathrm{e}_\mathcal{X} \in \mathcal{D}_\mathcal{R} \mathcal{E}(\mathcal{X})$, satisfying
\begin{equation}
    \mathrm{e}_\mathcal{C}(\mathrm{s}_\mathcal{C}) = \sum_{\substack{\mathrm{s}^\prime \in \mathcal{E}(\mathcal{X}) \\ \mathrm{s}^\prime|_\mathcal{C} = \mathrm{s}_\mathcal{C}}} \mathrm{e}_\mathcal{X}(\mathrm{s}^\prime),
    \label{eq:global-marginal}
\end{equation}
i.e., $\mathrm{e}_\mathcal{C}$ is the marginal of $\mathrm{e}_\mathcal{X}$. Hence, ``If such a global probability distribution exists, the model corresponds to a deterministic hidden‑variable scenario.''

The existence of a global section implies that the model can be described by a noncontextual hidden‑variable theory. One may ask whether, for a given empirical model, a global section exists. This question can be addressed using linear algebra. For a measurement scenario $\langle \mathcal{X}, \mathcal{M}, \mathcal{O} \rangle$ with $n$ global sections and $m$ local sections, the restriction map can be represented as an incidence matrix $[M]_{n\times m}$. Moreover, if a global distribution $\mathbf{d}\in \mathcal{D}_\mathcal{R} \mathcal{E}(\mathcal{X})$ exists, then the following linear system must have a solution over the semiring of non‑negative reals:
\begin{equation}
    \mathbf{M}\,\mathbf{d} = \mathbf{v}, \qquad \mathbf{d} \geq \mathbf{0},
    \label{eq:linear-system}
\end{equation}
where $\mathbf{M}$ is the incidence matrix, $\mathbf{d}$ is the global probability distribution, and $\mathbf{v}$ is the vector of local probability assignments. The problem reduces to determining whether a non‑negative solution $\mathbf{d}$ exists.

\begin{defn}[Contextual]
    A no‑signaling empirical model $\{\mathrm{e}_{\mathcal{C}}\}_{\mathcal{C}\in \mathcal{M}}$ is \emph{contextual} if and only if the linear system in Eq.~(\ref{eq:linear-system}) admits no solution over the non‑negative reals.
\end{defn}

If one replaces the semiring of non‑negative reals by the Boolean semiring, the probabilistic empirical model reduces to a \emph{possibilistic} model.

\begin{defn}[Strong contextuality]
    Given a possibilistic model derived from an empirical model, if there exists no global section $g$ such that $\forall \mathcal{C}\in\mathcal{M},\; \mathrm{e}_{\mathcal{C}}(g|_{\mathcal{C}}) > 0$, then the model is called \emph{strongly contextual} (SC).
    \label{def: Strong contextuality}
\end{defn}

Any empirical model can be decomposed into the following convex combination:
\begin{equation}
    M=pM^{\mathrm{NC}}+(1-p)M^{\mathrm{SC}},
    \label{eq:convex-decomp}
\end{equation}
where the noncontextual fraction $\mathrm{NCF}=p$ and the contextual fraction $\mathrm{CF}=1-p$. If $p=0$, the model is \emph{maximally contextual}; if $p=1$, the model is noncontextual. It has been shown that a model is strongly contextual if and only if it is maximally contextual \cite{CF}. 
The contextual fraction can be computed via linear programming (specifically, using the simplex method) \cite{lee2016first}. This involves setting up an objective function and linear constraints, which together define a multidimensional convex polytope. The simplex method provides an efficient algorithm for such optimization problems, relying on three main components: slack variables, a tableau (data table), and pivot operations. Thus, it is well suited for calculating the CF value.

Since nonlocality is a special case of contextuality, and strong contextuality represents the maximal form of contextuality, the correlations associated with it are referred to as maximal contextual correlations. The extension of such a correlation to a multipartite scenario has yet to be explored. 

 The correlations corresponding to maximal contextuality lie on the face of the nonlocal polytope, as they do not contain any local fractions \footnote{Any empirical model can be decomposed as a convex mixture of contextual and noncontextual models, with contextuality being a more general concept than nonlocality. The convex decomposition of a nonlocal correlation can be expressed as \[p(x_1,\ldots,x_n) = q^L p_{L}(x_1,\ldots,x_n) + q^{NL} p_{NL}(x_1,\ldots,x_n),
\] where (p{L}) denotes the local part and (p{NL}) the nonlocal part. This decomposition is a special case within the broader framework of contextuality. Hence, the condition that the contextual fraction $CF = 1$ corresponds to a special case where the nonlocal fraction is also maximal}. This is significant because such correlations also correspond to the general form of AVN proofs \cite{cohom_AVN_abramsky_et_al2015}. The term all versus nothing is used by Mermin \cite{Mermin_argument}, and later the equivalence of AVN and different forms of the face correlations is discussed \cite{FNSandallvsnothingequivalece}. In the probabilistic form of AVN, the presence of zeros determines the structure of these correlations; these zeros specify the facets that maximally reduce the dimension of the face while still satisfying the no-signaling and normalization conditions. We utilize this property in Sec.~\ref{sec 01} for construction purposes. While the concept of maximal entanglement is well established for quantum states, its direct analog in terms of correlations remains somewhat ambiguous. To clarify the notion of maximality in correlations, we take AME states \cite{AME} as a reference point, since their genuine bipartite marginals are maximally mixed. Accordingly, we focus on the face-no-signaling correlations that are not only maximally contextual but also exhibit maximal marginality. This approach leads to the identification of a new class of correlations and provides a precise definition of maximality in the context of correlations, which will be introduced in the next section.

\section{Absolutely maximal contextual correlations}
\label{sec 01}
The results and calculations presented here are obtained using the mathematical framework of sheaf theory \cite{abramsky2011sheaf}, which has been employed to define AMCCs. Contextuality—a fundamentally nonclassical property—serves as a key feature within this model and is quantified by the contextual fraction ($CF$), which ranges between $0$ and $1$. Accordingly, three classes of models can be distinguished: noncontextual $(\mathrm{CF} = 0)$, contextual $(0 < \mathrm{CF} < 1)$, and maximally contextual $(\mathrm{CF} = 1)$ \cite{CF}. Any empirical model can thus be expressed as a convex combination of a noncontextual and a maximally contextual component.

In this section, we utilize these concepts, as well as those discussed in the previous section, to introduce a new class of maximal contextual correlations based on the maximally contextual model. These correlations are of particular interest as they are the face nosignaling correlations, which generally satisfy the algebraic form of AVN proof \cite{cohom_AVN_abramsky_et_al2015}. The required mathematical definitions are defined below in this section. 

\begin{defn}[Maximal Marginals]
For any $(n,m,o)$ Bell scenario, the correlations are called \emph{maximal marginals} if all marginals satisfy
\begin{equation}
    e\mid_U = p(x_1,\dots,x_k \mid X_1,\dots,X_k) = \frac{1}{2^k} \quad \forall \; k < n,
\end{equation}
where $U = \{X_1,\dots,X_k\} \subset \mathcal{X}$ and each $x_i \in \{0,1\}$.
\label{def:max-mixed}
\end{defn}

\begin{defn}[Absolutely Maximally Contextual Correlations]
An $(n,m,o)$ correlation is called \emph{absolutely maximally contextual} if it is simultaneously maximally contextual and maximal marginal.
\label{def:amc}
\end{defn}

The bipartite case of AMCCs is represented by the PR-box \cite{popescu1994quantum} for the $(2,2,2)$ Bell scenario, for which the contextual fraction is $\mathrm{CF}=1$. The Bell polytope $P$ for this scenario has dimension $8$ and contains $24$ vertices: $16$ local vertices and $8$ PR-box vertices. The eight PR boxes are given by \cite{barrett2005nonlocal}
\begin{equation}
P_{\text{PR}}^{\alpha\beta\gamma} (x_1x_2\mid X_1X_2) =
\begin{cases}
\frac{1}{2}, &
\begin{aligned}
x_1 \oplus x_2 &= X_1X_2 \oplus \alpha X_1 \\ &\oplus \beta X_2
\oplus \gamma,
\end{aligned}\\
0, & \text{otherwise},
\end{cases}
\label{eq:pr-box}
\end{equation}
where $\alpha, \beta, \gamma \in \{0,1\}$ and $\oplus$ denotes addition modulo $2$. Defining $f = X_1X_2 \oplus \alpha X_1 \oplus \beta X_2  \oplus \gamma$, the PR box can also be expressed as 
\begin{equation}
    P_{\text{PR}}^{\alpha\beta\gamma} (x_1x_2\mid X_1X_2) = \frac{1}{2} \left( \delta_{0}^{x_1} \delta_{f}^{x_2} + \delta_{1}^{x_1} \delta_{f \oplus 1}^{x_2} \right),
    \label{eq:pr-box-delta}
\end{equation}
where $\delta_a^b$ is the Kronecker delta.
 The bipartite marginals for the correlation are maximal marginals:
\begin{equation}
\begin{split}
    &P (x_1\mid X_1) = \frac{1}{2} \left( \delta_{0}^{x_1} + \delta_{1}^{x_1}  \right),\\
    &P (x_2\mid X_2) = \frac{1}{2} \left( \delta_{0}^{x_2} + \delta_{1}^{x_2}  \right)
\end{split}
\label{eq:pr-box-delta-mrginals}
\end{equation}
Since the marginals corresponding to these PR box correlations are maximal marginals, and the contextual fraction is $\mathrm{CF}=1$, for the $(2,2,2)$-Bell scenario, the PR boxes are the AMCCs. The only maximal correlations in the $(2, 2, 2)$ scenario are PR boxes \cite{abramsky2011sheaf}; consequently, the PR boxes are the only possible AMCCs.

\subsection{$(3,2,2)$ Bell scenario}
Consider the three-party $(3,2,2)$-Bell scenario. The corresponding no-signaling polytope has dimension $26$ with $46$ extremal points, which can be grouped into three categories: local, two-way nonlocal, and three-way nonlocal \cite{barrett2005nonlocal}. We investigate which types of correlations in $(3,2,2)$ Bell scenarios give rise to AMCCs. The extremal correlations (vertices) \cite{Extremal_points_Pironio_2011} have been studied extensively for the $(3, 2, 2)$ scenario, which are maximally contextual, whereas the correlations we are introducing are the nosignaling face correlations that lie on the face having zero local fraction; in addition, they are maximally marginal.

In the $(3,2,2)$ scenario, there are $8$ measurement contexts, and each context admits $8$ possible outcome assignments (sections), yielding a total of $64$ local joint assignments. Over the full set of observables $\mathcal{X}=\{X_1, X_1^\prime, X_2, X_2^\prime, X_3, X_3^\prime\}$, there are $2^6 = 64$ possible global assignments. Whether such global sections exist can be checked using the linear programming approach, namely the simplex method, where the optimization process takes place over an objective function along with several inequalities. The objective function for the setup is a global subprobability distribution $\mathbf{b}$ that consists of non-negative entries. Thereafter, finding the subprobability distribution yields the value of the noncontextual fraction, namely $\text{NCF}=\mathbf{1.b}$ \cite{CF}, where $\mathbf{1}$ is a vector with each element $1$. Hence, the contextual fraction is $\text{CF}=1-\text{NCF}$.

As an example, consider the correlation obtained from the GHZ state 
\[
|\text{GHZ}\rangle = \frac{|000\rangle + |111\rangle}{\sqrt{2}}
\] 
with each party performing measurements in the $X$ or $Y$ basis (e.g., $X_1 = \sigma_x$, $X_1^\prime = \sigma_y$, and similarly for parties $2$ and $3$). The resulting probability distribution is given by
\begin{widetext}
\begin{equation}
P(x_1x_2x_3\mid X_1X_2X_3) = 
\begin{cases}
   \frac{1}{4}, & \text{if } x_1 \oplus x_2 \oplus x_3 = X_1X_2X_3 \oplus X_1X_2 \oplus X_2X_3 \oplus X_3X_1, \\[6pt]
   \frac{1}{8}, & \text{for } (X_1,X_2,X_3^\prime),\;(X_1,X_2^\prime,X_3),\;(X_1^\prime,X_2,X_3),\;(X_1^\prime,X_2^\prime,X_3^\prime), \\[6pt]
   0, & \text{otherwise},
\end{cases} 
\label{eq:ghz-dist}
\end{equation}
\end{widetext}
where $X_i,X_i^\prime \in \{0,1\}$ encodes the measurement choices (e.g., $0$ for $X$, $1$ for $Y$) and $x_i \in \{0,1\}$ are the outcomes. The bipartite marginals of this correlation are uniformly random:
\begin{equation}
\begin{split}
    p(x_1x_2\mid X_1X_2) &= \frac{1}{4} \left[ \delta_{0}^{x_1}\delta_{0}^{x_2} + \delta_{0}^{x_1}\delta_{1}^{x_2} + \delta_{1}^{x_1}\delta_{0}^{x_2} + \delta_{1}^{x_1}\delta_{1}^{x_2} \right], \\
    p(x_1x_3\mid X_1X_3) &= \frac{1}{4} \left[ \delta_{0}^{x_1}\delta_{0}^{x_3} + \delta_{0}^{x_1}\delta_{1}^{x_3} + \delta_{1}^{x_1}\delta_{0}^{x_3} + \delta_{1}^{x_1}\delta_{1}^{x_3} \right], \\
    p(x_2x_3\mid X_2X_3) &= \frac{1}{4} \left[ \delta_{0}^{x_2}\delta_{0}^{x_3} + \delta_{0}^{x_2}\delta_{1}^{x_3} + \delta_{1}^{x_2}\delta_{0}^{x_3} + \delta_{1}^{x_2}\delta_{1}^{x_3} \right],
\end{split}
\label{eq:ghz-marginals}
\end{equation}
where $\delta_a^b$ is the Kronecker delta. Since all bipartite marginals are maximally mixed and the contextual fraction is $\text{CF}=1$, because the subprobability distribution $\mathbf{b}$ after optimization is a null vector, hence $\text{NCF} = \mathbf{1.b}=0$, implying $CF=0$, which is also evident from the AVN type proof \cite{Mermin_argument, cohom_AVN_abramsky_et_al2015}, the GHZ correlation constitutes an AMCC. While in the $(2, 2, 2)$ scenario for Bell correlations, $\mathbf{b} = [0.125, 0.125, 0, 0, 0.125, 0, 0, 0, 0, 0, 0, 0.125, 0, 0, 0.125, 0.125]$ is a column vector, such that the $\text{NCF}=\mathbf{1}^\top.\mathbf{b}=0.75$ shows that Bell correlations are not maximally contextual. Furthermore, in the three-qubit case, the GHZ state is the only quantum state that exhibits maximal contextual correlations, which corresponds only to equatorial measurements. This has been proved in \cite{min_quntun_resources_abramsky2017}.

The three-way nonlocal correlations are given by:
\begin{equation}
    p(x_1x_2x_3\mid X_1X_2X_3) =
\begin{cases} 
    1/4, & \text{if } x_1 \oplus x_2 \oplus x_3 = X_1X_2X_3, \\
    0, & \text{otherwise.}
\end{cases}
\label{eq:three-way-nl}
\end{equation}
These correlations are maximally mixed, as can be verified by direct calculation of the marginals. Using the linear optimization method described earlier, we computed the CF for this distribution and obtained $\mathrm{CF}=1$. We also verified that the other three-way nonlocal correlations listed in Ref.~\cite{barrett2005nonlocal} are likewise AMCCs.

There exist cases in the $(3,2,2)$-Bell scenario that are maximally contextual but do not qualify as AMCCs, as they fail to satisfy the maximal marginals condition. For instance, when two parties share a PR box, and the third party obtains $0$ with certainty for both measurements \cite{Extremal_points_Pironio_2011}. Each context has two supports (nonzero values), and each support takes a value $1/2$. These correlations are maximally contextual but do not satisfy the maximal marginal condition. Moreover, the correlation is not a genuinely multipartite correlation. We further consider the correlations presented in Table~\ref{tab: non_AMCC_genuine}: their bipartite marginals are not uniformly random. This correlation is a genuine multipartite maximal nonlocal correlation, as it violates the Svetlichny inequality \cite{Extremal_points_Pironio_2011} by $7.5$, which is discussed in Sec~\ref{subsec: construction_AMCC_and_non_AMCC}.
\begin{widetext}
    \begin{table}[H]
\centering
\begin{tabularx}{\textwidth}{|>{\raggedright\arraybackslash}X|*{8}{>{\centering\arraybackslash}X|}}
\hline
Context and Section & $(0,0,0)$ & $(0,0,1)$ & $(0,1,0)$ & $(0,1,1)$ & $(1,0,0)$ & $(1,0,1)$ & $(1,1,0)$ & $(1,1,1)$  \\ \hline

$(0, 0, 0)$
 & $0$
 & $\tfrac{1}{4}$
 & $\tfrac{1}{4}$
 & $0$
 & $\tfrac{1}{4}$
 & $0$
 & $0$
 & $\tfrac{1}{4}$
\\ \hline

$(0, 0, 1)$
 & $\tfrac{1}{20}$
 & $\tfrac{1}{5}$
 & $\tfrac{1}{5}$
 & $\tfrac{1}{20}$
 & $\tfrac{1}{4}$
 & $0$
 & $0$
 & $\tfrac{1}{4}$
\\ \hline

$(0,1,0)$
   & $\tfrac{1}{20}$
 & $\tfrac{1}{5}$
 & $\tfrac{1}{5}$
 & $\tfrac{1}{20}$
 & $\tfrac{1}{4}$
 & $0$
 & $0$
 & $\tfrac{1}{4}$
\\ \hline

$(0, 1, 1)$
 & $\tfrac{1}{4}$
 & $0$
 & $0$
 & $\tfrac{1}{4}$
 & $0$
 & $\tfrac{1}{4}$
 & $\tfrac{1}{4}$
 & $0$
\\ \hline

$(1, 0, 0)$
 & $\tfrac{1}{20}$
 & $\tfrac{1}{4}$
 & $\tfrac{1}{4}$
 & $\tfrac{1}{20}$
 & $\tfrac{1}{5}$
 & $0$
 & $0$
 & $\tfrac{1}{5}$
\\ \hline

$(1, 0, 1)$
& $\tfrac{3}{10}$
 & $0$
 & $0$
 & $\tfrac{3}{10}$
 & $0$
 & $\tfrac{1}{5}$
 & $\tfrac{1}{5}$
 & $0$
\\ \hline

$(1, 1, 0)$
 & $\tfrac{3}{10}$
 & $0$
 & $0$
 & $\tfrac{3}{10}$
 & $0$
 & $\tfrac{1}{5}$
 & $\tfrac{1}{5}$
 & $0$
\\ \hline

$(1, 1, 1)$
 & $\tfrac{1}{4}$
 & $\tfrac{1}{20}$
 & $\tfrac{1}{20}$
 & $\tfrac{1}{4}$
 & $0$
 & $\tfrac{1}{5}$
 & $\tfrac{1}{5}$
 & $0$
\\ \hline

\end{tabularx}
\caption{Strongly contextual table for (3,2,2)- scenario but not AMCC.}
\label{tab: non_AMCC_genuine}
\end{table}
\end{widetext}

 We present additional instances of AMCC for the $(3,2,2)$ scenario here. We provide a general construction and will extend it to broader families in the next section. We begin by considering a parametrized family of no-signaling $(3,2,2)$ Bell scenarios that satisfy the normalization condition. The corresponding polytope has dimension 26 (reflecting 26 independent parameters), as detailed in Appendix~\ref{app: reduced table}. After further parameter reduction, the resulting parametric table is shown in Table~\ref{tab: reduced 8parameters ptable},
\begin{widetext}
    \begin{table}[H]
\centering
\begin{tabularx}{\textwidth}{|>{\raggedright\arraybackslash}X|*{8}{>{\centering\arraybackslash}X|}}
\hline
Context and Section & $(0,0,0)$ & $(0,0,1)$ & $(0,1,0)$ & $(0,1,1)$ & $(1,0,0)$ & $(1,0,1)$ & $(1,1,0)$ & $(1,1,1)$ \\ \hline

$(0, 0, 0)$
 & $p_1$
 & $\tfrac{1}{4} - p_1$
 & $\tfrac{1}{4} - p_1$
 & $p_1$
 & $\tfrac{1}{4} - p_1$
 & $p_1$
 & $p_1$
 & $\tfrac{1}{4} - p_1$
\\ \hline

$(0, 0, 1)$
 & $p_2$
 & $\tfrac{1}{4} - p_2$
 & $\tfrac{1}{4} - p_2$
 & $p_2$
 & $\tfrac{1}{4} - p_2$
 & $p_2$
 & $p_2$
 & $\tfrac{1}{4} - p_2$
\\ \hline

$(0, 1, 0)$
 & $p_{3}$
 & $\tfrac{1}{4} - p_{3}$
 & $\tfrac{1}{4} - p_{3}$
 & $p_{3}$
 & $\tfrac{1}{4} - p_{3}$
 & $p_{3}$
 & $p_{3}$
 & $\tfrac{1}{4} - p_{3}$
\\ \hline

$(0, 1, 1)$
 & $p_{4}$
 & $\tfrac{1}{4} - p_{4}$
 & $\tfrac{1}{4} - p_{4}$
 & $p_{4}$
 & $\tfrac{1}{4} - p_{4}$
 & $p_{4}$
 & $p_{4}$
 & $\tfrac{1}{4} - p_{4}$
\\ \hline

$(1, 0, 0)$
 & $p_{5}$
 & $\tfrac{1}{4} - p_{5}$
 & $\tfrac{1}{4} - p_{5}$
 & $p_{5}$
 & $\tfrac{1}{4} - p_{5}$
 & $p_{5}$
 & $p_{5}$
 & $\tfrac{1}{4} - p_{5}$
\\ \hline

$(1, 0, 1)$
 & $p_{6}$
 & $\tfrac{1}{4} - p_{6}$
 & $\tfrac{1}{4} - p_{6}$
 & $p_{6}$
 & $\tfrac{1}{4} - p_{6}$
 & $p_{6}$
 & $p_{6}$
 & $\tfrac{1}{4} - p_{6}$
\\ \hline

$(1, 1, 0)$
 & $p_{7}$
 & $\tfrac{1}{4} - p_{7}$
 & $\tfrac{1}{4} - p_{7}$
 & $p_{7}$
 & $\tfrac{1}{4} - p_{7}$
 & $p_{7}$
 & $p_{7}$
 & $\tfrac{1}{4} - p_{7}$
\\ \hline

$(1, 1, 1)$
 & $p_{8}$
 & $\tfrac{1}{4} - p_{8}$
 & $\tfrac{1}{4} - p_{8}$
 & $p_{8}$
 & $\tfrac{1}{4} - p_{8}$
 & $p_{8}$
 & $p_{8}$
 & $\tfrac{1}{4} - p_{8}$
\\ \hline

\end{tabularx}
\caption{Reduced (eight free parameters) probability table.}
\label{tab: reduced 8parameters ptable}
\end{table}
\end{widetext}
which now contains only eight free parameters. These parameters can only take values in $[0,1/4]$ to preserve the non-negativity of the probabilities. The empirical model corresponding to this table has maximal marginals for all parameter choices.

\begin{figure}[t]
    \centering
    \includegraphics[width=1\linewidth]{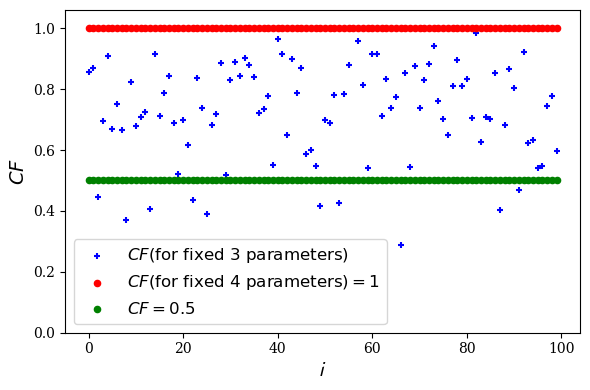}
    \caption{The contextual fraction (CF) is evaluated over $100$ random choices of $[p_1, p_2, p_3, p_4, p_5, p_6, p_7, p_8]$. }
    \label{fig: max_context_varification}
\end{figure}

We examine which parameter conditions yield a CF of $1$ using the optimization method described earlier. When we fix two of the eight parameters to $0.125$ or $0.25$ and set the remaining parameters to $0$, the possible combinations of these assignments lead to exactly three possible CF values: $\text{CF}\in \{0, 0.5, 1\}$. To search for the parametric condition that leads to $\text{CF}=1$, we fix the first few parameters to the specific values in the set $\{0, 0.125, 0.25\}$, while the remaining are allowed to take any values in $[0,0.25]$. Allowing more than four parameters to take any values in $[0,0.25]$ does not always lead to $\text{CF}=1$, whereas for four and fewer parameters, taking values in $[0, 0.25]$ yields $\text{CF}=1$ by fixing the remaining parameters to some specific values in $\{0, 0.125, 0.25\}$. This is demonstrated in Fig ~\ref{fig: max_context_varification}. The $\text{CF}$ value is shown with blue markers corresponding to $p1=0.25, p_2=0, p_3=0$ and $p_4, p_5, p_6, p_7, p_8 \in [0, 0.25]$, while red dots are for $p1=0.25, p_2=0, p_3=0, p_4=0$ and $p_5, p_6, p_7, p_8 \in [0, 0.25]$, which yield $\text{CF}=1$, and the green dots for $\text{CF=0.5}$ are evaluated on the choices $p_1=0.125, p_2=p_3=p_7=p_8=0$ and $p_4, p_5, p_6\in[0,0.125]$. The $i$ is the indexing of the $100$ random choices $[p_1, p_2, p_3, p_4, p_5, p_6, p_7, p_8]$.

 We generalize these results in the next section using the parity-check construction and the CSP approach.

\section{The Construction of AMC and non-AMC Correlations}
The evidence provided in the previous section indicates that there exist maximally contextual correlations (MCCs) that are maximally contextual but not maximally marginal (see Table~\ref{tab: non_AMCC_genuine}). Moreover, from the parametric Table~\ref{tab: reduced 8parameters ptable}, we have observed several examples of AMCCs by conditioning on the parameters. In this section, we present systematic constructions that generate both AMCC and non-AMCC \footnote{The correlation with the contextual fraction $\mathrm{CF}=1$ but not the maximal marginal} models in a unified framework. We also provide proofs for certain properties of the empirical models introduced in the previous section. The AME construction \cite{AME_construction, AME_construction_PPT, AME_construction_combi} is understood in several ways, but still, there is no general method of AME construction for any number of parties and any local dimension.

\subsection{Constraint satisfaction problem model}
To search for new classes of AMCCs, we must first understand the structural properties of correlations. A useful approach is to examine the possibilistic form of correlations, also known as logical contextual models \cite{santos2021conditions}. Given a scenario $\langle \mathcal{X}, \mathcal{M}, \mathcal{O} \rangle$, a distribution over the Boolean semiring $\mathbb{B}=(\{0,1\}, \vee, \wedge, \neg)$ is called a \emph{possibilistic distribution}. That is, a mapping $\bar{\mathcal{P}} : \mathcal{E} \to \mathbb{B}$ with $\sum_{x \in \mathcal{E}} \bar{\mathcal{P}}(x) = 1$ defines an element of $D_{\mathbb{B}}(\mathcal{E})$. For any probabilistic empirical model, the corresponding possibilistic collapse is obtained by taking the support:
\begin{equation}
    \bar{e}_C(s) =
\begin{cases} 
    1, & \text{if } e_C(s) > 0, \\
    0, & \text{otherwise},
\end{cases}
\qquad \forall C \in \mathcal{M}.
\end{equation}
The set of global sections compatible with the support of a probabilistic empirical model $e = \{e_C\}_{C\in\mathcal{M}}$ is
\begin{equation}
    S_e = \{\, s \in \mathcal{E}(\mathcal{X}) \mid \forall C \in \mathcal{M},\; s|_C \in \operatorname{supp}(e_C) \,\}.
\end{equation}
The model $e$ is strongly contextual if and only if $S_e = \varnothing$. Equivalently, there exists no global distribution $d \in D_{\mathbb{B}}(\mathcal{X})$ such that $d|_C \in \operatorname{supp}(e_C)$ for all $C \in \mathcal{M}$. From Definition~(\ref{def: Strong contextuality}), strong contextuality coincides with maximal contextuality.

This characterization can be mapped to a CSP, which for dichotomic measurements reduces to a Boolean satisfiability problem (SAT). Given an empirical model $e$, we associate a set of Boolean variables $\bar{\mathcal{X}}$ corresponding to the observables in the probabilistic model. For each context $C \in \mathcal{M}$ and each section $s \in S_e(C)$ in the support, we define a Boolean proposition. The overall propositional formula for the model is
\begin{equation}
    B_{e} \;:=\; \bigwedge_{C \in \mathcal{M}} 
   \Bigl( \; \bigvee_{s \in S_{e}(C)} b_{s} \;\Bigr),
   \label{eq: Be}
\end{equation}
where each $b_s$ is a Boolean statement encoding the assignment prescribed by the section $s$:
\begin{equation}
    b_{s} \;:=\;
\Bigl( \; \bigwedge_{\substack{X \in C \\ s(X) = 1}} X \;\Bigr)
\;\wedge\;
\Bigl( \; \bigwedge_{\substack{X \in C \\ s(X) = 0}} \neg X \;\Bigr).
\end{equation}
Thus $B_e$ is a conjunction over contexts of the disjunctions of all supported local assignments.

As an illustration, consider the $(2,2,2)$ PR‑box empirical model Eq.~(\ref{eq:pr-box-delta}). Its possibilistic collapse is
\begin{equation}
    \begin{array}{c|cccc}
    X_1X_2\backslash x_1x_2 & (0,0) & (0,1) & (1,0) & (1,1) \\ \hline
    (0,0) & 1 & 0 & 0 & 1 \\
    (0,1) & 1 & 0 & 0 & 1 \\
    (1,0) & 1 & 0 & 0 & 1 \\
    (1,1) & 0 & 1 & 1 & 0
    \end{array}
    \label{table:poss-PR}
\end{equation}
which yields the Boolean formulas
\begin{equation}
\begin{aligned}
B_{1} &= (X_1 \wedge X_2) \;\vee\; (\neg X_1 \wedge \neg X_2), \\
B_{2} &= (X_1 \wedge X_2') \;\vee\; (\neg X_1 \wedge \neg X_2'), \\
B_{3} &= (X_1' \wedge X_2) \;\vee\; (\neg X_1' \wedge \neg X_2), \\
B_{4} &= (\neg X_1' \wedge X_2') \;\vee\; (X_1' \wedge \neg X_2').
\end{aligned}
\label{eq:PR-boolean}
\end{equation}
The first three formulas impose correlations ($x_1 = x_2$, $x_1 = x_2'$, $x_1' = x_2$), while the fourth imposes an anticorrelation ($x_1' \neq x_2'$). This set of conditions is logically inconsistent: the first three together imply $x_1' = x_2'$, which contradicts the fourth condition. Thus, no assignment in the set $\{(x_1, x_2, x_1', x_2') \mid x_1, x_2, x_1', x_2' \in \{0,1\}\}$ simultaneously satisfies all four Boolean formulas, implying $S_e = \varnothing$. Hence, the model is strongly contextual; consequently, $\mathrm{CF}=1$.

We now introduce a construction that produces only symmetric AMCCs. By symmetric, we mean that in each row of the probability table, exactly four entries are $0$ and the remaining four are $1/4$.\\

\subsection{Parity check}
This mathematical construction is based on Mermin's type of proof \cite{Mermin_argument} and generates only symmetric AMCCs.
In this construction, we consider a set of equations, each involving a modulo~$2$ sum of observables. Each equation takes the form $\sum X = P \;(\mathrm{mod}\;2)$, where $X\in\mathcal{X}$ and the parity variable $P\in\{0,1\}$. The equation is of even parity if $P=0$ and odd parity if $P=1$. For example, in the $(2,2,2)$ scenario, the equations are:
\begin{equation}
\begin{aligned}
 X_1 + X_2 &= P_1 \quad (\mathrm{mod}\;2),\\
X_1 + X_2' &= P_2 \quad (\mathrm{mod}\;2),\\
X_1' + X_2 &= P_3 \quad (\mathrm{mod}\;2),\\
X_1' + X_2' &= P_4 \quad (\mathrm{mod}\;2).
\end{aligned}
\label{eq:parity-2part}
\end{equation}
There are $2^4 = 16$ possible combinations of parity choices $(P_1,P_2,P_3,P_4)\in\{0,1\}^4$. The Boolean statements $b_s$ are selected according to the parity of each equation. Each assignment yields a corresponding set of Boolean formulas $B_i$. 

Consider the specific choices $P_1=P_2=P_3=0$ and $P_4=1$:
\begin{equation}
\begin{aligned}
 X_1 + X_2 &= 0 \quad (\mathrm{mod}\;2),\\
X_1 + X_2' &= 0 \quad (\mathrm{mod}\;2),\\
X_1' + X_2 &= 0 \quad (\mathrm{mod}\;2),\\
X_1' + X_2' &= 1 \quad (\mathrm{mod}\;2).
\end{aligned}
\label{eq:parity-PR}
\end{equation}
{ Upon summing up all the equations, we obtain
\begin{equation}
    2X_1+2X_2+2X_1^\prime+2X_2^\prime \equiv 1 \pmod{2},
    \label{eq: 222 parity check}
\end{equation}
which simplifies to $0 \equiv 1$, a contradiction. Therefore, it follows that no assignment can simultaneously satisfy all four equations. To determine the corresponding Boolean formulas, the choice of Boolean statements is distinguished into two sets: for even parity ($P=0$) the allowed choices are $\{(\neg X_1 \wedge \neg X_2), (X_1 \wedge X_2)\}$, while for odd parity ($P=1$) they are $\{(\neg X_1 \wedge X_2), (X_1 \wedge \neg X_2)\}$. This set of choices is the same for $X_1^\prime$ and $X_2^\prime$. Here, the Boolean formula corresponding to $X_1+X_2=0 \pmod{2}$ is \[B_{1} = (X_1 \wedge X_2) \;\vee\; (\neg X_1 \wedge \neg X_2).\] This mapping is consistent since, for binary valuation, the modulo $2$ operation and the logical OR operation are identical. Here, this parity equation is satisfiable only when both variables are the same. After mapping to the Boolean form, $B_1$ is satisfiable when both Boolean variables are the same. Similarly, the other Boolean formulas, $B_1$, $B_2$, and $B_3$, can be obtained. The resulting set of Boolean formulas corresponds to Eq.~(\ref{eq:PR-boolean}), which describes a strongly contextual model and an AMCC for the $(2,2,2)$ scenario. The logical contradiction persists from the parity equation to the Boolean formula mapping because the reduced equation Eq.~(\ref{eq: 222 parity check}) and the Boolean equation Eq.~(\ref{eq: Be}) are identical. For instance, 
\begin{equation}
    B_e= B_1\;\wedge\;B_2 \;\wedge\; B_3 \;\wedge\; B_4
    \label{eq: reduced Be}
\end{equation}
If any of the Boolean formulas are false, then $B_e$ will be false. Namely, the first parity equation is satisfied by only two assignments $(0,0)$ and $(1,1)$ out of four possible assignments, and the corresponding Boolean formula $B_1$ does not satisfy the last parity equation, as well as the corresponding Boolean formula $B_4$. Hence, the reduced parity equation Eq.~(\ref{eq: 222 parity check}) shows the contradiction as well as the corresponding Boolean equation Eq.~(\ref{eq: reduced Be}). Accordingly, if no assignment $(x1, x2, x1', x2') \in \{0,1\}^4$ satisfies all parity equations simultaneously, the associated Boolean formulas are unsatisfiable. Consequently, the empirical model exhibits strong contextuality. The possibilistic table corresponding to the Boolean equations associated with parity equations consists of zeros that depict the absent Boolean terms in the set of Boolean equations. While substituting these zeros into the parametric table with eight parameters, the probability table can be obtained under the normalization and no-signaling conditions, yielding maximally contextual correlations with certain bounds on the parameters, to ensure positivity, ensuring the maximal marginality by the symmetry of the correlations certainly leads to each nonzero entry as $1/2$.

Since the model is strongly contextual, the corresponding probabilistic model is also maximally contextual, provided the parameter bounds are satisfied. Furthermore, maximal marginality is ensured by the symmetric structure of the possibilistic model. Thus, the resulting correlations constitute an AMCC.

For the $(3,2,2)$ scenario, let $X\in \{X_1,X_2,X_3,X_1',X_2',X_3'\}$ with each $X\in\{0,1\}$. The parity equations are:
\begin{equation}
\begin{aligned}
 X_1 + X_2 + X_3 &= P_1 \quad (\mathrm{mod}\;2),\\
X_1 + X_2 + X_3' &= P_2 \quad (\mathrm{mod}\;2),\\
X_1 + X_2' + X_3 &= P_3 \quad (\mathrm{mod}\;2),\\
X_1 + X_2' + X_3' &= P_4 \quad (\mathrm{mod}\;2),\\
 X_1' + X_2 + X_3 &= P_5 \quad (\mathrm{mod}\;2),\\
  X_1' + X_2 + X_3' &= P_6 \quad (\mathrm{mod}\;2),\\
  X_1' + X_2' + X_3 &= P_7 \quad (\mathrm{mod}\;2),\\
  X_1' + X_2' + X_3' &= P_8 \quad (\mathrm{mod}\;2).
\end{aligned}
\label{eq:parity-3part}
\end{equation}
There are $2^8$ possible parity assignments. The set of even‑parity Boolean statements (for three variables) is $
\{(\neg X_1 \wedge \neg X_2 \wedge \neg X_3), (\neg X_1 \wedge X_2 \wedge X_3), (X_1 \wedge \neg X_2 \wedge X_3), (X_1 \wedge X_2 \wedge \neg X_3)\}, $ and the set of odd‑parity statements is$ \{(\neg X_1 \wedge \neg X_2 \wedge X_3), (\neg X_1 \wedge X_2 \wedge \neg X_3), (X_1 \wedge \neg X_2 \wedge \neg X_3), (X_1 \wedge X_2 \wedge X_3)\}$, same follows for Boolean variables $X_1^\prime$, $ X_2^\prime$, and $X_3^\prime$. 
As an illustration, consider the model in Table~\ref{tab: reduced 8parameters ptable} with parameter $p_1=0.25$ and all others set to $0$, which yields $\mathrm{CF}=1$. This corresponds to the parity assignment $P_1=0$ and $P_2=P_3=P_4=P_5=P_6=P_7=P_8=1$. The equations become:
\begin{equation}
\begin{aligned}
 X_1 + X_2 + X_3 &= 0 \quad (\mathrm{mod}\;2),\\
X_1 + X_2 + X_3' &= 1 \quad (\mathrm{mod}\;2),\\
X_1 + X_2' + X_3 &= 1 \quad (\mathrm{mod}\;2),\\
X_1 + X_2' + X_3' &= 1 \quad (\mathrm{mod}\;2),\\
 X_1' + X_2 + X_3 &= 1 \quad (\mathrm{mod}\;2),\\
  X_1' + X_2 + X_3' &= 1 \quad (\mathrm{mod}\;2),\\
  X_1' + X_2' + X_3 &= 1 \quad (\mathrm{mod}\;2),\\
  X_1' + X_2' + X_3' &= 1 \quad (\mathrm{mod}\;2).
\end{aligned}
\label{eq:parity-example}
\end{equation}
Summing all eight equations modulo~2 gives
\begin{equation}
  4X_1 + 4X_1' + 4X_2 + 4X_2' + 4X_3 + 4X_3' \equiv 7 \pmod{2}, 
  \label{eq: 322 reduced parity}
\end{equation}
which reduces to $0 \equiv 1 \pmod{2}$, a logical contradiction. The corresponding Boolean equation
\begin{equation}
    B_e = B_1\;\wedge \cdots\wedge\; B_8,
    \label{eq: 322 reduced Be}
\end{equation}
where $B_1,\cdots, B_8$ are all Boolean formulas constructed from Eq.~(\ref{eq:parity-example}). Namely, the first parity equation 
\[ X_1 + X_2 + X_3 = 0 \pmod{2},\]
can be mapped to the Boolean formula
\[
\begin{aligned}
B_{1} =\;& (\neg X_1 \wedge \neg X_2 \wedge \neg X_3)\;\vee\;
(\neg X_1 \wedge X_2 \wedge X_3) \\
&\vee\; (X_1 \wedge \neg X_2 \wedge X_3)\;\vee\;
(X_1 \wedge X_2 \wedge \neg X_3)\,.
\end{aligned}
\]
The construction relies on the fact that the parity equation is satisfied by only four assignments $(0,0,0), (0,1,1), (1,0,1)$ and $(1,1,0)$ out of a total of eight possible assignments. Accordingly, the corresponding Boolean formula is satisfiable only for these assignments. The persistence of the contradiction from the parity equations to the Boolean formulas can be demonstrated as follows. Consider the first and second parity equations. The second parity equation, \[ X_1 + X_2 + X_3^\prime = 1 \pmod{2}\] implies that $(X_1 + X_2 \neq X_3^\prime \pmod{2})$, whereas from the first equation (i.e., $X_1 + X_2 + X_3 = 0$) we get $(X_1 + X_2 = X_3^\prime \pmod{2})$. Now, from both conditions, we have $X_3\neq X_3^{\prime}$. Similarly, if we take equations $X_1^\prime + X_2^\prime + X_3 = 1 \pmod{2}$ and $X_1^\prime + X_2^\prime + X_3^\prime = 1 \pmod{2}$, then we get $X_3=X_3^\prime$, which is a contradiction. Specifically, for the assignments that satisfy the first parity equation and, thus, the corresponding Boolean formula, the second parity equation, together with its associated Boolean formula, becomes unsatisfiable. Hence, this leads to the contradiction shown in Eq.~(\ref{eq: 322 reduced parity}), which, in turn, results in the unsatisfiability of the corresponding Boolean equation, Eq.~(\ref{eq: 322 reduced Be}).
Hence, no consistent assignment exists, i.e., $S_e = \varnothing$ for the possibilistic model. Therefore, the empirical model is strongly contextual. The possibilistic table associated with the set of Boolean equations consists of zeros corresponding to the absent Boolean terms in the Boolean formulas. Substituting these zeros in Table~\ref{tab: 26_parameter_table} having $26$ free parameters, will provide the parametric equations equal to zero. Solving these equations and substituting the solutions back into the table yielded a reduced parametric table that depicts the maximal contextual correlations associated with the maximal contextual model under the parametric bound, thereby maintaining positivity. Thereafter, implying the maximal marginal condition yields a further reduced parametric family of correlations, which will be AMCC under certain parametric bound conditions. This construction is further discussed in detail in the next section, along with the CSP construction.

\subsection{Construction of AMCC and non-AMCC}
\label{subsec: construction_AMCC_and_non_AMCC}
In this section, we construct both AMCC (symmetric and asymmetric) and non‑AMCC models using the parity‑check construction and the CSP approach. We begin with the parity‑check construction. Depending on the chosen parity values, we select the corresponding Boolean statements and assemble the resulting Boolean formula. We then test whether the set of parity equations is simultaneously unsatisfiable for all possible assignments. If the parity equations are unsatisfiable, then the associated Boolean formulas are also unsatisfiable for every assignment; as a result, the corresponding possibilistic model is strongly contextual.

Numerical analysis shows that, out of the $256$ possible parity assignments, exactly $240$ give rise to AMCCs, provided the construction procedure discussed above. As an example, consider the parity choice $P_1=P_2=P_3=P_4=P_5=P_6=P_7=0,\qquad P_8=1.$
The corresponding Boolean formulas are
\begin{widetext}
    \begin{equation}
\begin{aligned}
B_{1} &= (\neg X_1 \wedge \neg X_2 \wedge \neg X_3) \;\vee\; (\neg X_1 \wedge X_2 \wedge X_3) \;\vee\; (X_1 \wedge \neg X_2 \wedge X_3) \;\vee\; (X_1 \wedge X_2 \wedge \neg X_3) \;, \\
B_{2} &= (\neg X_1 \wedge \neg X_2 \wedge \neg X_3') \;\vee\; (\neg X_1 \wedge X_2 \wedge X_3') \;\vee\; (X_1 \wedge \neg X_2 \wedge X_3') \;\vee\; (X_1 \wedge X_2 \wedge \neg X_3') \;, \\
B_{3} &= (\neg X_1 \wedge \neg X_2' \wedge \neg X_3) \;\vee\; (\neg X_1 \wedge X_2' \wedge X_3) \;\vee\; (X_1 \wedge \neg X_2' \wedge X_3) \;\vee\; (X_1 \wedge X_2' \wedge \neg X_3) \;, \\
B_{4} &= (\neg X_1 \wedge \neg X_2' \wedge \neg X_3') \;\vee\; (\neg X_1 \wedge X_2' \wedge X_3') \;\vee\; (X_1 \wedge \neg X_2' \wedge X_3') \;\vee\; (X_1 \wedge X_2' \wedge \neg X_3') \;, \\ 
B_{5} &= (\neg X_1' \wedge \neg X_2 \wedge \neg X_3) \;\vee\; (\neg X_1' \wedge X_2 \wedge X_3) \;\vee\; (X_1' \wedge \neg X_2 \wedge X_3) \;\vee\; (X_1' \wedge X_2 \wedge \neg X_3) \;, \\
B_{6} &= (\neg X_1' \wedge \neg X_2 \wedge \neg X_3') \;\vee\; (\neg X_1' \wedge X_2 \wedge X_3') \;\vee\; (X_1' \wedge \neg X_2 \wedge X_3') \;\vee\; (X_1' \wedge X_2 \wedge \neg X_3') \;, \\
B_{7} &= (\neg X_1' \wedge \neg X_2' \wedge \neg X_3) \;\vee\; (\neg X_1' \wedge X_2' \wedge X_3) \;\vee\; (X_1' \wedge \neg X_2' \wedge X_3) \;\vee\; (X_1' \wedge X_2' \wedge \neg X_3) \;, \\
B_{8} &= (\neg X_1' \wedge \neg X_2' \wedge X_3') \;\vee\; (\neg X_1' \wedge X_2' \wedge \neg X_3') \;\vee\; (X_1' \wedge \neg X_2' \wedge \neg X_3') \;\vee\; (X_1' \wedge X_2' \wedge X_3') \;, \\ 
\end{aligned}
\label{eq:symmetric_equn_AMCC}
\end{equation}
\end{widetext}
This set of Boolean equations can be mapped to the possibilistic table consisting of four zeros in each row corresponding to the missing Boolean terms. Substituting it in Table~\ref{tab: 26_parameter_table} directly yields each nonzero entry to $1/4$. In addition, it satisfies the maximally marginal condition. Hence, the correlation corresponding to Eq.(\ref{eq:symmetric_equn_AMCC}), which is associated with the parity choices $P_1=P_2=P_3=P_4=P_5=P_6=P_7=0,\qquad P_8=1$, exhibits AMCC. While out of $240$ unsatisfiable set of parity equations, $192$ directly yield AMCC, after following the construction procedure, while $48$ among them leads to one-parametric-family non-AMCC under the parametric bound $p\in[0,1/4]$. However, the maximal marginal condition ensures that these correlations become AMCC only for the parametric condition $p=1/4$. Consequently, the parity check construction is not just the mere search for AMCC. Instead, it provides a procedure that ensures that the symmetric structure of the Boolean equations constructed using parity equations easily exhibits the AMCC.

In contrast, Tables~(\ref{tab: non_AMCC_genuine}) is asymmetric. We now turn to non‑AMCC and asymmetric empirical models, which, under appropriate conditions, lead to the asymmetric AMCC. To construct such models, we adopt a more general approach by freely selecting Boolean statements $b_s$. If we consider all possible combinations of Boolean statements across the eight contexts, there are $255^8$ possibilities (since each context can contain any nonempty subset of its eight possible local assignments). Previously, the parity condition provided a simple rule for selecting statements, and the number of allowed combinations was small. Now, scanning all combinations by brute force is intractable.

To overcome this difficulty, we start by fixing one Boolean statement in each context and then systematically varying the remaining choices. In particular, we work with the following set of Boolean formulas (constraints):
\begin{widetext}
    \begin{equation}
\begin{aligned}
B_{1} &= (\neg X_1 \wedge \neg X_2 \wedge X_3) \;\vee\; (\neg X_1 \wedge X_2 \wedge \neg X_3) \;\vee\; (X_1 \wedge \neg X_2 \wedge \neg X_3) \;\vee\; (X_1 \wedge X_2 \wedge X_3) \;, \\
B_{2} &= (\neg X_1 \wedge \neg X_2 \wedge \neg X_3') \;\vee\; (\neg X_1 \wedge \neg X_2 \wedge X_3') \;\vee\; (\neg X_1 \wedge X_2 \wedge \neg X_3') \;\vee\; (\neg X_1 \wedge X_2 \wedge X_3') \;, \\
B_{3} &= (\neg X_1 \wedge \neg X_2' \wedge \neg X_3) \;\vee\; (\neg X_1 \wedge \neg X_2' \wedge X_3) \;\vee\; (\neg X_1 \wedge X_2' \wedge \neg X_3) \;\vee\; (\neg X_1 \wedge X_2' \wedge X_3) \;, \\
B_{4} &= (\neg X_1 \wedge \neg X_2' \wedge \neg X_3') \;\vee\; (\neg X_1 \wedge X_2' \wedge X_3') \;\vee\; (X_1 \wedge \neg X_2' \wedge X_3') \;\vee\; (X_1 \wedge X_2' \wedge \neg X_3') \;, \\ 
B_{5} &= (\neg X_1' \wedge \neg X_2 \wedge \neg X_3) \;\vee\; (\neg X_1' \wedge \neg X_2 \wedge X_3) \;\vee\; (\neg X_1' \wedge X_2 \wedge \neg X_3) \;\vee\; (\neg X_1' \wedge X_2 \wedge X_3) \;, \\
B_{6} &= (\neg X_1' \wedge \neg X_2 \wedge \neg X_3') \;\vee\; (\neg X_1' \wedge X_2 \wedge X_3') \;\vee\; (X_1' \wedge \neg X_2 \wedge X_3') \;\vee\; (X_1' \wedge X_2 \wedge \neg X_3') \;, \\
B_{7} &= (\neg X_1' \wedge \neg X_2' \wedge \neg X_3) \;\vee\; (\neg X_1' \wedge X_2' \wedge X_3) \;\vee\; (X_1' \wedge \neg X_2' \wedge X_3) \;\vee\; (X_1' \wedge X_2' \wedge \neg X_3) \;, \\
B_{8} &= (\neg X_1' \wedge \neg X_2' \wedge \neg X_3') \;\vee\; (\neg X_1' \wedge \neg X_2' \wedge X_3') \;\vee\; (\neg X_1' \wedge X_2' \wedge \neg X_3') \;\vee\; (\neg X_1' \wedge X_2' \wedge X_3') \;, \\ 
\end{aligned}
\label{eq:40}
\end{equation}
\end{widetext}

One could in principle start with any set of admissible Boolean formulas; we choose the set above for convenience. For each Boolean formula, there are four remaining terms (local assignments) that can be added. Considering all combinations of these extra terms would yield approximately $4.2 \times 10^9$ possibilities, which is still too large to handle exhaustively.
We therefore reduce the complexity by only adding terms that are \emph{absent} from the formulas for $B_2, B_3, B_5$ and $B_8$. With this restriction, the number of combinations drops to $65,536$. Among these, only $2,401$ Boolean formulas satisfy the Boolean no‑signaling condition (Appendix~\ref{app: Boolean no-signaling}) and are simultaneously unsatisfiable as CSP instances. Consequently, all $2,401$ such sets of Boolean formulas correspond to strongly contextual empirical models. One of them is as follows:
\begin{widetext}
    \begin{equation}
\begin{aligned}
B_1 &= (\neg X_1 \wedge \neg X_2 \wedge X_3) \vee (\neg X_1 \wedge X_2 \wedge \neg X_3)\vee(X_1 \wedge \neg X_2 \wedge \neg X_3) \vee (X_1 \wedge X_2 \wedge X_3), \\
B_2 &= (\neg X_1 \wedge \neg X_2 \wedge \neg X_3') \vee (\neg X_1 \wedge \neg X_2 \wedge X_3') \vee (\neg X_1 \wedge X_2 \wedge \neg X_3') \vee (\neg X_1 \wedge X_2 \wedge X_3')\\ 
&\vee (X_1 \wedge \neg X_2 \wedge \neg X_3') \vee (X_1 \wedge X_2 \wedge X_3'), \\
B_3 &= (\neg X_1 \wedge \neg X_2' \wedge \neg X_3) \vee (\neg X_1 \wedge \neg X_2' \wedge X_3) \vee (\neg X_1 \wedge X_2' \wedge \neg X_3) \vee (\neg X_1 \wedge X_2' \wedge X_3)\\ 
&\vee (X_1 \wedge \neg X_2' \wedge \neg X_3) \vee (X_1 \wedge X_2' \wedge X_3), \\
B_4 &= (\neg X_1 \wedge \neg X_2' \wedge \neg X_3') \vee (\neg X_1 \wedge X_2' \wedge X_3') \vee (X_1 \wedge \neg X_2' \wedge X_3') \vee (X_1 \wedge X_2' \wedge \neg X_3'), \\ 
B_5 &= (\neg X_1' \wedge \neg X_2 \wedge \neg X_3) \vee (\neg X_1' \wedge \neg X_2 \wedge X_3) \vee (\neg X_1' \wedge X_2 \wedge \neg X_3) \vee (\neg X_1' \wedge X_2 \wedge X_3)\\ &\vee (X_1' \wedge \neg X_2 \wedge \neg X_3) \vee (X_1' \wedge X_2 \wedge X_3), \\
B_6 &= (\neg X_1' \wedge \neg X_2 \wedge \neg X_3') \vee (\neg X_1' \wedge X_2 \wedge X_3') \vee (X_1' \wedge \neg X_2 \wedge X_3') \vee (X_1' \wedge X_2 \wedge \neg X_3'), \\
B_7 &= (\neg X_1' \wedge \neg X_2' \wedge \neg X_3) \vee (\neg X_1' \wedge X_2' \wedge X_3) \vee (X_1' \wedge \neg X_2' \wedge X_3) \vee (X_1' \wedge X_2' \wedge \neg X_3), \\
B_8 &= (\neg X_1' \wedge \neg X_2' \wedge \neg X_3') \vee (\neg X_1' \wedge \neg X_2' \wedge X_3') \vee (\neg X_1' \wedge X_2' \wedge \neg X_3') \vee (\neg X_1' \wedge X_2' \wedge X_3')\\ &\vee (X_1' \wedge \neg X_2' \wedge \neg X_3') \vee (X_1' \wedge X_2' \wedge X_3'), \\ 
\end{aligned}
\label{eq:Bln_eqn_CSP_tab3}
\end{equation}
\end{widetext}
Substituting zeros from the possibilistic table corresponding to Eq.(\ref{eq:Bln_eqn_CSP_tab3}) in Table~\ref{tab: 26_parameter_table} gives rise to parametric constraints equal to $0$, for which further solving and substituting back the solution in the same table results in Table~\ref{tab: three_parameter_tab}, which consists of three free parameters.  This reduction to three parameters corresponds only to Eq~(\ref{eq:Bln_eqn_CSP_tab3}); for other possible Boolean sets of equations, it might lead to a different parametric family.
\begin{widetext}
    \begin{table}[H]
    \centering
\small 
\begin{tabularx}{\textwidth}{|>{\raggedright\arraybackslash}X|*{8}{>{\centering\arraybackslash}X |}}
\hline
\textbf{Context and Section} & \textbf{(0,0,0)} & \textbf{(0,0,1)} & \textbf{(0,1,0)} & \textbf{(0,1,1)} & \textbf{(1,0,0)} & \textbf{(1,0,1)} & \textbf{(1,1,0)} & \textbf{(1,1,1)} \\ \hline

{\centering$(0, 0, 0)$\par}
 & {\centering $0$\par}
 & {\centering$p_1$\par}
 & {\centering$p_1$\par}
 & {\centering$0$\par}
 & {\centering$1/2-p_1$\par}
 & {\centering$0$\par}
 & {\centering$0$\par}
 & {\centering$1/2-p_1$\par}
\\ \hline

{\centering$(0, 0, 1)$\par}
 & {\centering$p_2$\par}
 &  {\centering$p_1-p_2$\par}
 & {\centering$p_{3}$\par}
  & {\centering$p_1-p_{3}$\par} 
  & {\centering$1/2-p_1$\par}
  & {\centering$0$\par}
  & {\centering$0$\par}
&{\centering$1/2-p_1$\par}
\\ \hline

{\centering$(0, 1, 0)$\par}
 & {\centering$p_{2}$\par}
 & {\centering$p_{3}$\par}
  & {\centering$p_1-p_{2}$\par} 
  & {\centering$p_1-p_{3}$\par}
  & {\centering$1/2-p_1$\par}
  & {\centering$0$\par} 
  & {\centering$0$\par}
  &{\centering$1/2-p_1$\par}
\\ \hline

{\centering$(0, 1, 1)$\par}
 & {\centering$p_{2}+p_{3}$\par}
 & {\centering$0$\par} 
  & {\centering$0$\par}
  & {\centering$2p_1-p_2-p_{3}$\par}
  & {\centering$0$\par}
  & {\centering$1/2-p_1$\par}
  & {\centering$1/2-p_1$\par}
  &{\centering$0$\par}
\\ \hline

{\centering$(1, 0, 0)$\par}
 & {\centering$1/2-2p_1+p_{2}$\par}
 & {\centering$p_{1}$\par} 
  & {\centering$p_{1}$\par} 
  & {\centering$1/2-p_1-p_{3}$\par}
  & {\centering$p_1-p_{2}$\par}
  & {\centering$0$\par}
  & {\centering$0$\par}
  &{\centering$p_{3}$\par}
\\ \hline

{\centering$(1, 0, 1)$\par}
 & {\centering$1/2-p_1+p_2$\par}
 & {\centering$0$\par}
  & {\centering$0$\par} 
  & {\centering$1/2-p_{3}$\par}
  & {\centering$0$\par}
  & {\centering$p_1-p_2$\par} 
  & {\centering$p_{3}$\par}
&{\centering$0$\par}
\\ \hline

{\centering$(1, 1, 0)$\par}
 & {\centering$1/2-p_1+p_2$\par}
 & {\centering$0$\par}
  & {\centering$0$\par}
  & {\centering$1/2-p_{3}$\par}
  & {\centering$0$\par}
  & {\centering$p_{3}$\par}
  & {\centering$p_1-p_2$\par}
  & {\centering$0$\par}
\\ \hline

{\centering$(1, 1, 1)$\par}
 & {\centering$p_{2}$\par}
 & {\centering$1/2-p_1$\par}
  & {\centering$1/2-p_1$\par}
  & {\centering$p_{1}-p_{3}$\par} 
  & {\centering$p_{3}$\par} 
  & {\centering$0$\par} 
  & {\centering$0$\par}
  & {\centering$p_1-p_{2}$\par}
\\ \hline

\end{tabularx}
\caption{Maximum contextual table $(3,2,2)$-scenario.}
\label{tab: three_parameter_tab}
\end{table}
\end{widetext}
The correlation in Table~\ref{tab: three_parameter_tab} is strongly contextual, provided that the parameters satisfy the bounds}
\begin{equation}
\begin{aligned}
 0 \leq p_1 \leq \tfrac{1}{2}, \quad 
 \max\!\bigl(0,\ 2p_1-\tfrac{1}{2}\bigr) \leq p_2 \leq p_1,\\ \qquad 
0 \leq p_{3} \leq \min\!\bigl(p_1,\ \tfrac{1}{2}-p_1).\\
\end{aligned}
\label{eq:boundp1p2p3}
\end{equation}

This model is a non‑AMCC because it does not satisfy the maximal marginals condition of Definition~\ref{def:max-mixed}.

The same model becomes an AMCC when $p_1 = p_{3} = 1/4$ and $p_2 = 0$. In the parity‑check construction, determining strong contextuality reduces to checking the parity equations, which is computationally straightforward. In contrast, the general CSP‑based construction requires verifying both the no‑signaling condition and the unsatisfiability of the Boolean constraints, adding extra complexity. The CSP construction provides the construction of a strong contextual model that, under specific parameter conditions, leads to AMCC or non-AMCC. For instance, the table \ref{tab: three_parameter_tab} shows the strongly contextual correlation under the conditions in Eq~(\ref{eq:boundp1p2p3}) and is a symmetric AMCC for $p_1 = p_{3} = 1/4$ and $p_2 = 0$. To generate the asymmetric AMCC, among $2401$, we take the following set of Boolean formulas:
\begin{widetext}
    \begin{equation}
\begin{aligned}
B_1 &= (\neg X_1 \wedge \neg X_2 \wedge X_3) \vee (\neg X_1 \wedge X_2 \wedge \neg X_3)\vee(X_1 \wedge \neg X_2 \wedge \neg X_3) \vee (X_1 \wedge X_2 \wedge X_3), \\
B_2 &= (\neg X_1 \wedge \neg X_2 \wedge \neg X_3') \vee (\neg X_1 \wedge \neg X_2 \wedge X_3') \vee (\neg X_1 \wedge X_2 \wedge \neg X_3') \vee (\neg X_1 \wedge X_2 \wedge X_3')\\ 
&\vee (X_1 \wedge \neg X_2 \wedge \neg X_3') \vee (X_1 \wedge X_2 \wedge X_3'), \\
B_3 &= (\neg X_1 \wedge \neg X_2' \wedge \neg X_3) \vee (\neg X_1 \wedge \neg X_2' \wedge X_3) \vee (\neg X_1 \wedge X_2' \wedge \neg X_3) \vee (\neg X_1 \wedge X_2' \wedge X_3)\\ 
&\vee (X_1 \wedge \neg X_2' \wedge \neg X_3) \vee (X_1 \wedge X_2' \wedge X_3), \\
B_4 &= (\neg X_1 \wedge \neg X_2' \wedge \neg X_3') \vee (\neg X_1 \wedge X_2' \wedge X_3') \vee (X_1 \wedge \neg X_2' \wedge X_3') \vee (X_1 \wedge X_2' \wedge \neg X_3'), \\ 
B_5 &= (\neg X_1' \wedge \neg X_2 \wedge \neg X_3) \vee (\neg X_1' \wedge \neg X_2 \wedge X_3) \vee (\neg X_1' \wedge X_2 \wedge \neg X_3) \vee (\neg X_1' \wedge X_2 \wedge X_3)\\ &\vee (X_1' \wedge \neg X_2 \wedge \neg X_3) \vee (X_1' \wedge \neg X_2 \wedge X_3) \vee (X_1' \wedge X_2 \wedge \neg X_3) \vee (X_1' \wedge X_2 \wedge X_3), \\
B_6 &= (\neg X_1' \wedge \neg X_2 \wedge \neg X_3') \vee (\neg X_1' \wedge X_2 \wedge X_3') \vee (X_1' \wedge \neg X_2 \wedge X_3') \vee (X_1' \wedge X_2 \wedge \neg X_3'), \\
B_7 &= (\neg X_1' \wedge \neg X_2' \wedge \neg X_3) \vee (\neg X_1' \wedge X_2' \wedge X_3) \vee (X_1' \wedge \neg X_2' \wedge X_3) \vee (X_1' \wedge X_2' \wedge \neg X_3), \\
B_8 &= (\neg X_1' \wedge \neg X_2' \wedge \neg X_3') \vee (\neg X_1' \wedge \neg X_2' \wedge X_3') \vee (\neg X_1' \wedge X_2' \wedge \neg X_3') \vee (\neg X_1' \wedge X_2' \wedge X_3')\\ &\vee (X_1' \wedge \neg X_2' \wedge \neg X_3') \vee (X_1' \wedge \neg X_2' \wedge X_3') \vee (X_1' \wedge X_2' \wedge \neg X_3') \vee (X_1' \wedge X_2' \wedge X_3'). \\ 
\end{aligned}
\label{eq:maxcontextualmodel}
\end{equation}
\end{widetext}
The Boolean equations satisfy the Boolean nosignaling condition (Appendix~\ref{app: Boolean no-signaling}). Substituting the zeros obtained from the possibilistic model corresponding to the unsatisfiable Boolean formulas in Eqs~(\ref{eq:maxcontextualmodel}) into the parametric Table~\ref{tab: 26_parameter_table}, and then solving the zero constraints obtained from it follows the maximally contextual Table~\ref{tab:sixparametertable}, which consists of six independent parameters that are bound by the conditions in Eqs.~(\ref{eq: bounds}). Subsequently, by taking the maximal marginal conditions on the parameters, we construct the following parametric table, which consists of two free parameters, 
\begin{widetext}
    \begin{table}[H]
    \centering
\begin{tabularx}{\textwidth}{|>{\raggedright\arraybackslash}X|*{8}{>{\centering\arraybackslash}X |}}
\hline
\textbf{Context and Section} & \textbf{(0,0,0)} & \textbf{(0,0,1)} & \textbf{(0,1,0)} & \textbf{(0,1,1)} & \textbf{(1,0,0)} & \textbf{(1,0,1)} & \textbf{(1,1,0)} & \textbf{(1,1,1)} \\ \hline

{\centering$(0, 0, 0)$\par}
 & {\centering $0$\par}
 & {\centering$\frac{1}{4}$\par}
 & {\centering$\frac{1}{4}$\par}
 & {\centering$0$\par}
 & {\centering$\frac{1}{4}$\par}
 & {\centering$0$\par}
 & {\centering$0$\par}
 & {\centering$\frac{1}{4}$\par}
\\ \hline

{\centering$(0, 0, 1)$\par}
 & {\centering $0$\par}
 & {\centering$\frac{1}{4}$\par}
 & {\centering$\frac{1}{4}$\par}
 & {\centering$0$\par}
 & {\centering$\frac{1}{4}$\par}
 & {\centering$0$\par}
 & {\centering$0$\par}
 & {\centering$\frac{1}{4}$\par}
\\ \hline

{\centering$(0, 1, 0)$\par}
 & {\centering $0$\par}
 & {\centering$\frac{1}{4}$\par}
 & {\centering$\frac{1}{4}$\par}
 & {\centering$0$\par}
 & {\centering$\frac{1}{4}$\par}
 & {\centering$0$\par}
 & {\centering$0$\par}
 & {\centering$\frac{1}{4}$\par}
\\ \hline

{\centering$(0, 1, 1)$\par}
 & {\centering$\frac{1}{4}$\par}
 & {\centering$0$\par} 
  & {\centering$0$\par}
  & {\centering$\frac{1}{4}$\par}
  & {\centering$0$\par}
  & {\centering$\frac{1}{4}$\par}
  & {\centering$\frac{1}{4}$\par}
  &{\centering$0$\par}
\\ \hline

{\centering$(1, 0, 0)$\par}
 & {\centering$ p_1 $\par}
 & {\centering$\frac{1}{4} - p_1$\par} 
  & {\centering$\frac{1}{4} - p_1$\par} 
  & {\centering$p_{1}$\par}
  & {\centering$\frac{1}{4} - p_1$\par}
  & {\centering$p_1$\par}
  & {\centering$p_1$\par}
  &{\centering$\frac{1}{4} - p_1$\par}
\\ \hline

{\centering$(1, 0, 1)$\par}
 & {\centering$\frac{1}{4}$\par}
 & {\centering$0$\par}
  & {\centering$0$\par} 
  & {\centering$\frac{1}{4}$\par}
  & {\centering$0$\par}
  & {\centering$\frac{1}{4}$\par} 
  & {\centering$\frac{1}{4}$\par}
&{\centering$0$\par}
\\ \hline

{\centering$(1, 1, 0)$\par}
 & {\centering$\frac{1}{4}$\par}
 & {\centering$0$\par}
  & {\centering$0$\par}
  & {\centering$\frac{1}{4}$\par}
  & {\centering$0$\par}
  & {\centering$\frac{1}{4}$\par}
  & {\centering$\frac{1}{4}$\par}
  & {\centering$0$\par}
\\ \hline

{\centering$(1, 1, 1)$\par}
 & {\centering$p_{2}$\par}
 & {\centering$\frac{1}{4} - p_2$\par}
 & {\centering$\frac{1}{4} - p_2$\par}
  & {\centering$p_2$\par}
  & {\centering$\frac{1}{4} - p_2$\par} 
  & {\centering$p_2$\par} 
  & {\centering$p_2$\par} 
  & {\centering$\frac{1}{4} - p_2$\par}
\\ \hline

\end{tabularx}
\caption{Asymmetric AMCC for $(3,2,2)$ scenario.}
\label{tab:asymmetricAMCC}
\end{table}
\end{widetext}

Consequently, for parametric values $p_1, p_2 \in [0,\tfrac{1}{4}]$, the resulting correlation table corresponds to an AMCC. Under the bounded parametric conditions, the CSP construction yields both symmetric and asymmetric AMCCs, subject to the no-signaling and maximal marginal constraints. Furthermore, the correlation presented in Table~\ref{tab:asymmetricAMCC} represents a particular instance of the parametric Table~\ref{tab: reduced 8parameters ptable}, which can be obtained by setting $p_1 = p_2 = p_3 = 0$ and $p_4 = p_6 = p_7 = \tfrac{1}{4}$.

Given these correlations, genuine multipartite nonlocality can be tested via the Svetlichny inequality $\lvert E_{\text{Svet}} \rvert \leq 4$ \cite{Extremal_points_Pironio_2011}, where the Svetlichny parameter $E_{\text{Svet}}$ is defined as
\begin{equation}
\begin{aligned}
    E_{\text{Svet}} = -2 \sum_{X_1,X_2,X_3}
    &(-1)^{X_1X_2 + X_2X_3 + X_1X_3}\, p(x_1 \oplus  \\
    &x_2 \oplus x_3 = 1 \mid X_1 X_2 X_3),
\end{aligned}
\end{equation}
and can attain a maximum absolute violation of up to $8$. The correlation specified in Table~(\ref{tab: non_AMCC_genuine}) constitutes a genuine tripartite nonlocal correlation, achieving a violation of the Svetlichny inequality up to $7.2$.

\begin{figure*}[t]
    \centering
    \begin{subfigure}{0.48\textwidth}
        \centering
        \includegraphics[width=\linewidth]{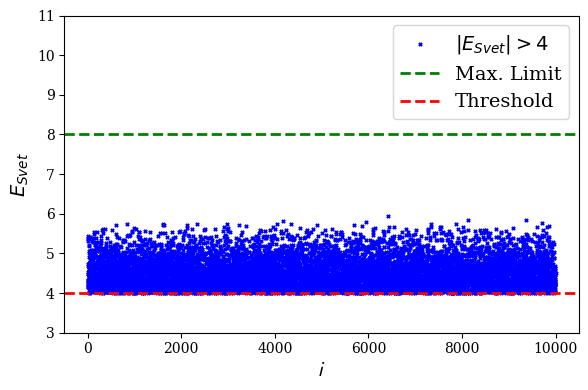}
        \caption{}
    \end{subfigure}
    \hfill
    \begin{subfigure}{0.48\textwidth}
        \centering
        \includegraphics[width=\linewidth]{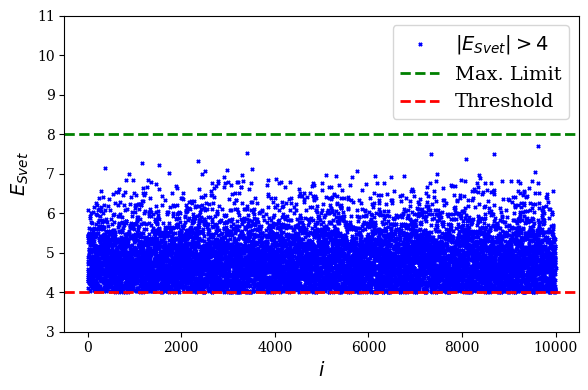}
        \caption{}
    \end{subfigure}
    \caption{The Sevtlichny inequality violation: (a) The Sevtlichny value $E_{Svet}$ evaluated corresponding to $10,000$ random choices of $[p_1, p_2, p_3]$, where $i$ is the associated index for each choice. (b)  The Svetlichny value $E_{\text{Svet}}$ evaluated corresponding to $10,000$ random choices of $[p_1, p_2, p_3, p_4, p_5, p_6]$, where $i$ is the associated index for each choice.}
    \label{fig:E_value_table4and6}
\end{figure*}

The maximal contextual correlations presented in Tables~\ref{tab: three_parameter_tab} and \ref{tab:sixparametertable} demonstrate that random choices of parametric values within the bounds given by Eqs.~(\ref{eq:boundp1p2p3}) and Eqs.~(\ref{eq: bounds}) can also lead to violations of the Svetlichny inequality and thus to genuine tripartite maximally contextual correlations, as illustrated in Fig.~\ref{fig:E_value_table4and6}. For all $10\,000$ random parameter choices considered, the Svetlichny inequality is violated.

\subsection{$(n,2,2)$ GHZ correlations}
In this section, we present a proof that the \((n,2,2)\)-GHZ correlations constitute an AMCC under suitably chosen local measurement settings. Consider the \(n\)-partite GHZ state
\begin{equation}
    \ket{\text{GHZ}}_n=\frac{\ket{00\dots0}+\ket{11\dots1}}{\sqrt{2}}.
\end{equation}
Each party performs local projective measurements in the Pauli \(X\) and \(Y\) bases (for instance, \(X_1=\sigma_x, X^\prime_1=\sigma_y\), and analogously for the remaining parties). The measurement setting at site \(i\in\{1,2,\dots,n\}\) is encoded by a binary variable \(X_i\in\{0,1\}\), where, for example, the value \(0\) corresponds to a measurement in the \(X\) basis and the value \(1\) to a measurement in the \(Y\) basis. The corresponding measurement outcomes are denoted by \(x_i\in\{0,1\}\). The proof in Ref.~\cite{abramsky2011sheaf} establishes that the $(n,2,2)$ GHZ models exhibit strong contextuality for all $n \geq 3$. The proof presented here demonstrates, in addition, that the corresponding correlations constitute maximal marginals.

The support of the correlations associated with a given context (i.e., the set of outcomes that occur with nonzero probability) depends on the parity of both the inputs and the outcomes. For input settings with odd parity, the correlations have full support and are uniformly distributed over all possible outcomes (Appendix~\ref{app: GHZ_correlation}),
\begin{equation}
    e_C(x) = 1/2^{n}, \quad \forall x=(x_1,x_2,\dots,x_n)\in\{0,1\}^n,
    \label{eq: odd_support}
\end{equation}
where $C$ is the context with an odd number of $Y$ measurements. For input settings with even parity, the correlations have support on exactly half of the possible outcomes (Appendix~\ref{app: GHZ_correlation}), namely
\begin{equation}
    e_C(x) =
\begin{cases} 
    1/2^{(n-1)}, &  \bigoplus_{i=1}^nx_i=P_C\\
    0, & \text{otherwise},
\end{cases}
\label{eq: eventsupport}
\end{equation}
where $P_C\in\{0,1\}$ is a parameter associated with the context $C$. Specifically, $P_C = 0$ for those contexts $C$ whose number of $Y$ measurements satisfies \[ N_1(Y)=\{4K \mid K=0,1,\dots,\lfloor n/4\rfloor\}, \] and, analogously, $P_C = 1$ for those contexts $C$ whose number of $Y$ measurements satisfies \[ N_2(Y)=\{4K+2 \mid K=0,1,\dots,\lfloor (n-2)/4\rfloor\}. \]

Given a context \(C\), we obtain the marginal for the \(j\)th party by fixing the outcomes of the other \(k=n-1\) parties. We denote the corresponding fixed outcome string by \(w\in\{0,1\}^{\,n-1}\). For $N_1(Y)$ or $N_2(Y)$ conditions, the outcome of the $j$th party is given by
\begin{equation}
x_j = P_C \bigoplus_{\substack{i=1 \\ i \neq j}}^{k} w_i,
\end{equation}
and thus can take only two possible values, $x_j = 0$ or $x_j = 1$. However, for a chosen context $C$, the corresponding $P_C$  value is fixed. In addition, by fixing the $k$ outcomes, only one of the values of $x_j$ is compatible with the outcome constraint $\bigoplus_{i=1}^nx_i=P_C$, which means the compatible outcome has a nonzero value, from Eqs.~(\ref{eq: eventsupport}). Consequently, for any $U \subset C$, the marginal
\begin{equation}
    e_C|_U=\sum_{x_j=\{0,1\}}e_C(x_1,x_2,\dots,x_j,\dots,x_n),
\end{equation}
only one term is nonzero; therefore, $e_C|_U=1/2^{n-1}$.
If the input parity is odd, then for the corresponding context $C$, the correlations have full support, i.e. $e_C(x)=1/2^n\quad \forall x=(x_1,\dots,x_n)$. Hence, the marginal over $j^{\mathrm{th}}$ party 
\begin{equation}
\begin{aligned}
    e_C|_U=&\sum_{x_j=\{0,1\}}e_C(x_1,x_2,\dots,x_j,\dots,x_n)\\
    &=\frac{1}{2^n}+\frac{1}{2^n}\\
    &=\frac{1}{2^{n-1}}.
    \end{aligned}
\end{equation}
This shows that, for $k=n-1$ parties, the correlations are maximal marginals. Since, for $k=n-1$ parties, all the supports are $1/2^{n-1}$; therefore, it follows that for $k<n-1$,
\begin{equation}
    e_C|_U= 2^{(n-1)-k}\cdot 2^{-(n-1)}=2^{-k}
\end{equation}
Hence, for any $k\leq n-1$, the marginals are $2^{-k}$, and the GHZ correlations for the $(n,2,2)$-scenario are AMCC.

\bla 
\section{Application}
\label{02}
In this section, we will discuss how AMCC is involved in various applications, such as secret sharing schemes and randomness extraction.
\subsection{Secret sharing}
In this section, we describe a hybrid secret‑sharing protocol that combines classical and quantum components \cite{Secretsharingmaxcontext}. The Hillery-Bužek-Berthiaume (HBB) scheme \cite{HBBscheme} involves a dealer distributing an entangled quantum state among several parties and then communicating the plaintext via a classical channel, with the actual secret encoded within it. The original secret can only be recovered when all players cooperate.

This scheme can be extended using a generalized Mermin‑type argument, which corresponds precisely to the set of parity equations we employed to generate AMCCs. Suppose a dealer (Alice) wishes to share a secret $S$. She first prepares a plaintext by combining the secret $S$ with a random pad $q$, i.e., $p_t = S \oplus q$ — this constitutes the classical part of the protocol. In the quantum part, the resource state is shared so that each player holds $k$ shares, and Alice retains $n-k$ shares. Thus, each party can be regarded as possessing $k$ measurement devices; in each round $r$, the inputs are $i_r \in \{0,\dots,m\}$ and the outputs are $o_r \in P(\bullet)$. The inputs are chosen using independent classical randomness sources, with Alice’s source determining whether a given round is a “secret” round or a “test” round with nonzero probability. All parties also have access to an authenticated classical channel. To maintain device independence, the quantum resource may be any entangled state.

Alice generates a key only if the protocol succeeds. Her key in round $r$ is obtained from the measurement outcomes on her device as
\begin{equation}
    a_{\text{dealer}}^r = \bigoplus_{j=k+1}^{n} a_j^r .
\end{equation}
The players’ inputs are communicated to Alice, who uses them to verify whether the observed correlations form a valid measurement context. This verification checks for perfect correlations that satisfy the parity equations. If the parity conditions are not satisfied, she aborts the protocol. If the correlations are perfect and correspond to a valid context, she treats the round as a test round, shares her ciphertext $q^r = p_t^r \oplus S^r$, and also broadcasts the associated phase value $ph^r \in \{0,\dots,s\}$ corresponding to the measurement context. By combining their outcomes, the ciphertext, and the phase information, the players can jointly reconstruct Alice’s original secret.
The parity conditions employed in the protocol are generalized forms of Mermin-type arguments, meaning that the parity equations cannot be simultaneously satisfied by any global assignment. Consequently, the empirical model associated with the protocol is maximally contextual. As previously discussed, the Boolean formulas corresponding to these parity conditions can be mapped to a probabilistic form through the application of zero constraints. Due to the symmetric structure of the Boolean formulas, the resulting probabilistic model exhibits maximal marginality. Hence, the correlations associated with the parity conditions are AMCCs.

\subsection{Randomness extraction}
Random number generation is a challenging task both in practice and in mathematical characterization. Random number generators (RNGs) play a crucial role in device‑independent (DI) protocols, including cryptographic applications. However, such devices are vulnerable to adversarial manipulation, making it essential to obtain randomness that is certified in a DI manner. This certification can be achieved by violating a Bell inequality: correlations obtained from the inputs and outputs of a device that violate a Bell inequality cannot be explained by local hidden variables or predetermined strategies. Such a guarantee is therefore called \emph{randomness certified by Bell’s theorem} \cite{randomnessextraction}.

The amount of randomness is quantified by the conditional min‑entropy
\begin{equation}
    H_{\min}(x_1x_2 \mid X_1X_2) = -\log_2 \max_{x_1,x_2} P(x_1x_2 \mid X_1X_2).
\end{equation}
A lower bound for this quantity can be expressed as
\begin{equation}
    H_{\min}(x_1x_2 \mid X_1X_2) \ge H_{\min}(x_1 \mid X_2),
\end{equation}
where the \emph{local min‑entropy} is defined by $H_{\min}(x \mid X) = -\log_2 \max_{x,X} P(x \mid X)$.

For an AMCC, every reduced bipartition exhibits maximally random correlations. Consequently, an adversary trying to guess the outcomes $x_1,\dots, x_k$ given the inputs $X_1,\dots, X_k$ faces the probability
\begin{equation}
    p_{\mathrm{guess}} = \max_{x_1,\dots, x_k} P(x_1,\dots, x_k \mid X_1,\dots, X_k) = \frac{1}{2^k} \qquad \forall\; k < n.
\end{equation}
The corresponding min‑entropy is therefore
\begin{equation}
\begin{aligned}
    H_{\min}(x_1,\dots, x_k \mid X_1,\dots, X_k) &= -\log_2 p_{\mathrm{guess}} \\
    &= \log_2 2^k = k .
\end{aligned}
\end{equation}
Thus, we obtain the lower bound on the global min‑entropy
\begin{equation}
    H_{\min}(xy \mid XY) \ge k .
\end{equation}
This bound shows that the local randomness is maximal: an adversary gains no information about Alice’s outcome and cannot simulate it using any preshared local randomness.

\bla 

\section{Conclusion}
Nonlocal correlations have become an integral part of studies in all aspects of information processing involving quantum resources. While empirical models based on correlations are well studied for the $(2,2,2)$ scenario, extending these concepts to multipartite systems remains an active area of research. Notably, the characterization of various forms of maximality in the multipartite scenario, similar to that of multipartite maximal entanglement, has remained unclear.

To address this, we introduce a class of contextual correlations in the multipartite setting, specifically motivated by absolutely maximally entangled (AME) states \cite{AME} and maximal contextual correlations \cite{FNSandallvsnothingequivalece}. These correlations, which we define as AMCCs, lie on the face of a polytope and are both maximally contextual and maximally marginal. The distinction introduced by AMCCs is particularly noteworthy: while AME states for two parties are the Bell states \cite{AME}, the two-party, two-input, two-output AMCCs correspond to PR boxes, which do not have any quantum realization. The three-qubit AME states correspond to three-qubit GHZ states \cite{AME}, and in terms of correlations, the GHZ correlations represent the $(3,2,2)$ AMCCs. Additionally, there exist infinitely many such correlations, as demonstrated in Table~(\ref{tab: reduced 8parameters ptable}), which features eight free parameters. Furthermore, although four-qubit AME states do not exist, the parity argument demonstrates that four-party AMCCs are possible \cite{four_party_AMCC}. We have also introduced non-AMCC correlations that are maximally contextual but not maximally marginal, which are constructed using the CSP construction. This construction provides the asymmetric AMCCs, whereas the constructions also provide the genuine-multipartite correlations. It is interesting to treat and explore the relation of $\text{CF} = 1$ correlations with multipartite non-local scenarios by invoking LOSR/WCCPI structures in future work.  In this work, we also demonstrated that for the specific measurement choices, in $(n,2,2)$ settings, the GHZ correlations are AMCC.

The results presented here can be extended in future work to include the findings of Ref.~\cite{aravinda2013complementarity}, where the complementarity trade-off between no-signaling and local indeterminacy for quantum correlations in the $(2,2,2)$ scenario is introduced. Their work presents a stronger version of Bell's theorem and provides a quantitative definition of a bound on randomness. This scope is justified by the fact that the violation of a generalized Bell inequality up to its algebraic bound occurs if and only if the model is strongly contextual \cite{min_quntun_resources_abramsky2017}. The very existence of non-AMCCs, which are maximally contextual but not maximally marginal, provides a departure from the complementarity seen in the bipartite scenario, thereby offering insights into the deeper relationship between nonlocality, randomness, and other nonclassical features in multipartite correlations.

\appendix 

\section{ $(n ,2, 2)$ GHZ correlations}
\label{app: GHZ_correlation}
The $n$-party GHZ state
\begin{equation}
    \ket{GHZ}_n=\frac{\ket{00\dots0}+\ket{11\dots1}}{\sqrt{2}}.
\end{equation}
The local measurements are in the Pauli $X$ and $Y$ bases encoded in the measurement variable $X_j\in\{0,1\}$, such that $X_j=X=\sigma_x$ and $X^{\prime}_j=Y=\sigma_y$, and the corresponding basis choices are $\{(\ket{0}+\ket{1})/\sqrt{2}, (\ket{0}-\ket{1})/\sqrt{2}\}$ and $\{(\ket{0}+i\ket{1})/\sqrt{2}, (\ket{0}-i\ket{1})/\sqrt{2}\}$, where $j\in\{1,2,\dots,n\}$. For a context $C=\{X_1,X_2,\dots,X_n\}$ and outcome string $x=(x_1,x_2,\dots,x_n)$, where $x_j\in\{0,1\}$, the joint probability
\begin{equation}
    p(x\mid C)=\lvert \langle GHZ | v(C,x) \rangle \rvert^{2},
\end{equation}
where, $\ket{v(C,x)}$ is the product of the basis vectors,
\begin{equation}
    \ket{v(C,x)}=\bigotimes_{j=1}^{n}\ket{v_j},
    \label{eq: product_basis}
\end{equation}
such that the local basis measurement is $\ket{v_j}=(\ket{0}+a_j\ket{1})/\sqrt{2}$, where the local phase $a_j$ is,
\begin{equation}
    a_j =
\begin{cases} 
    (-1)^{x_j}, &  X_j=X\\
    i(-1)^{x_j}, & X_j=Y.
\end{cases}
\label{eq: phase_of _pauli_basis}
\end{equation}
Thereafter, the product basis in Eqs.~(\ref{eq: product_basis}) can be written as
\begin{equation}
    \ket{v(C,x)}=\bigotimes_{j=1}^{n}\left(\frac{\ket{0}+a_j\ket{1}}{\sqrt{2}}\right).
\end{equation}
As a result, the joint probability is,
\begin{widetext}
    \begin{equation}
    \begin{aligned}
    p(x\mid C)&=\lvert \langle GHZ | v(C,x) \rangle \rvert^{2},\\
    &=\frac{1}{2}\lvert \langle 00\dots0 | v(C,x) \rangle +\langle 11\dots1 | v(C,x) \rangle\rvert^{2},\\
    &=\frac{1}{2}\left\lvert \langle 00\dots0 | \bigotimes_{j=1}^{n}\left(\frac{\ket{0}+a_j\ket{1}}{\sqrt{2}}\right) +\langle 11\dots1 | \bigotimes_{j=1}^{n}\left(\frac{\ket{0}+a_j\ket{1}}{\sqrt{2}}\right)\right\rvert^{2},\\
    &=\frac{1}{2}\left\lvert \prod^{n}_{j=1}\frac{1}{\sqrt{2}}+\prod^{n}_{j=1}\frac{a_j}{\sqrt{2}}\right\rvert^{2},\\
    &=\frac{1}{2}\left\lvert \frac{1}{2^{n/2}}+\frac{\prod^{n}_{j=1}a_j}{2^{n/2}}\right\rvert^{2},\\
    &=\frac{1}{2^{n+1}}\left\lvert \left(1+\prod_{j=1}^{n}a_j\right)\right\rvert^{2}.
    \end{aligned}
\end{equation}
\end{widetext}
Now substituting $a_j$ for Eqs.~(\ref{eq: phase_of _pauli_basis}),
\begin{equation}
     p(x\mid C)=\frac{1}{2^{n+1}}\left\lvert \left(1+\left(\prod_{j=1}^{n}(-1)^{x_j}\right)i^{N}\right)\right\rvert^{2},
\end{equation}
where $N$ is the number of $Y$'s in the context; therefore,
\begin{equation}
     p(x\mid C)=\frac{1}{2^{n+1}}\left\lvert \left(1+(-1)^{x}i^{N}\right)\right\rvert^{2},
\end{equation}
where $x=\oplus_{j=1}^{n}x_j$. If the number of $Y$'s is odd in the context, then $i^{N}=\pm i$ therefore,
\begin{equation}
     \begin{aligned}
         p(x\mid C)&=\frac{1}{2^{n+1}}\left\lvert \left(1\pm(-1)^{x}i\right)\right\rvert^{2},\\
         &=\frac{1}{2^{n+1}}(1+(-1)^{2x}),\\
         &=\frac{1}{2^{n+1}}(1+1)=\frac{1}{2^n}.
     \end{aligned}
\end{equation}
For the context consisting of an even number of $Y$, there are two cases. The first is for $N\in\{4K \mid K=0,1,\dots,\lfloor n/4\rfloor\}$,
\begin{equation}
     \begin{aligned}
         p(x\mid C)&=\frac{1}{2^{n+1}}\left\lvert \left(1+(-1)^{x}i^{N}\right)\right\rvert^{2},\\
         &=\frac{1}{2^{n+1}}\left\lvert \left(1+(-1)^{x}i^{4K}\right)\right\rvert^{2},\\
         &=\frac{1}{2^{n+1}}\left\lvert \left(1+(-1)^{x}\right)\right\rvert^{2}.
     \end{aligned}
\end{equation}
Since the condition $N\in\{4K \mid K=0,1,\dots,\lfloor n/4\rfloor\}$ corresponds to $P_{C}=0$, which provides the outcome parity condition as $x=\oplus_{j=1}^{n}x_j=0$ for nonzero support, consequently,
\begin{equation}
     \begin{aligned}
         p(x\mid C)&=\frac{1}{2^{n+1}}\left\lvert \left(1+(-1)^{x}\right)\right\rvert^{2},\\
         &=\frac{1}{2^{n+1}}\left\lvert \left(1+1\right)\right\rvert^{2},\\
         &=\frac{1}{2^{n+1}}\cdot2^2=\frac{1}{2^{n-1}}.
     \end{aligned}
\end{equation}
If, $P_{C}=1$, then the outcome condition is $x=\oplus_{j=1}^{n}x_j=1$, which lead to,
\begin{equation}
     \begin{aligned}
         p(x\mid C)&=\frac{1}{2^{n+1}}\left\lvert \left(1+(-1)^{x}\right)\right\rvert^{2},\\
         &=\frac{1}{2^{n+1}}\left\lvert \left(1-1\right)\right\rvert^{2}=0,
     \end{aligned}
\end{equation}

Taking the condition $N\in \{4K+2 \mid K=0,1,\dots,\lfloor (n-2)/4\rfloor\}$, lead to nonzero support for $P_{C}=1$, such that the output condition becomes $x=\oplus_{j=1}^{n}x_j=1$, as a result,
\begin{equation}
     \begin{aligned}
         p(x\mid C)&=\frac{1}{2^{n+1}}\left\lvert \left(1+(-1)^{x}i^{N}\right)\right\rvert^{2},\\
         &=\frac{1}{2^{n+1}}\left\lvert \left(1+(-1)^{x}i^{4K+2}\right)\right\rvert^{2},\\
         &=\frac{1}{2^{n+1}}\left\lvert \left(1-(-1)^{x}\right)\right\rvert^{2},\\
         &=\frac{1}{2^{n+1}}\left\lvert \left(1+1\right)\right\rvert^{2},\\
         &=\frac{1}{2^{n+1}}\cdot2^2=\frac{1}{2^{n-1}}.
     \end{aligned}
\end{equation}
If $P_{C}=0$, the output condition becomes $x=\oplus_{j=1}^{n}x_j=0$, therefore,
\begin{equation}
     \begin{aligned}
         p(x\mid C)&=\frac{1}{2^{n+1}}\left\lvert \left(1-(-1)^{x}\right)\right\rvert^{2},\\
         &=\frac{1}{2^{n+1}}\left\lvert \left(1-1\right)\right\rvert^{2}=0.
     \end{aligned}
\end{equation}
Hence, for the condition when contexts consist of an odd number of $Y$ measurements, the correlations satisfy the condition in Eqs.~(\ref{eq: odd_support}), and for contexts having an even number of $Y$ measurements, the correlations satisfy the condition provided in Eqs.~(\ref{eq: eventsupport}).

\section{ Proper faces and AVN}
\label{app: prope facet}
The proper face $F$ \cite{pironio2005lifting, AshutoshRaiGeometryofquantumset2019}, with dimension $\dim F \leq \dim P - 1$, can be defined by imposing zero constraints:
\begin{equation}
F = \{\mathbf{p} \in \mathbf{NS}\;|\; p = 0\},
\end{equation} 
where $p$ is an element of $\mathbf{p} = {p(x_1, \cdots, x_i \mid X_1, \cdots, X_i)_{i \leq \dim P - 1}})$, and $\mathbf{NS} \subset \mathbb{R}^r$ denotes the set of no-signaling correlations. For example, in the $(2,2,2)$ scenario, after imposing the no-signaling and normalization constraints, the maximal dimension is reduced to $\dim P-1=8$. If we set $p(01\mid00)$ as a zero constraint, one parameter is fixed to zero while the others remain free, resulting in a lower-dimensional face. Adding another zero constraint, such as $p(10\mid00)=0$, further reduces the dimensionality of the face. Thus, for the input $(X_1,X_2)=(0,0)$, we obtain
\begin{equation}
    p(00\mid00)+p(11\mid00)=1
\end{equation}
which characterizes a lower-dimensional proper face:
\begin{equation}
    F(X_1=0,X_2=0)=\{\mathbf{p}\mid p(01\mid00)+p(10\mid00)=0\}
\end{equation}
Applying the same procedure for other inputs yields correlations such that, for given inputs $X_1, X_2$ and outputs $x_1,x_2$, the logical formula $x_1\oplus x_2=f(X_1, X_2)$ must hold with certainty, i.e., $p(x_1\oplus x_2=f(X_1, X_2)\mid X_1, X_2)=1$. Any probabilities violating this logical formula are forced to zero by the imposed constraints. Hence, the zero constraints directly correspond to the logical formula and lead to an AvN proof \cite{cohom_AVN_abramsky_et_al2015, FNSandallvsnothingequivalece}. These zeros define the face of the polytope with maximally reduced dimension.

\section{ Parametric table for $(3,2,2)$ scenario}
\label{app: reduced table}
The reduced parametric table for the $(3,2,2)$ scenario is obtained using no-signaling and the normalization condition mentioned in Sec.~\ref{sec 00}.
\begin{widetext}
    \begin{table}[H]
    \centering
\small 
\begin{tabularx}{\textwidth}{|>{\raggedright\arraybackslash}X|*{8}{>{\centering\arraybackslash}X |}}
\hline
\textbf{Context and Section} & \textbf{(0,0,0)} & \textbf{(0,0,1)} & \textbf{(0,1,0)} & \textbf{(0,1,1)} & \textbf{(1,0,0)} & \textbf{(1,0,1)} & \textbf{(1,1,0)} & \textbf{(1,1,1)} \\ \hline

{\centering$(0, 0, 0)$\par}
 & {\centering $p_1$\par}
 & {\centering$p_2$\par}
 & {\centering$p_3$\par}
 & {\centering$p_4$\par}
 & {\centering$p_5$\par}
 & {\centering$p_6$\par}
 & {\centering$p_7$\par}
 & $1-(p_1+p_2+p_3+p_4+p_5+p_6+p_7)$
\\ \hline

{\centering$(0, 0, 1)$\par}
 & {\centering$p_9$\par}
 &  {\centering$p_1+p_2-p_9$\par}
 & {\centering$p_{11}$\par}
  & {\centering$p_3+p_4-p_{11}$\par} 
  & {\centering$p_{13}$\par}
  & {\centering$p_5+p_6-p_{13}$\par}
  & {\centering$p_{15}$\par}
&{\centering$1-(p_1+p_2+p_3+p_4+p_5+p_6+p_{15})$\par}
\\ \hline

{\centering$(0, 1, 0)$\par}
 & {\centering$p_{17}$\par}
 & {\centering$p_{18}$\par}
  & {\centering$p_1+p_3-p_{17}$\par} 
  & {\centering$p_2+p_4-p_{18}$\par}
  & {\centering$p_{21}$\par}
  & {\centering$p_{22}$\par} 
  & {\centering$p_5+p_7-p_{21}$\par}
  &{\centering$1-(p_1+p_2+p_3+p_4+p_5+p_7+p_{22})$\par}
\\ \hline

{\centering$(0, 1, 1)$\par}
 & {\centering$p_{25}$\par}
 & {\centering$p_{17}+p_{18}-p_{25}$\par} 
  & {\centering$p_9+p_{11}-p_{25}$\par}
  & {\centering$p_1+p_2+p_3+p_4-p_9-p_{11}-p_{17}-p_{18}+p_{25}$\par}
  & {\centering$p_{8}$\par}
  & {\centering$p_{21}+p_{22}-p_{8}$\par}
  & {\centering$p_{13}+p_{15}-p_{8}$\par}
  &{\centering$1-(p_1+p_2+p_3+p_4+p_{21}+p_{22}+p_{13}+p_{15}-p_{8})$\par}
\\ \hline

{\centering$(1, 0, 0)$\par}
 & {\centering$p_{10}$\par}
 & {\centering$p_{12}$\par} 
  & {\centering$p_{14}$\par} 
  & {\centering$p_{16}$\par}
  & {\centering$p_1+p_5-p_{10}$\par}
  & {\centering$p_2+p_6-p_{12}$\par}
  & {\centering$p_3+p_7-p_{14}$\par}
  &{\centering$1-(p_1+p_2+p_3+p_5+p_6+p_7+p_{16})$\par}
\\ \hline

{\centering$(1, 0, 1)$\par}
 & {\centering$p_{19}$\par}
 & {\centering$p_{10}+p_{12}-p_{19}$\par}
  & {\centering$p_{20}$\par} 
  & {\centering$p_{14}+p_{16}-p_{20}$\par}
  & {\centering$p_9+p_{13}-p_{19}$\par}
  & {\centering$p_1+p_2+p_5+p_6-p_9-p_{13}-p_{10}-p_{12}+p_{19}$\par} 
  & {\centering$p_{11}+p_{15}-p_{20}$\par}
&{\centering$1-(p_1+p_2+p_5+p_6+p_{11}+p_{14}+p_{15}+p_{16}-p_{20})$\par}
\\ \hline

{\centering$(1, 1, 0)$\par}
 & {\centering$p_{23}$\par}
 & {\centering$p_{24}$\par}
  & {\centering$p_{10}+p_{14}-p_{23}$\par}
  & {\centering$p_{12}+p_{16}-p_{24}$\par}
  & {\centering$p_{17}+p_{21}-p_{23}$\par}
  & {\centering$p_{18}+p_{22}-p_{24}$\par}
  & {\centering$p_1+p_3+p_5+p_7-p_{17}-p_{21}-p_{10}-p_{14}+p_{23}$\par}
  & {\centering$1-(p_1+p_3+p_5+p_7+p_{18}+p_{22}+p_{12}+p_{16}-p_{24})$\par}
\\ \hline

{\centering$(1, 1, 1)$\par}
 & {\centering$p_{26}$\par}
 & {\centering$p_{23}+p_{24}-p_{26}$\par}
  & {\centering$p_{19}+p_{20}-p_{26}$\par}
  & {\centering$p_{10}+p_{12}+p_{14}+p_{16}-p_{19}-p_{20}-p_{23}-p_{24}+p_{26}$\par} 
  & {\centering$p_{25}+p_{8}-p_{26}$\par} 
  & {\centering$p_{17}+p_{18}+p_{21}+p_{22}-p_{25}-p_{8}-p_{23}-p_{24}+p_{26}$\par} 
  & {\centering$p_9+p_{11}+p_{13}+p_{15}-p_{25}-p_{8}-p_{19}-p_{20}+p_{26}$\par}
  & {\centering$1-(p_9+p_{11}+p_{13}+p_{15}+p_{17}+p_{18}+p_{21}+p_{22}-p_{25}-p_{8}+p_{10}+p_{12}+p_{14}+p_{16}-p_{19}-p_{20}-p_{23}-p_{24}+p_{26})$\par}
\\ \hline

\end{tabularx}
\caption{Reduced (26 free parameters) probability table.}
\label{tab: 26_parameter_table}
\end{table}
\end{widetext}

\section{Boolean no-signaling condition}
\label{app: Boolean no-signaling}
The Boolean no-signaling condition is the possibilistic form of Eq.~(\ref{eq: no-signaling condition}), such that the probability parameters are replaced by Boolean formulas ($B_{i}^\prime s$) and the summation operation is replaced with the Boolean OR operation.
\begin{widetext}
    \begin{equation}
        \bigvee_{x_j}B(x_1,\cdots,x_j,\cdots,x_n\mid X_1,\cdots,X_j,\cdots,X_n)=
\bigvee_{x_j}B(x_1,\cdots,x_j,\cdots,x_n \mid X_1,\cdots,X_j',\cdots,X_n)
\label{eq; no-signaling condition}
    \end{equation}
\end{widetext}
for all \(j\), all outcomes \(x_1,\dots,x_{j-1},x_{j+1},\dots,x_n\), and all settings \(X_1,\dots,X_{j-1},X_j,X_j',X_{j+1},\dots,X_n\).

\section{Parametric maximal contextual table}
The maximal contextual Table~ \ref{tab:sixparametertable} corresponding to Eqs.~(\ref{eq:maxcontextualmodel}), which is constructed by substituting zeros from the possibilistic model obtained from Eqs.~(\ref{eq:maxcontextualmodel}) into Table~\ref{tab: 26_parameter_table}.
\begin{widetext}
\begin{table}[H]
\centering
\small
\begin{tabularx}{\textwidth}{|>{\raggedright\arraybackslash}X|*{8}{>{\centering\arraybackslash}X|}}
\hline
\textbf{Context and Section} & \textbf{(0,0,0)} & \textbf{(0,0,1)} & \textbf{(0,1,0)} & \textbf{(0,1,1)} & \textbf{(1,0,0)} & \textbf{(1,0,1)} & \textbf{(1,1,0)} & \textbf{(1,1,1)} \\ 
\hline

{\centering$(0,0,0)$\par}
& {\centering $0$\par}
& {\centering $p_1 + p_2 + p_3 + p_4 - 2p_5$\par}
& {\centering $p_1 + p_2 + p_3 - p_4$\par}
& {\centering $0$\par}
& {\centering $-p_1 - p_2 - p_3 + p_5 + 1/2$\par}
& {\centering $0$\par}
& {\centering $0$\par}
& {\centering $-p_1 - p_2 - p_3 + p_5 + 1/2$\par}
\\ \hline

{\centering$(0,0,1)$\par}
& {\centering $p_1 + p_2 + p_3 + p_4 - p_5 - 1/2$\par}
& {\centering $1/2 - p_5$\par}
& {\centering $-p_2 - p_3 - p_4 + p_5 + 1/2$\par}
& {\centering $p_1 + 2p_2 + 2p_3 - p_5 - 1/2$\par}
& {\centering $-p_1 - p_2 - p_3 + p_5 + 1/2$\par}
& {\centering $0$\par}
& {\centering $0$\par}
& {\centering $-p_1 - p_2 - p_3 + p_5 + 1/2$\par}
\\ \hline

{\centering$(0,1,0)$\par}
& {\centering $p_1 + p_2 + p_3 - 1/2$\par}
& {\centering $-p_2 - p_3 + 1/2$\par}
& {\centering $1/2 - p_4$\par}
& {\centering $p_1 + 2p_2 + 2p_3 + p_4 - 2p_5 - 1/2$\par}
& {\centering $-p_1 - p_2 - p_3 + p_5 + 1/2$\par}
& {\centering $0$\par}
& {\centering $0$\par}
& {\centering $-p_1 - p_2 - p_3 + p_5 + 1/2$\par}
\\ \hline

{\centering$(0,1,1)$\par}
& {\centering $p_1$\par}
& {\centering $0$\par}
& {\centering $0$\par}
& {\centering $p_1 + 2p_2 + 2p_3 - 2p_5$\par}
& {\centering $0$\par}
& {\centering $-p_1 - p_2 - p_3 + p_5 + 1/2$\par}
& {\centering $-p_1 - p_2 - p_3 + p_5 + 1/2$\par}
& {\centering $0$\par}
\\ \hline

{\centering$(1,0,0)$\par}
& {\centering $-p_2 + p_5$\par}
& {\centering $p_2 + p_4 - p_5$\par}
& {\centering $p_2$\par}
& {\centering $p_3$\par}
& {\centering $-p_1 - p_3 + 1/2$\par}
& {\centering $p_1 + p_3 - p_5$\par}
& {\centering $p_1 + p_3 - p_4$\par}
& {\centering $-p_1 - p_2 - 2p_3 + p_5 + 1/2$\par}
\\ \hline

{\centering$(1,0,1)$\par}
& {\centering $p_4$\par}
& {\centering $0$\par}
& {\centering $0$\par}
& {\centering $p_2 + p_3$\par}
& {\centering $0$\par}
& {\centering $1/2 - p_5$\par}
& {\centering $-p_2 - p_3 - p_4 + p_5 + 1/2$\par}
& {\centering $0$\par}
\\ \hline

{\centering$(1,1,0)$\par}
& {\centering $p_5$\par}
& {\centering $0$\par}
& {\centering $0$\par}
& {\centering $p_2 + p_3 + p_4 - p_5$\par}
& {\centering $0$\par}
& {\centering $-p_2 - p_3 + 1/2$\par}
& {\centering $1/2 - p_4$\par}
& {\centering $0$\par}
\\ \hline

{\centering$(1,1,1)$\par}
& {\centering $p_6$\par}
& {\centering $p_5 - p_6$\par}
& {\centering $p_4 - p_6$\par}
& {\centering $p_2 + p_3 - p_5 + p_6$\par}
& {\centering $p_1 - p_6$\par}
& {\centering $-p_1 - p_2 - p_3 + p_6 + 1/2$\par}
& {\centering $-p_1 - p_2 - p_3 - p_4 + p_5 + p_6 + 1/2$\par}
& {\centering $p_1 + p_2 + p_3 - p_5 - p_6$\par}
\\ \hline

\end{tabularx}
\caption{Maximum contextual table corresponding to Eq.~(\ref{eq:maxcontextualmodel}) for the $(3,2,2)$ scenario.}
\label{tab:sixparametertable}
\end{table}
\end{widetext}
The table consists of six free parameters; it is maximal contextual only for the following bounds on the parameters
\begin{widetext}
\begin{equation}
\begin{aligned}
0\leq p_{3}\leq \frac{1}{2},\qquad
0\leq p_{2}\leq \frac{1}{2}-p_{3},\qquad
\max\!\left(\frac{1}{2}-p_{2}-p_{3},\, p_{2}-p_{3}\right)\leq p_{1}\leq \frac{1}{2}-p_{3},
\\[4pt]
\max\!\left(p_{2},\, p_{1}+p_{2}+2p_{3}-\frac{1}{2}\right)\leq p_{5}\leq
\min\!\left(p_{1}+p_{3},\, p_{2}+p_{3}+\frac{p_{1}}{2},\, p_{1}+2p_{2}+2p_{3}-\frac{1}{2}\right),
\\[4pt]
\max\!\left(
0,\,
p_{5}-p_{2},\,
\frac{1}{2}+p_{5}-p_{1}-p_{2}-p_{3},\,
2p_{5}-p_{1}-2p_{2}-2p_{3}+\frac{1}{2}
\right)
\leq p_{4}\leq
\min\!\left(
p_{1}+p_{3},\,
\frac{1}{2}+p_{5}-p_{2}-p_{3}
\right),
\\[4pt]
\max\!\left(
0,\,
p_{5}-p_{2}-p_{3},\,
p_{1}+p_{2}+p_{3}+p_{4}-p_{5}-\frac{1}{2},\,
p_{1}+p_{2}+p_{3}-\frac{1}{2}
\right)
\leq p_{6}\leq
\min\!\left(
p_{1},\, p_{4},\, p_{5},\, p_{1}+p_{2}+p_{3}-p_{5}
\right).
\end{aligned}
\label{eq: bounds}
\end{equation}
\end{widetext}

\newpage


\begin{thebibliography}{74}%
\makeatletter
\providecommand \@ifxundefined [1]{%
 \@ifx{#1\undefined}
}%
\providecommand \@ifnum [1]{%
 \ifnum #1\expandafter \@firstoftwo
 \else \expandafter \@secondoftwo
 \fi
}%
\providecommand \@ifx [1]{%
 \ifx #1\expandafter \@firstoftwo
 \else \expandafter \@secondoftwo
 \fi
}%
\providecommand \natexlab [1]{#1}%
\providecommand \enquote  [1]{``#1''}%
\providecommand \bibnamefont  [1]{#1}%
\providecommand \bibfnamefont [1]{#1}%
\providecommand \citenamefont [1]{#1}%
\providecommand \href@noop [0]{\@secondoftwo}%
\providecommand \href [0]{\begingroup \@sanitize@url \@href}%
\providecommand \@href[1]{\@@startlink{#1}\@@href}%
\providecommand \@@href[1]{\endgroup#1\@@endlink}%
\providecommand \@sanitize@url [0]{\catcode `\\12\catcode `\$12\catcode
  `\&12\catcode `\#12\catcode `\^12\catcode `\_12\catcode `\%12\relax}%
\providecommand \@@startlink[1]{}%
\providecommand \@@endlink[0]{}%
\providecommand \url  [0]{\begingroup\@sanitize@url \@url }%
\providecommand \@url [1]{\endgroup\@href {#1}{\urlprefix }}%
\providecommand \urlprefix  [0]{URL }%
\providecommand \Eprint [0]{\href }%
\providecommand \doibase [0]{http://dx.doi.org/}%
\providecommand \selectlanguage [0]{\@gobble}%
\providecommand \bibinfo  [0]{\@secondoftwo}%
\providecommand \bibfield  [0]{\@secondoftwo}%
\providecommand \translation [1]{[#1]}%
\providecommand \BibitemOpen [0]{}%
\providecommand \bibitemStop [0]{}%
\providecommand \bibitemNoStop [0]{.\EOS\space}%
\providecommand \EOS [0]{\spacefactor3000\relax}%
\providecommand \BibitemShut  [1]{\csname bibitem#1\endcsname}%
\let\auto@bib@innerbib\@empty
\bibitem [{\citenamefont {Einstein}\ \emph {et~al.}(1935)\citenamefont
  {Einstein}, \citenamefont {Podolsky},\ and\ \citenamefont
  {Rosen}}]{PhysRev.47.777}%
  \BibitemOpen
  \bibfield  {author} {\bibinfo {author} {\bibfnamefont {A.}~\bibnamefont
  {Einstein}}, \bibinfo {author} {\bibfnamefont {B.}~\bibnamefont {Podolsky}},
  \ and\ \bibinfo {author} {\bibfnamefont {N.}~\bibnamefont {Rosen}},\
  }\bibfield  {title} {\enquote {\bibinfo {title} {Can quantum-mechanical
  description of physical reality be considered complete?}}\ }\href {\doibase
  10.1103/PhysRev.47.777} {\bibfield  {journal} {\bibinfo  {journal} {Phys.
  Rev.}\ }\textbf {\bibinfo {volume} {47}},\ \bibinfo {pages} {777--780}
  (\bibinfo {year} {1935})}\BibitemShut {NoStop}%
\bibitem [{\citenamefont {Bell}(1964)}]{bell1964einstein}%
  \BibitemOpen
  \bibfield  {author} {\bibinfo {author} {\bibfnamefont {John~S}\ \bibnamefont
  {Bell}},\ }\bibfield  {title} {\enquote {\bibinfo {title} {On the einstein
  podolsky rosen paradox},}\ }\href@noop {} {\bibfield  {journal} {\bibinfo
  {journal} {Physics Physique Fizika}\ }\textbf {\bibinfo {volume} {1}},\
  \bibinfo {pages} {195} (\bibinfo {year} {1964})}\BibitemShut {NoStop}%
\bibitem [{\citenamefont {Horodecki}\ \emph {et~al.}(2009)\citenamefont
  {Horodecki}, \citenamefont {Horodecki}, \citenamefont {Horodecki},\ and\
  \citenamefont {Horodecki}}]{RevModPhys.81.865}%
  \BibitemOpen
  \bibfield  {author} {\bibinfo {author} {\bibfnamefont {Ryszard}\ \bibnamefont
  {Horodecki}}, \bibinfo {author} {\bibfnamefont {Pawe\l{}}\ \bibnamefont
  {Horodecki}}, \bibinfo {author} {\bibfnamefont {Micha\l{}}\ \bibnamefont
  {Horodecki}}, \ and\ \bibinfo {author} {\bibfnamefont {Karol}\ \bibnamefont
  {Horodecki}},\ }\bibfield  {title} {\enquote {\bibinfo {title} {Quantum
  entanglement},}\ }\href {\doibase 10.1103/RevModPhys.81.865} {\bibfield
  {journal} {\bibinfo  {journal} {Rev. Mod. Phys.}\ }\textbf {\bibinfo {volume}
  {81}},\ \bibinfo {pages} {865--942} (\bibinfo {year} {2009})}\BibitemShut
  {NoStop}%
\bibitem [{\citenamefont {Popescu}\ and\ \citenamefont
  {Rohrlich}(1997)}]{PhysRevA.56.R3319}%
  \BibitemOpen
  \bibfield  {author} {\bibinfo {author} {\bibfnamefont {Sandu}\ \bibnamefont
  {Popescu}}\ and\ \bibinfo {author} {\bibfnamefont {Daniel}\ \bibnamefont
  {Rohrlich}},\ }\bibfield  {title} {\enquote {\bibinfo {title} {Thermodynamics
  and the measure of entanglement},}\ }\href {\doibase
  10.1103/PhysRevA.56.R3319} {\bibfield  {journal} {\bibinfo  {journal} {Phys.
  Rev. A}\ }\textbf {\bibinfo {volume} {56}},\ \bibinfo {pages} {R3319--R3321}
  (\bibinfo {year} {1997})}\BibitemShut {NoStop}%
\bibitem [{\citenamefont {Nielsen}(1999)}]{PhysRevLett.83.436}%
  \BibitemOpen
  \bibfield  {author} {\bibinfo {author} {\bibfnamefont {M.~A.}\ \bibnamefont
  {Nielsen}},\ }\bibfield  {title} {\enquote {\bibinfo {title} {Conditions for
  a class of entanglement transformations},}\ }\href {\doibase
  10.1103/PhysRevLett.83.436} {\bibfield  {journal} {\bibinfo  {journal} {Phys.
  Rev. Lett.}\ }\textbf {\bibinfo {volume} {83}},\ \bibinfo {pages} {436--439}
  (\bibinfo {year} {1999})}\BibitemShut {NoStop}%
\bibitem [{\citenamefont {D\"ur}\ \emph {et~al.}(2000)\citenamefont {D\"ur},
  \citenamefont {Vidal},\ and\ \citenamefont {Cirac}}]{PhysRevA.62.062314}%
  \BibitemOpen
  \bibfield  {author} {\bibinfo {author} {\bibfnamefont {W.}~\bibnamefont
  {D\"ur}}, \bibinfo {author} {\bibfnamefont {G.}~\bibnamefont {Vidal}}, \ and\
  \bibinfo {author} {\bibfnamefont {J.~I.}\ \bibnamefont {Cirac}},\ }\bibfield
  {title} {\enquote {\bibinfo {title} {Three qubits can be entangled in two
  inequivalent ways},}\ }\href {\doibase 10.1103/PhysRevA.62.062314} {\bibfield
   {journal} {\bibinfo  {journal} {Phys. Rev. A}\ }\textbf {\bibinfo {volume}
  {62}},\ \bibinfo {pages} {062314} (\bibinfo {year} {2000})}\BibitemShut
  {NoStop}%
\bibitem [{\citenamefont {Verstraete}\ \emph {et~al.}(2002)\citenamefont
  {Verstraete}, \citenamefont {Dehaene}, \citenamefont {De~Moor},\ and\
  \citenamefont {Verschelde}}]{verstraete2002four}%
  \BibitemOpen
  \bibfield  {author} {\bibinfo {author} {\bibfnamefont {Frank}\ \bibnamefont
  {Verstraete}}, \bibinfo {author} {\bibfnamefont {Jeroen}\ \bibnamefont
  {Dehaene}}, \bibinfo {author} {\bibfnamefont {Bart}\ \bibnamefont {De~Moor}},
  \ and\ \bibinfo {author} {\bibfnamefont {Henri}\ \bibnamefont {Verschelde}},\
  }\bibfield  {title} {\enquote {\bibinfo {title} {Four qubits can be entangled
  in nine different ways},}\ }\href@noop {} {\bibfield  {journal} {\bibinfo
  {journal} {Physical Review A}\ }\textbf {\bibinfo {volume} {65}},\ \bibinfo
  {pages} {052112} (\bibinfo {year} {2002})}\BibitemShut {NoStop}%
\bibitem [{\citenamefont {Kraus}(2010)}]{kraus2010local}%
  \BibitemOpen
  \bibfield  {author} {\bibinfo {author} {\bibfnamefont {B}~\bibnamefont
  {Kraus}},\ }\bibfield  {title} {\enquote {\bibinfo {title} {Local unitary
  equivalence of multipartite pure states},}\ }\href@noop {} {\bibfield
  {journal} {\bibinfo  {journal} {Physical review letters}\ }\textbf {\bibinfo
  {volume} {104}},\ \bibinfo {pages} {020504} (\bibinfo {year}
  {2010})}\BibitemShut {NoStop}%
\bibitem [{\citenamefont {de~Vicente}\ \emph {et~al.}(2013)\citenamefont
  {de~Vicente}, \citenamefont {Spee},\ and\ \citenamefont
  {Kraus}}]{de2013maximally}%
  \BibitemOpen
  \bibfield  {author} {\bibinfo {author} {\bibfnamefont {Julio~I}\ \bibnamefont
  {de~Vicente}}, \bibinfo {author} {\bibfnamefont {Cornelia}\ \bibnamefont
  {Spee}}, \ and\ \bibinfo {author} {\bibfnamefont {Barbara}\ \bibnamefont
  {Kraus}},\ }\bibfield  {title} {\enquote {\bibinfo {title} {Maximally
  entangled set of multipartite quantum states},}\ }\href@noop {} {\bibfield
  {journal} {\bibinfo  {journal} {Physical review letters}\ }\textbf {\bibinfo
  {volume} {111}},\ \bibinfo {pages} {110502} (\bibinfo {year}
  {2013})}\BibitemShut {NoStop}%
\bibitem [{\citenamefont {Miyake}(2003)}]{miyake2003classification}%
  \BibitemOpen
  \bibfield  {author} {\bibinfo {author} {\bibfnamefont {Akimasa}\ \bibnamefont
  {Miyake}},\ }\bibfield  {title} {\enquote {\bibinfo {title} {Classification
  of multipartite entangled states by multidimensional determinants},}\
  }\href@noop {} {\bibfield  {journal} {\bibinfo  {journal} {Physical Review
  A}\ }\textbf {\bibinfo {volume} {67}},\ \bibinfo {pages} {012108} (\bibinfo
  {year} {2003})}\BibitemShut {NoStop}%
\bibitem [{\citenamefont {Gour}\ and\ \citenamefont
  {Wallach}(2011)}]{gour2011necessary}%
  \BibitemOpen
  \bibfield  {author} {\bibinfo {author} {\bibfnamefont {Gilad}\ \bibnamefont
  {Gour}}\ and\ \bibinfo {author} {\bibfnamefont {Nolan~R}\ \bibnamefont
  {Wallach}},\ }\bibfield  {title} {\enquote {\bibinfo {title} {Necessary and
  sufficient conditions for local manipulation of multipartite pure quantum
  states},}\ }\href@noop {} {\bibfield  {journal} {\bibinfo  {journal} {New
  Journal of Physics}\ }\textbf {\bibinfo {volume} {13}},\ \bibinfo {pages}
  {073013} (\bibinfo {year} {2011})}\BibitemShut {NoStop}%
\bibitem [{\citenamefont {Spee}\ \emph {et~al.}(2017)\citenamefont {Spee},
  \citenamefont {de~Vicente}, \citenamefont {Sauerwein},\ and\ \citenamefont
  {Kraus}}]{spee2017entangled}%
  \BibitemOpen
  \bibfield  {author} {\bibinfo {author} {\bibfnamefont {Cornelia}\
  \bibnamefont {Spee}}, \bibinfo {author} {\bibfnamefont {Julio~I}\
  \bibnamefont {de~Vicente}}, \bibinfo {author} {\bibfnamefont {David}\
  \bibnamefont {Sauerwein}}, \ and\ \bibinfo {author} {\bibfnamefont {Barbara}\
  \bibnamefont {Kraus}},\ }\bibfield  {title} {\enquote {\bibinfo {title}
  {Entangled pure state transformations via local operations assisted by
  finitely many rounds of classical communication},}\ }\href@noop {} {\bibfield
   {journal} {\bibinfo  {journal} {Physical Review Letters}\ }\textbf {\bibinfo
  {volume} {118}},\ \bibinfo {pages} {040503} (\bibinfo {year}
  {2017})}\BibitemShut {NoStop}%
\bibitem [{\citenamefont {Li}\ \emph {et~al.}(2024)\citenamefont {Li},
  \citenamefont {Spee}, \citenamefont {Hebenstreit}, \citenamefont
  {De~Vicente},\ and\ \citenamefont {Kraus}}]{li2024identifying}%
  \BibitemOpen
  \bibfield  {author} {\bibinfo {author} {\bibfnamefont {Nicky Kai~Hong}\
  \bibnamefont {Li}}, \bibinfo {author} {\bibfnamefont {Cornelia}\ \bibnamefont
  {Spee}}, \bibinfo {author} {\bibfnamefont {Martin}\ \bibnamefont
  {Hebenstreit}}, \bibinfo {author} {\bibfnamefont {Julio~I}\ \bibnamefont
  {De~Vicente}}, \ and\ \bibinfo {author} {\bibfnamefont {Barbara}\
  \bibnamefont {Kraus}},\ }\bibfield  {title} {\enquote {\bibinfo {title}
  {Identifying families of multipartite states with non-trivial local
  entanglement transformations},}\ }\href@noop {} {\bibfield  {journal}
  {\bibinfo  {journal} {Quantum}\ }\textbf {\bibinfo {volume} {8}},\ \bibinfo
  {pages} {1270} (\bibinfo {year} {2024})}\BibitemShut {NoStop}%
\bibitem [{\citenamefont {Gunn}\ \emph {et~al.}(2023)\citenamefont {Gunn},
  \citenamefont {Hebenstreit}, \citenamefont {Spee}, \citenamefont
  {de~Vicente},\ and\ \citenamefont {Kraus}}]{gunn2023approximate}%
  \BibitemOpen
  \bibfield  {author} {\bibinfo {author} {\bibfnamefont {David}\ \bibnamefont
  {Gunn}}, \bibinfo {author} {\bibfnamefont {Martin}\ \bibnamefont
  {Hebenstreit}}, \bibinfo {author} {\bibfnamefont {Cornelia}\ \bibnamefont
  {Spee}}, \bibinfo {author} {\bibfnamefont {Julio~I}\ \bibnamefont
  {de~Vicente}}, \ and\ \bibinfo {author} {\bibfnamefont {Barbara}\
  \bibnamefont {Kraus}},\ }\bibfield  {title} {\enquote {\bibinfo {title}
  {Approximate and ensemble local entanglement transformations for multipartite
  states},}\ }\href@noop {} {\bibfield  {journal} {\bibinfo  {journal}
  {Physical Review A}\ }\textbf {\bibinfo {volume} {108}},\ \bibinfo {pages}
  {052401} (\bibinfo {year} {2023})}\BibitemShut {NoStop}%
\bibitem [{\citenamefont {Ma}\ \emph {et~al.}(2011)\citenamefont {Ma},
  \citenamefont {Chen}, \citenamefont {Chen}, \citenamefont {Spengler},
  \citenamefont {Gabriel},\ and\ \citenamefont {Huber}}]{ma2011measure}%
  \BibitemOpen
  \bibfield  {author} {\bibinfo {author} {\bibfnamefont {Zhi-Hao}\ \bibnamefont
  {Ma}}, \bibinfo {author} {\bibfnamefont {Zhi-Hua}\ \bibnamefont {Chen}},
  \bibinfo {author} {\bibfnamefont {Jing-Ling}\ \bibnamefont {Chen}}, \bibinfo
  {author} {\bibfnamefont {Christoph}\ \bibnamefont {Spengler}}, \bibinfo
  {author} {\bibfnamefont {Andreas}\ \bibnamefont {Gabriel}}, \ and\ \bibinfo
  {author} {\bibfnamefont {Marcus}\ \bibnamefont {Huber}},\ }\bibfield  {title}
  {\enquote {\bibinfo {title} {Measure of genuine multipartite entanglement
  with computable lower bounds},}\ }\href@noop {} {\bibfield  {journal}
  {\bibinfo  {journal} {Physical Review A—Atomic, Molecular, and Optical
  Physics}\ }\textbf {\bibinfo {volume} {83}},\ \bibinfo {pages} {062325}
  (\bibinfo {year} {2011})}\BibitemShut {NoStop}%
\bibitem [{\citenamefont {Xie}\ and\ \citenamefont
  {Eberly}(2021)}]{xie2021triangle}%
  \BibitemOpen
  \bibfield  {author} {\bibinfo {author} {\bibfnamefont {Songbo}\ \bibnamefont
  {Xie}}\ and\ \bibinfo {author} {\bibfnamefont {Joseph~H}\ \bibnamefont
  {Eberly}},\ }\bibfield  {title} {\enquote {\bibinfo {title} {Triangle measure
  of tripartite entanglement},}\ }\href@noop {} {\bibfield  {journal} {\bibinfo
   {journal} {Physical Review Letters}\ }\textbf {\bibinfo {volume} {127}},\
  \bibinfo {pages} {040403} (\bibinfo {year} {2021})}\BibitemShut {NoStop}%
\bibitem [{\citenamefont {Pironio}\ \emph {et~al.}(2010)\citenamefont
  {Pironio}, \citenamefont {Ac{\'{\i}}n}, \citenamefont {Massar}, \citenamefont
  {de~la Giroday}, \citenamefont {Matsukevich}, \citenamefont {Maunz},
  \citenamefont {Olmschenk}, \citenamefont {Hayes}, \citenamefont {Luo},
  \citenamefont {Manning},\ and\ \citenamefont
  {Monroe}}]{randomnessextraction}%
  \BibitemOpen
  \bibfield  {author} {\bibinfo {author} {\bibfnamefont {S.}~\bibnamefont
  {Pironio}}, \bibinfo {author} {\bibfnamefont {A.}~\bibnamefont
  {Ac{\'{\i}}n}}, \bibinfo {author} {\bibfnamefont {S.}~\bibnamefont {Massar}},
  \bibinfo {author} {\bibfnamefont {A.~Boyer}\ \bibnamefont {de~la Giroday}},
  \bibinfo {author} {\bibfnamefont {D.~N.}\ \bibnamefont {Matsukevich}},
  \bibinfo {author} {\bibfnamefont {P.}~\bibnamefont {Maunz}}, \bibinfo
  {author} {\bibfnamefont {S.}~\bibnamefont {Olmschenk}}, \bibinfo {author}
  {\bibfnamefont {D.}~\bibnamefont {Hayes}}, \bibinfo {author} {\bibfnamefont
  {L.}~\bibnamefont {Luo}}, \bibinfo {author} {\bibfnamefont {T.~A.}\
  \bibnamefont {Manning}}, \ and\ \bibinfo {author} {\bibfnamefont
  {C.}~\bibnamefont {Monroe}},\ }\bibfield  {title} {\enquote {\bibinfo {title}
  {Random numbers certified by bell's theorem},}\ }\href {\doibase
  10.1038/nature09008} {\bibfield  {journal} {\bibinfo  {journal} {Nature}\
  }\textbf {\bibinfo {volume} {464}},\ \bibinfo {pages} {1021--1024} (\bibinfo
  {year} {2010})}\BibitemShut {NoStop}%
\bibitem [{\citenamefont {Moreno}\ \emph {et~al.}(2020)\citenamefont {Moreno},
  \citenamefont {Brito}, \citenamefont {Nery},\ and\ \citenamefont
  {Chaves}}]{DImoreno2020device}%
  \BibitemOpen
  \bibfield  {author} {\bibinfo {author} {\bibfnamefont {MGM}\ \bibnamefont
  {Moreno}}, \bibinfo {author} {\bibfnamefont {Samura{\'\i}}\ \bibnamefont
  {Brito}}, \bibinfo {author} {\bibfnamefont {Ranieri~V}\ \bibnamefont {Nery}},
  \ and\ \bibinfo {author} {\bibfnamefont {Rafael}\ \bibnamefont {Chaves}},\
  }\bibfield  {title} {\enquote {\bibinfo {title} {Device-independent secret
  sharing and a stronger form of bell nonlocality},}\ }\href@noop {} {\bibfield
   {journal} {\bibinfo  {journal} {Physical Review A}\ }\textbf {\bibinfo
  {volume} {101}},\ \bibinfo {pages} {052339} (\bibinfo {year}
  {2020})}\BibitemShut {NoStop}%
\bibitem [{\citenamefont {Rai}\ \emph {et~al.}(2019)\citenamefont {Rai},
  \citenamefont {Duarte}, \citenamefont {Brito},\ and\ \citenamefont
  {Chaves}}]{AshutoshRaiGeometryofquantumset2019}%
  \BibitemOpen
  \bibfield  {author} {\bibinfo {author} {\bibfnamefont {Ashutosh}\
  \bibnamefont {Rai}}, \bibinfo {author} {\bibfnamefont {Cristhiano}\
  \bibnamefont {Duarte}}, \bibinfo {author} {\bibfnamefont {Samura\'{\i}}\
  \bibnamefont {Brito}}, \ and\ \bibinfo {author} {\bibfnamefont {Rafael}\
  \bibnamefont {Chaves}},\ }\bibfield  {title} {\enquote {\bibinfo {title}
  {Geometry of the quantum set on no-signaling faces},}\ }\href {\doibase
  10.1103/PhysRevA.99.032106} {\bibfield  {journal} {\bibinfo  {journal} {Phys.
  Rev. A}\ }\textbf {\bibinfo {volume} {99}},\ \bibinfo {pages} {032106}
  (\bibinfo {year} {2019})}\BibitemShut {NoStop}%
\bibitem [{\citenamefont {Cirel'son}(1980)}]{cirel1980quantum}%
  \BibitemOpen
  \bibfield  {author} {\bibinfo {author} {\bibfnamefont {Boris~S}\ \bibnamefont
  {Cirel'son}},\ }\bibfield  {title} {\enquote {\bibinfo {title} {Quantum
  generalizations of bell's inequality},}\ }\href@noop {} {\bibfield  {journal}
  {\bibinfo  {journal} {Letters in Mathematical Physics}\ }\textbf {\bibinfo
  {volume} {4}},\ \bibinfo {pages} {93--100} (\bibinfo {year}
  {1980})}\BibitemShut {NoStop}%
\bibitem [{\citenamefont {Popescu}\ and\ \citenamefont
  {Rohrlich}(1994)}]{popescu1994quantum}%
  \BibitemOpen
  \bibfield  {author} {\bibinfo {author} {\bibfnamefont {Sandu}\ \bibnamefont
  {Popescu}}\ and\ \bibinfo {author} {\bibfnamefont {Daniel}\ \bibnamefont
  {Rohrlich}},\ }\bibfield  {title} {\enquote {\bibinfo {title} {Quantum
  nonlocality as an axiom},}\ }\href@noop {} {\bibfield  {journal} {\bibinfo
  {journal} {Foundations of Physics}\ }\textbf {\bibinfo {volume} {24}},\
  \bibinfo {pages} {379--385} (\bibinfo {year} {1994})}\BibitemShut {NoStop}%
\bibitem [{\citenamefont {Schmid}\ \emph {et~al.}(2023)\citenamefont {Schmid},
  \citenamefont {Fraser}, \citenamefont {Kunjwal}, \citenamefont {Sainz},
  \citenamefont {Wolfe},\ and\ \citenamefont
  {Spekkens}}]{schmid2023understanding}%
  \BibitemOpen
  \bibfield  {author} {\bibinfo {author} {\bibfnamefont {David}\ \bibnamefont
  {Schmid}}, \bibinfo {author} {\bibfnamefont {Thomas~C}\ \bibnamefont
  {Fraser}}, \bibinfo {author} {\bibfnamefont {Ravi}\ \bibnamefont {Kunjwal}},
  \bibinfo {author} {\bibfnamefont {Ana~Belen}\ \bibnamefont {Sainz}}, \bibinfo
  {author} {\bibfnamefont {Elie}\ \bibnamefont {Wolfe}}, \ and\ \bibinfo
  {author} {\bibfnamefont {Robert~W}\ \bibnamefont {Spekkens}},\ }\bibfield
  {title} {\enquote {\bibinfo {title} {Understanding the interplay of
  entanglement and nonlocality: motivating and developing a new branch of
  entanglement theory},}\ }\href@noop {} {\bibfield  {journal} {\bibinfo
  {journal} {Quantum}\ }\textbf {\bibinfo {volume} {7}},\ \bibinfo {pages}
  {1194} (\bibinfo {year} {2023})}\BibitemShut {NoStop}%
\bibitem [{\citenamefont {Schmid}\ \emph {et~al.}(2020)\citenamefont {Schmid},
  \citenamefont {Rosset},\ and\ \citenamefont
  {Buscemi}}]{LOSRSchmid2020typeindependent}%
  \BibitemOpen
  \bibfield  {author} {\bibinfo {author} {\bibfnamefont {David}\ \bibnamefont
  {Schmid}}, \bibinfo {author} {\bibfnamefont {Denis}\ \bibnamefont {Rosset}},
  \ and\ \bibinfo {author} {\bibfnamefont {Francesco}\ \bibnamefont
  {Buscemi}},\ }\bibfield  {title} {\enquote {\bibinfo {title} {The
  type-independent resource theory of local operations and shared
  randomness},}\ }\href {\doibase 10.22331/q-2020-04-30-262} {\bibfield
  {journal} {\bibinfo  {journal} {{Quantum}}\ }\textbf {\bibinfo {volume}
  {4}},\ \bibinfo {pages} {262} (\bibinfo {year} {2020})}\BibitemShut {NoStop}%
\bibitem [{\citenamefont {Rosset}\ \emph {et~al.}(2020)\citenamefont {Rosset},
  \citenamefont {Schmid},\ and\ \citenamefont {Buscemi}}]{LOSRTRosset2020}%
  \BibitemOpen
  \bibfield  {author} {\bibinfo {author} {\bibfnamefont {Denis}\ \bibnamefont
  {Rosset}}, \bibinfo {author} {\bibfnamefont {David}\ \bibnamefont {Schmid}},
  \ and\ \bibinfo {author} {\bibfnamefont {Francesco}\ \bibnamefont
  {Buscemi}},\ }\bibfield  {title} {\enquote {\bibinfo {title}
  {Type-independent characterization of spacelike separated resources},}\
  }\href {\doibase 10.1103/PhysRevLett.125.210402} {\bibfield  {journal}
  {\bibinfo  {journal} {Phys. Rev. Lett.}\ }\textbf {\bibinfo {volume} {125}},\
  \bibinfo {pages} {210402} (\bibinfo {year} {2020})}\BibitemShut {NoStop}%
\bibitem [{\citenamefont {Gallego}\ \emph {et~al.}(2012)\citenamefont
  {Gallego}, \citenamefont {W{\"u}rflinger}, \citenamefont {Ac{\'\i}n},\ and\
  \citenamefont {Navascu{\'e}s}}]{gallego2012operational}%
  \BibitemOpen
  \bibfield  {author} {\bibinfo {author} {\bibfnamefont {Rodrigo}\ \bibnamefont
  {Gallego}}, \bibinfo {author} {\bibfnamefont {Lars~Erik}\ \bibnamefont
  {W{\"u}rflinger}}, \bibinfo {author} {\bibfnamefont {Antonio}\ \bibnamefont
  {Ac{\'\i}n}}, \ and\ \bibinfo {author} {\bibfnamefont {Miguel}\ \bibnamefont
  {Navascu{\'e}s}},\ }\bibfield  {title} {\enquote {\bibinfo {title}
  {Operational framework for nonlocality},}\ }\href@noop {} {\bibfield
  {journal} {\bibinfo  {journal} {Physical review letters}\ }\textbf {\bibinfo
  {volume} {109}},\ \bibinfo {pages} {070401} (\bibinfo {year}
  {2012})}\BibitemShut {NoStop}%
\bibitem [{\citenamefont {Dutta}\ \emph {et~al.}(2020)\citenamefont {Dutta},
  \citenamefont {Mukherjee},\ and\ \citenamefont
  {Banik}}]{Amit2020Operationalcharacterization}%
  \BibitemOpen
  \bibfield  {author} {\bibinfo {author} {\bibfnamefont {Sagnik}\ \bibnamefont
  {Dutta}}, \bibinfo {author} {\bibfnamefont {Amit}\ \bibnamefont {Mukherjee}},
  \ and\ \bibinfo {author} {\bibfnamefont {Manik}\ \bibnamefont {Banik}},\
  }\bibfield  {title} {\enquote {\bibinfo {title} {Operational characterization
  of multipartite nonlocal correlations},}\ }\href {\doibase
  10.1103/PhysRevA.102.052218} {\bibfield  {journal} {\bibinfo  {journal}
  {Phys. Rev. A}\ }\textbf {\bibinfo {volume} {102}},\ \bibinfo {pages}
  {052218} (\bibinfo {year} {2020})}\BibitemShut {NoStop}%
\bibitem [{\citenamefont {Bancal}\ \emph {et~al.}(2011)\citenamefont {Bancal},
  \citenamefont {Barrett}, \citenamefont {Gisin},\ and\ \citenamefont
  {Pironio}}]{bancal2011definition}%
  \BibitemOpen
  \bibfield  {author} {\bibinfo {author} {\bibfnamefont {Jean-Daniel}\
  \bibnamefont {Bancal}}, \bibinfo {author} {\bibfnamefont {Jonathan}\
  \bibnamefont {Barrett}}, \bibinfo {author} {\bibfnamefont {Nicolas}\
  \bibnamefont {Gisin}}, \ and\ \bibinfo {author} {\bibfnamefont {Stefano}\
  \bibnamefont {Pironio}},\ }\bibfield  {title} {\enquote {\bibinfo {title}
  {The definition of multipartite nonlocality},}\ }\href@noop {} {\bibfield
  {journal} {\bibinfo  {journal} {arXiv preprint arXiv:1112.2626}\ } (\bibinfo
  {year} {2011})}\BibitemShut {NoStop}%
\bibitem [{\citenamefont {Ghosh}\ \emph {et~al.}(2024)\citenamefont {Ghosh},
  \citenamefont {Chowdhury}, \citenamefont {Kar}, \citenamefont {Roy},
  \citenamefont {Guha},\ and\ \citenamefont {Banik}}]{ghosh2024quantum}%
  \BibitemOpen
  \bibfield  {author} {\bibinfo {author} {\bibfnamefont {Subhendu~B}\
  \bibnamefont {Ghosh}}, \bibinfo {author} {\bibfnamefont {Snehasish~Roy}\
  \bibnamefont {Chowdhury}}, \bibinfo {author} {\bibfnamefont {Guruprasad}\
  \bibnamefont {Kar}}, \bibinfo {author} {\bibfnamefont {Arup}\ \bibnamefont
  {Roy}}, \bibinfo {author} {\bibfnamefont {Tamal}\ \bibnamefont {Guha}}, \
  and\ \bibinfo {author} {\bibfnamefont {Manik}\ \bibnamefont {Banik}},\
  }\bibfield  {title} {\enquote {\bibinfo {title} {Quantum nonlocality:
  Multicopy resource interconvertibility and their asymptotic inequivalence},}\
  }\href@noop {} {\bibfield  {journal} {\bibinfo  {journal} {Physical Review
  Letters}\ }\textbf {\bibinfo {volume} {132}},\ \bibinfo {pages} {250205}
  (\bibinfo {year} {2024})}\BibitemShut {NoStop}%
\bibitem [{\citenamefont {Abramsky}\ and\ \citenamefont
  {Brandenburger}(2011)}]{abramsky2011sheaf}%
  \BibitemOpen
  \bibfield  {author} {\bibinfo {author} {\bibfnamefont {Samson}\ \bibnamefont
  {Abramsky}}\ and\ \bibinfo {author} {\bibfnamefont {Adam}\ \bibnamefont
  {Brandenburger}},\ }\bibfield  {title} {\enquote {\bibinfo {title} {The
  sheaf-theoretic structure of non-locality and contextuality},}\ }\href@noop
  {} {\bibfield  {journal} {\bibinfo  {journal} {New Journal of Physics}\
  }\textbf {\bibinfo {volume} {13}},\ \bibinfo {pages} {113036} (\bibinfo
  {year} {2011})}\BibitemShut {NoStop}%
\bibitem [{\citenamefont {Kochen}\ and\ \citenamefont
  {Specker}(2011)}]{kS2011problem}%
  \BibitemOpen
  \bibfield  {author} {\bibinfo {author} {\bibfnamefont {Simon}\ \bibnamefont
  {Kochen}}\ and\ \bibinfo {author} {\bibfnamefont {Ernst~P}\ \bibnamefont
  {Specker}},\ }\bibfield  {title} {\enquote {\bibinfo {title} {The problem of
  hidden variables in quantum mechanics},}\ }in\ \href@noop {} {\emph {\bibinfo
  {booktitle} {Ernst Specker Selecta}}}\ (\bibinfo  {publisher} {Springer},\
  \bibinfo {year} {2011})\ pp.\ \bibinfo {pages} {235--263}\BibitemShut
  {NoStop}%
\bibitem [{\citenamefont {Mermin}(1990{\natexlab{a}})}]{mermin1990simple}%
  \BibitemOpen
  \bibfield  {author} {\bibinfo {author} {\bibfnamefont {N~David}\ \bibnamefont
  {Mermin}},\ }\bibfield  {title} {\enquote {\bibinfo {title} {Simple unified
  form for the major no-hidden-variables theorems},}\ }\href@noop {} {\bibfield
   {journal} {\bibinfo  {journal} {Physical review letters}\ }\textbf {\bibinfo
  {volume} {65}},\ \bibinfo {pages} {3373} (\bibinfo {year}
  {1990}{\natexlab{a}})}\BibitemShut {NoStop}%
\bibitem [{\citenamefont {Cabello}\ \emph {et~al.}(1996)\citenamefont
  {Cabello}, \citenamefont {Estebaranz},\ and\ \citenamefont
  {Garc{\'\i}a-Alcaine}}]{18vectorcabello1996bell}%
  \BibitemOpen
  \bibfield  {author} {\bibinfo {author} {\bibfnamefont {Ad{\'a}n}\
  \bibnamefont {Cabello}}, \bibinfo {author} {\bibfnamefont {Jos{\'e}M}\
  \bibnamefont {Estebaranz}}, \ and\ \bibinfo {author} {\bibfnamefont
  {Guillermo}\ \bibnamefont {Garc{\'\i}a-Alcaine}},\ }\bibfield  {title}
  {\enquote {\bibinfo {title} {Bell-kochen-specker theorem: A proof with 18
  vectors},}\ }\href@noop {} {\bibfield  {journal} {\bibinfo  {journal}
  {Physics Letters A}\ }\textbf {\bibinfo {volume} {212}},\ \bibinfo {pages}
  {183--187} (\bibinfo {year} {1996})}\BibitemShut {NoStop}%
\bibitem [{\citenamefont {Yu}\ and\ \citenamefont
  {Oh}(2012)}]{13vector2012state}%
  \BibitemOpen
  \bibfield  {author} {\bibinfo {author} {\bibfnamefont {Sixia}\ \bibnamefont
  {Yu}}\ and\ \bibinfo {author} {\bibfnamefont {Choo~Hiap}\ \bibnamefont
  {Oh}},\ }\bibfield  {title} {\enquote {\bibinfo {title} {State-independent
  proof of kochen-specker theorem with 13 rays},}\ }\href@noop {} {\bibfield
  {journal} {\bibinfo  {journal} {Physical review letters}\ }\textbf {\bibinfo
  {volume} {108}},\ \bibinfo {pages} {030402} (\bibinfo {year}
  {2012})}\BibitemShut {NoStop}%
\bibitem [{\citenamefont {Fine}(1982)}]{fine1982hidden}%
  \BibitemOpen
  \bibfield  {author} {\bibinfo {author} {\bibfnamefont {Arthur}\ \bibnamefont
  {Fine}},\ }\bibfield  {title} {\enquote {\bibinfo {title} {Hidden variables,
  joint probability, and the bell inequalities},}\ }\href@noop {} {\bibfield
  {journal} {\bibinfo  {journal} {Physical Review Letters}\ }\textbf {\bibinfo
  {volume} {48}},\ \bibinfo {pages} {291} (\bibinfo {year} {1982})}\BibitemShut
  {NoStop}%
\bibitem [{\citenamefont {Clauser}\ \emph {et~al.}(1969)\citenamefont
  {Clauser}, \citenamefont {Horne}, \citenamefont {Shimony},\ and\
  \citenamefont {Holt}}]{clauser1969proposed}%
  \BibitemOpen
  \bibfield  {author} {\bibinfo {author} {\bibfnamefont {John~F}\ \bibnamefont
  {Clauser}}, \bibinfo {author} {\bibfnamefont {Michael~A}\ \bibnamefont
  {Horne}}, \bibinfo {author} {\bibfnamefont {Abner}\ \bibnamefont {Shimony}},
  \ and\ \bibinfo {author} {\bibfnamefont {Richard~A}\ \bibnamefont {Holt}},\
  }\bibfield  {title} {\enquote {\bibinfo {title} {Proposed experiment to test
  local hidden-variable theories},}\ }\href@noop {} {\bibfield  {journal}
  {\bibinfo  {journal} {Physical review letters}\ }\textbf {\bibinfo {volume}
  {23}},\ \bibinfo {pages} {880} (\bibinfo {year} {1969})}\BibitemShut
  {NoStop}%
\bibitem [{\citenamefont {Klyachko}\ \emph {et~al.}(2008)\citenamefont
  {Klyachko}, \citenamefont {Can}, \citenamefont {Binicio{\u{g}}lu},\ and\
  \citenamefont {Shumovsky}}]{kcbs2008simple}%
  \BibitemOpen
  \bibfield  {author} {\bibinfo {author} {\bibfnamefont {Alexander~A}\
  \bibnamefont {Klyachko}}, \bibinfo {author} {\bibfnamefont {M~Ali}\
  \bibnamefont {Can}}, \bibinfo {author} {\bibfnamefont {Sinem}\ \bibnamefont
  {Binicio{\u{g}}lu}}, \ and\ \bibinfo {author} {\bibfnamefont {Alexander~S}\
  \bibnamefont {Shumovsky}},\ }\bibfield  {title} {\enquote {\bibinfo {title}
  {Simple test for hidden variables in spin-1 systems},}\ }\href@noop {}
  {\bibfield  {journal} {\bibinfo  {journal} {Physical review letters}\
  }\textbf {\bibinfo {volume} {101}},\ \bibinfo {pages} {020403} (\bibinfo
  {year} {2008})}\BibitemShut {NoStop}%
\bibitem [{\citenamefont {Abramsky}\ \emph {et~al.}(2017)\citenamefont
  {Abramsky}, \citenamefont {Barbosa},\ and\ \citenamefont {Mansfield}}]{CF}%
  \BibitemOpen
  \bibfield  {author} {\bibinfo {author} {\bibfnamefont {Samson}\ \bibnamefont
  {Abramsky}}, \bibinfo {author} {\bibfnamefont {Rui~Soares}\ \bibnamefont
  {Barbosa}}, \ and\ \bibinfo {author} {\bibfnamefont {Shane}\ \bibnamefont
  {Mansfield}},\ }\bibfield  {title} {\enquote {\bibinfo {title} {Contextual
  fraction as a measure of contextuality},}\ }\href {\doibase
  10.1103/PhysRevLett.119.050504} {\bibfield  {journal} {\bibinfo  {journal}
  {Phys. Rev. Lett.}\ }\textbf {\bibinfo {volume} {119}},\ \bibinfo {pages}
  {050504} (\bibinfo {year} {2017})}\BibitemShut {NoStop}%
\bibitem [{\citenamefont {Anders}\ and\ \citenamefont
  {Browne}(2009)}]{comppower}%
  \BibitemOpen
  \bibfield  {author} {\bibinfo {author} {\bibfnamefont {Janet}\ \bibnamefont
  {Anders}}\ and\ \bibinfo {author} {\bibfnamefont {Dan~E.}\ \bibnamefont
  {Browne}},\ }\bibfield  {title} {\enquote {\bibinfo {title} {Computational
  power of correlations},}\ }\href {\doibase 10.1103/PhysRevLett.102.050502}
  {\bibfield  {journal} {\bibinfo  {journal} {Phys. Rev. Lett.}\ }\textbf
  {\bibinfo {volume} {102}},\ \bibinfo {pages} {050502} (\bibinfo {year}
  {2009})}\BibitemShut {NoStop}%
\bibitem [{\citenamefont {Raussendorf}(2013)}]{CMBQC}%
  \BibitemOpen
  \bibfield  {author} {\bibinfo {author} {\bibfnamefont {Robert}\ \bibnamefont
  {Raussendorf}},\ }\bibfield  {title} {\enquote {\bibinfo {title}
  {Contextuality in measurement-based quantum computation},}\ }\href {\doibase
  10.1103/PhysRevA.88.022322} {\bibfield  {journal} {\bibinfo  {journal} {Phys.
  Rev. A}\ }\textbf {\bibinfo {volume} {88}},\ \bibinfo {pages} {022322}
  (\bibinfo {year} {2013})}\BibitemShut {NoStop}%
\bibitem [{\citenamefont {Okay}\ \emph {et~al.}(2017)\citenamefont {Okay},
  \citenamefont {Roberts}, \citenamefont {Bartlett},\ and\ \citenamefont
  {Raussendorf}}]{topologicalproofscontextualityquantum}%
  \BibitemOpen
  \bibfield  {author} {\bibinfo {author} {\bibfnamefont {Cihan}\ \bibnamefont
  {Okay}}, \bibinfo {author} {\bibfnamefont {Sam}\ \bibnamefont {Roberts}},
  \bibinfo {author} {\bibfnamefont {Stephen~D.}\ \bibnamefont {Bartlett}}, \
  and\ \bibinfo {author} {\bibfnamefont {Robert}\ \bibnamefont {Raussendorf}},\
  }\href {https://arxiv.org/abs/1701.01888} {\enquote {\bibinfo {title}
  {Topological proofs of contextuality in quantum mechanics},}\ } (\bibinfo
  {year} {2017}),\ \Eprint {http://arxiv.org/abs/1701.01888} {arXiv:1701.01888
  [quant-ph]} \BibitemShut {NoStop}%
\bibitem [{\citenamefont {Abramsky}\ \emph {et~al.}(2012)\citenamefont
  {Abramsky}, \citenamefont {Mansfield},\ and\ \citenamefont
  {Barbosa}}]{cohomandnonlocalityandcontextuality}%
  \BibitemOpen
  \bibfield  {author} {\bibinfo {author} {\bibfnamefont {Samson}\ \bibnamefont
  {Abramsky}}, \bibinfo {author} {\bibfnamefont {Shane}\ \bibnamefont
  {Mansfield}}, \ and\ \bibinfo {author} {\bibfnamefont {Rui~Soares}\
  \bibnamefont {Barbosa}},\ }\bibfield  {title} {\enquote {\bibinfo {title}
  {The cohomology of non-locality and contextuality},}\ }\href {\doibase
  10.4204/eptcs.95.1} {\bibfield  {journal} {\bibinfo  {journal} {Electronic
  Proceedings in Theoretical Computer Science}\ }\textbf {\bibinfo {volume}
  {95}},\ \bibinfo {pages} {1–14} (\bibinfo {year} {2012})}\BibitemShut
  {NoStop}%
\bibitem [{\citenamefont
  {Raussendorf}(2019)}]{cohomologicalframeworkcontextualquantum}%
  \BibitemOpen
  \bibfield  {author} {\bibinfo {author} {\bibfnamefont {Robert}\ \bibnamefont
  {Raussendorf}},\ }\href {https://arxiv.org/abs/1602.04155} {\enquote
  {\bibinfo {title} {Cohomological framework for contextual quantum
  computations},}\ } (\bibinfo {year} {2019}),\ \Eprint
  {http://arxiv.org/abs/1602.04155} {arXiv:1602.04155 [quant-ph]} \BibitemShut
  {NoStop}%
\bibitem [{\citenamefont {Frembs}\ \emph {et~al.}(2018)\citenamefont {Frembs},
  \citenamefont {Roberts},\ and\ \citenamefont
  {Bartlett}}]{ContextualityasresourceinMBQC}%
  \BibitemOpen
  \bibfield  {author} {\bibinfo {author} {\bibfnamefont {Markus}\ \bibnamefont
  {Frembs}}, \bibinfo {author} {\bibfnamefont {Sam}\ \bibnamefont {Roberts}}, \
  and\ \bibinfo {author} {\bibfnamefont {Stephen~D}\ \bibnamefont {Bartlett}},\
  }\bibfield  {title} {\enquote {\bibinfo {title} {Contextuality as a resource
  for measurement-based quantum computation beyond qubits},}\ }\href {\doibase
  10.1088/1367-2630/aae3ad} {\bibfield  {journal} {\bibinfo  {journal} {New
  Journal of Physics}\ }\textbf {\bibinfo {volume} {20}},\ \bibinfo {pages}
  {103011} (\bibinfo {year} {2018})}\BibitemShut {NoStop}%
\bibitem [{\citenamefont {Walleghem}\ \emph {et~al.}(2024)\citenamefont
  {Walleghem}, \citenamefont {Barbosa}, \citenamefont {Pusey},\ and\
  \citenamefont {Weigert}}]{walleghem2024refinedfrauchigerrennerparadoxbased}%
  \BibitemOpen
  \bibfield  {author} {\bibinfo {author} {\bibfnamefont {Laurens}\ \bibnamefont
  {Walleghem}}, \bibinfo {author} {\bibfnamefont {Rui~Soares}\ \bibnamefont
  {Barbosa}}, \bibinfo {author} {\bibfnamefont {Matthew}\ \bibnamefont
  {Pusey}}, \ and\ \bibinfo {author} {\bibfnamefont {Stefan}\ \bibnamefont
  {Weigert}},\ }\href {https://arxiv.org/abs/2409.05491} {\enquote {\bibinfo
  {title} {A refined frauchiger--renner paradox based on strong
  contextuality},}\ } (\bibinfo {year} {2024}),\ \Eprint
  {http://arxiv.org/abs/2409.05491} {arXiv:2409.05491 [quant-ph]} \BibitemShut
  {NoStop}%
\bibitem [{\citenamefont {Raussendorf}\ \emph {et~al.}(2023)\citenamefont
  {Raussendorf}, \citenamefont {Yang},\ and\ \citenamefont
  {Adhikary}}]{stringordermeasurementbased}%
  \BibitemOpen
  \bibfield  {author} {\bibinfo {author} {\bibfnamefont {Robert}\ \bibnamefont
  {Raussendorf}}, \bibinfo {author} {\bibfnamefont {Wang}\ \bibnamefont
  {Yang}}, \ and\ \bibinfo {author} {\bibfnamefont {Arnab}\ \bibnamefont
  {Adhikary}},\ }\bibfield  {title} {\enquote {\bibinfo {title}
  {Measurement-based quantum computation in finite one-dimensional systems:
  string order implies computational power},}\ }\href {\doibase
  10.22331/q-2023-12-28-1215} {\bibfield  {journal} {\bibinfo  {journal}
  {{Quantum}}\ }\textbf {\bibinfo {volume} {7}},\ \bibinfo {pages} {1215}
  (\bibinfo {year} {2023})}\BibitemShut {NoStop}%
\bibitem [{\citenamefont {Okay}\ and\ \citenamefont
  {Raussendorf}(2020)}]{homotopicalapproach}%
  \BibitemOpen
  \bibfield  {author} {\bibinfo {author} {\bibfnamefont {Cihan}\ \bibnamefont
  {Okay}}\ and\ \bibinfo {author} {\bibfnamefont {Robert}\ \bibnamefont
  {Raussendorf}},\ }\bibfield  {title} {\enquote {\bibinfo {title} {Homotopical
  approach to quantum contextuality},}\ }\href {\doibase
  10.22331/q-2020-01-05-217} {\bibfield  {journal} {\bibinfo  {journal}
  {{Quantum}}\ }\textbf {\bibinfo {volume} {4}},\ \bibinfo {pages} {217}
  (\bibinfo {year} {2020})}\BibitemShut {NoStop}%
\bibitem [{\citenamefont {Xu}\ \emph {et~al.}(2020)\citenamefont {Xu},
  \citenamefont {Chen},\ and\ \citenamefont
  {G\"uhne}}]{Peresconjectureforcontextuality}%
  \BibitemOpen
  \bibfield  {author} {\bibinfo {author} {\bibfnamefont {Zhen-Peng}\
  \bibnamefont {Xu}}, \bibinfo {author} {\bibfnamefont {Jing-Ling}\
  \bibnamefont {Chen}}, \ and\ \bibinfo {author} {\bibfnamefont {Otfried}\
  \bibnamefont {G\"uhne}},\ }\bibfield  {title} {\enquote {\bibinfo {title}
  {Proof of the peres conjecture for contextuality},}\ }\href {\doibase
  10.1103/PhysRevLett.124.230401} {\bibfield  {journal} {\bibinfo  {journal}
  {Phys. Rev. Lett.}\ }\textbf {\bibinfo {volume} {124}},\ \bibinfo {pages}
  {230401} (\bibinfo {year} {2020})}\BibitemShut {NoStop}%
\bibitem [{\citenamefont {Saha}\ \emph {et~al.}(2019)\citenamefont {Saha},
  \citenamefont {Horodecki},\ and\ \citenamefont
  {Pawłowski}}]{contextualityadvancesonewaycommunication}%
  \BibitemOpen
  \bibfield  {author} {\bibinfo {author} {\bibfnamefont {Debashis}\
  \bibnamefont {Saha}}, \bibinfo {author} {\bibfnamefont {Paweł}\ \bibnamefont
  {Horodecki}}, \ and\ \bibinfo {author} {\bibfnamefont {Marcin}\ \bibnamefont
  {Pawłowski}},\ }\bibfield  {title} {\enquote {\bibinfo {title} {State
  independent contextuality advances one-way communication},}\ }\href {\doibase
  10.1088/1367-2630/ab4149} {\bibfield  {journal} {\bibinfo  {journal} {New
  Journal of Physics}\ }\textbf {\bibinfo {volume} {21}},\ \bibinfo {pages}
  {093057} (\bibinfo {year} {2019})}\BibitemShut {NoStop}%
\bibitem [{\citenamefont {Beer}\ and\ \citenamefont
  {Osborne}(2018)}]{CinBundlediagram}%
  \BibitemOpen
  \bibfield  {author} {\bibinfo {author} {\bibfnamefont {Kerstin}\ \bibnamefont
  {Beer}}\ and\ \bibinfo {author} {\bibfnamefont {Tobias~J.}\ \bibnamefont
  {Osborne}},\ }\bibfield  {title} {\enquote {\bibinfo {title} {Contextuality
  and bundle diagrams},}\ }\href {\doibase 10.1103/PhysRevA.98.052124}
  {\bibfield  {journal} {\bibinfo  {journal} {Phys. Rev. A}\ }\textbf {\bibinfo
  {volume} {98}},\ \bibinfo {pages} {052124} (\bibinfo {year}
  {2018})}\BibitemShut {NoStop}%
\bibitem [{\citenamefont {Kirchmair}\ \emph {et~al.}(2009)\citenamefont
  {Kirchmair}, \citenamefont {Z{\"a}hringer}, \citenamefont {Gerritsma},
  \citenamefont {Kleinmann}, \citenamefont {G{\"u}hne}, \citenamefont
  {Cabello}, \citenamefont {Blatt},\ and\ \citenamefont
  {Roos}}]{Merminsquareexperiment2009}%
  \BibitemOpen
  \bibfield  {author} {\bibinfo {author} {\bibfnamefont {G.}~\bibnamefont
  {Kirchmair}}, \bibinfo {author} {\bibfnamefont {F.}~\bibnamefont
  {Z{\"a}hringer}}, \bibinfo {author} {\bibfnamefont {R.}~\bibnamefont
  {Gerritsma}}, \bibinfo {author} {\bibfnamefont {M.}~\bibnamefont
  {Kleinmann}}, \bibinfo {author} {\bibfnamefont {O.}~\bibnamefont
  {G{\"u}hne}}, \bibinfo {author} {\bibfnamefont {A.}~\bibnamefont {Cabello}},
  \bibinfo {author} {\bibfnamefont {R.}~\bibnamefont {Blatt}}, \ and\ \bibinfo
  {author} {\bibfnamefont {C.~F.}\ \bibnamefont {Roos}},\ }\bibfield  {title}
  {\enquote {\bibinfo {title} {State-independent experimental test of quantum
  contextuality},}\ }\href {\doibase 10.1038/nature08172} {\bibfield  {journal}
  {\bibinfo  {journal} {Nature}\ }\textbf {\bibinfo {volume} {460}},\ \bibinfo
  {pages} {494--497} (\bibinfo {year} {2009})}\BibitemShut {NoStop}%
\bibitem [{\citenamefont {Li}\ \emph {et~al.}(2017)\citenamefont {Li},
  \citenamefont {Zeng}, \citenamefont {Song},\ and\ \citenamefont
  {Zhang}}]{KCBSexperiment2017}%
  \BibitemOpen
  \bibfield  {author} {\bibinfo {author} {\bibfnamefont {Tao}\ \bibnamefont
  {Li}}, \bibinfo {author} {\bibfnamefont {Qiang}\ \bibnamefont {Zeng}},
  \bibinfo {author} {\bibfnamefont {Xinbing}\ \bibnamefont {Song}}, \ and\
  \bibinfo {author} {\bibfnamefont {Xiangdong}\ \bibnamefont {Zhang}},\
  }\bibfield  {title} {\enquote {\bibinfo {title} {Experimental contextuality
  in classical light},}\ }\href {\doibase 10.1038/srep44467} {\bibfield
  {journal} {\bibinfo  {journal} {Scientific Reports}\ }\textbf {\bibinfo
  {volume} {7}},\ \bibinfo {pages} {44467} (\bibinfo {year}
  {2017})}\BibitemShut {NoStop}%
\bibitem [{\citenamefont {Zhao}\ \emph {et~al.}(2025)\citenamefont {Zhao},
  \citenamefont {Liew}, \citenamefont {Ho}, \citenamefont {Liu},\ and\
  \citenamefont {Bulchandani}}]{scalabletestsquantumcontextuality}%
  \BibitemOpen
  \bibfield  {author} {\bibinfo {author} {\bibfnamefont {Wanbing}\ \bibnamefont
  {Zhao}}, \bibinfo {author} {\bibfnamefont {H.~W.~Shawn}\ \bibnamefont
  {Liew}}, \bibinfo {author} {\bibfnamefont {Wen~Wei}\ \bibnamefont {Ho}},
  \bibinfo {author} {\bibfnamefont {Chunxiao}\ \bibnamefont {Liu}}, \ and\
  \bibinfo {author} {\bibfnamefont {Vir~B.}\ \bibnamefont {Bulchandani}},\
  }\href {https://arxiv.org/abs/2512.16654} {\enquote {\bibinfo {title}
  {Scalable tests of quantum contextuality from stabilizer-testing nonlocal
  games},}\ } (\bibinfo {year} {2025}),\ \Eprint
  {http://arxiv.org/abs/2512.16654} {arXiv:2512.16654 [quant-ph]} \BibitemShut
  {NoStop}%
\bibitem [{\citenamefont {Hart}\ \emph {et~al.}(2025)\citenamefont {Hart},
  \citenamefont {Stephen}, \citenamefont {Wickenden},\ and\ \citenamefont
  {Nandkishore}}]{manybodycontextualityselftestingquantum}%
  \BibitemOpen
  \bibfield  {author} {\bibinfo {author} {\bibfnamefont {Oliver}\ \bibnamefont
  {Hart}}, \bibinfo {author} {\bibfnamefont {David~T.}\ \bibnamefont
  {Stephen}}, \bibinfo {author} {\bibfnamefont {Evan}\ \bibnamefont
  {Wickenden}}, \ and\ \bibinfo {author} {\bibfnamefont {Rahul}\ \bibnamefont
  {Nandkishore}},\ }\href {https://arxiv.org/abs/2512.16886} {\enquote
  {\bibinfo {title} {Many-body contextuality and self-testing quantum matter
  via nonlocal games},}\ } (\bibinfo {year} {2025}),\ \Eprint
  {http://arxiv.org/abs/2512.16886} {arXiv:2512.16886 [quant-ph]} \BibitemShut
  {NoStop}%
\bibitem [{\citenamefont {et~al.}(2025)}]{kumar2025qcsep}%
  \BibitemOpen
  \bibfield  {author} {\bibinfo {author} {\bibfnamefont {Kumar}\ \bibnamefont
  {et~al.}},\ }\href@noop {} {\enquote {\bibinfo {title} {Quantum--classical
  separation in bounded-resource tasks arising from measurement
  contextuality},}\ } (\bibinfo {year} {2025}),\ \Eprint
  {http://arxiv.org/abs/2512.02284} {2512.02284} \BibitemShut {NoStop}%
\bibitem [{\citenamefont {Liu}\ and\ \citenamefont
  {Ramanathan}(2025)}]{contextualityandrandomness}%
  \BibitemOpen
  \bibfield  {author} {\bibinfo {author} {\bibfnamefont {Yuan}\ \bibnamefont
  {Liu}}\ and\ \bibinfo {author} {\bibfnamefont {Ravishankar}\ \bibnamefont
  {Ramanathan}},\ }\bibfield  {title} {\enquote {\bibinfo {title} {Optimal and
  feasible contextuality-based randomness generation},}\ }\href {\doibase
  10.1103/cn3m-j743} {\bibfield  {journal} {\bibinfo  {journal} {Phys. Rev.
  Lett.}\ }\textbf {\bibinfo {volume} {135}},\ \bibinfo {pages} {170805}
  (\bibinfo {year} {2025})}\BibitemShut {NoStop}%
\bibitem [{\citenamefont {Liu}\ \emph {et~al.}(2024)\citenamefont {Liu},
  \citenamefont {Chung}, \citenamefont {Cruzeiro}, \citenamefont
  {Gonzales-Ureta}, \citenamefont {Ramanathan},\ and\ \citenamefont
  {Cabello}}]{FNSandallvsnothingequivalece}%
  \BibitemOpen
  \bibfield  {author} {\bibinfo {author} {\bibfnamefont {Yuan}\ \bibnamefont
  {Liu}}, \bibinfo {author} {\bibfnamefont {Ho~Yiu}\ \bibnamefont {Chung}},
  \bibinfo {author} {\bibfnamefont {Emmanuel~Zambrini}\ \bibnamefont
  {Cruzeiro}}, \bibinfo {author} {\bibfnamefont {Junior~R.}\ \bibnamefont
  {Gonzales-Ureta}}, \bibinfo {author} {\bibfnamefont {Ravishankar}\
  \bibnamefont {Ramanathan}}, \ and\ \bibinfo {author} {\bibfnamefont {Ad\'an}\
  \bibnamefont {Cabello}},\ }\bibfield  {title} {\enquote {\bibinfo {title}
  {Equivalence between face nonsignaling correlations, full nonlocality,
  all-versus-nothing proofs, and pseudotelepathy},}\ }\href {\doibase
  10.1103/PhysRevResearch.6.L042035} {\bibfield  {journal} {\bibinfo  {journal}
  {Phys. Rev. Res.}\ }\textbf {\bibinfo {volume} {6}},\ \bibinfo {pages}
  {L042035} (\bibinfo {year} {2024})}\BibitemShut {NoStop}%
\bibitem [{\citenamefont {Helwig}\ \emph {et~al.}(2012)\citenamefont {Helwig},
  \citenamefont {Cui}, \citenamefont {Latorre}, \citenamefont {Riera},\ and\
  \citenamefont {Lo}}]{AME}%
  \BibitemOpen
  \bibfield  {author} {\bibinfo {author} {\bibfnamefont {Wolfram}\ \bibnamefont
  {Helwig}}, \bibinfo {author} {\bibfnamefont {Wei}\ \bibnamefont {Cui}},
  \bibinfo {author} {\bibfnamefont {Jos\'e~Ignacio}\ \bibnamefont {Latorre}},
  \bibinfo {author} {\bibfnamefont {Arnau}\ \bibnamefont {Riera}}, \ and\
  \bibinfo {author} {\bibfnamefont {Hoi-Kwong}\ \bibnamefont {Lo}},\ }\bibfield
   {title} {\enquote {\bibinfo {title} {Absolute maximal entanglement and
  quantum secret sharing},}\ }\href {\doibase 10.1103/PhysRevA.86.052335}
  {\bibfield  {journal} {\bibinfo  {journal} {Phys. Rev. A}\ }\textbf {\bibinfo
  {volume} {86}},\ \bibinfo {pages} {052335} (\bibinfo {year}
  {2012})}\BibitemShut {NoStop}%
\bibitem [{\citenamefont {Pironio}(2005)}]{pironio2005lifting}%
  \BibitemOpen
  \bibfield  {author} {\bibinfo {author} {\bibfnamefont {Stefano}\ \bibnamefont
  {Pironio}},\ }\bibfield  {title} {\enquote {\bibinfo {title} {Lifting bell
  inequalities},}\ }\href@noop {} {\bibfield  {journal} {\bibinfo  {journal}
  {Journal of mathematical physics}\ }\textbf {\bibinfo {volume} {46}}
  (\bibinfo {year} {2005})}\BibitemShut {NoStop}%
\bibitem [{\citenamefont {Abramsky}\ \emph {et~al.}(2015)\citenamefont
  {Abramsky}, \citenamefont {Soares~Barbosa}, \citenamefont {Kishida},
  \citenamefont {Lal},\ and\ \citenamefont
  {Mansfield}}]{cohom_AVN_abramsky_et_al2015}%
  \BibitemOpen
  \bibfield  {author} {\bibinfo {author} {\bibfnamefont {Samson}\ \bibnamefont
  {Abramsky}}, \bibinfo {author} {\bibfnamefont {Rui}\ \bibnamefont
  {Soares~Barbosa}}, \bibinfo {author} {\bibfnamefont {Kohei}\ \bibnamefont
  {Kishida}}, \bibinfo {author} {\bibfnamefont {Raymond}\ \bibnamefont {Lal}},
  \ and\ \bibinfo {author} {\bibfnamefont {Shane}\ \bibnamefont {Mansfield}},\
  }\bibfield  {title} {\enquote {\bibinfo {title} {Contextuality, cohomology
  and paradox},}\ }\href {\doibase 10.4230/LIPIcs.CSL.2015.211} {\bibfield
  {journal} {\bibinfo  {journal} {Leibniz International Proceedings in
  Informatics (LIPIcs)}\ }\textbf {\bibinfo {volume} {41}},\ \bibinfo {pages}
  {211--228} (\bibinfo {year} {2015})}\BibitemShut {NoStop}%
\bibitem [{\citenamefont {Lee}(2016)}]{lee2016first}%
  \BibitemOpen
  \bibfield  {author} {\bibinfo {author} {\bibfnamefont {Jon}\ \bibnamefont
  {Lee}},\ }\href@noop {} {\enquote {\bibinfo {title} {A first course in linear
  optimization},}\ } (\bibinfo {year} {2016})\BibitemShut {NoStop}%
\bibitem [{Note1()}]{Note1}%
  \BibitemOpen
  \bibinfo {note} {Any empirical model can be decomposed as a convex mixture of
  contextual and noncontextual models, with contextuality being a more general
  concept than nonlocality. The convex decomposition of a nonlocal correlation
  can be expressed as \protect \[p(x_1,\protect \ldots ,x_n) = q^L
  p_{L}(x_1,\protect \ldots ,x_n) + q^{NL} p_{NL}(x_1,\protect \ldots ,x_n),
  \protect \] where (p{L}) denotes the local part and (p{NL}) the nonlocal
  part. This decomposition is a special case within the broader framework of
  contextuality. Hence, the condition that the contextual fraction $CF = 1$
  corresponds to a special case where the nonlocal fraction is also
  maximal}\BibitemShut {NoStop}%
\bibitem [{\citenamefont {Mermin}(1990{\natexlab{b}})}]{Mermin_argument}%
  \BibitemOpen
  \bibfield  {author} {\bibinfo {author} {\bibfnamefont {N.~David}\
  \bibnamefont {Mermin}},\ }\bibfield  {title} {\enquote {\bibinfo {title}
  {Extreme quantum entanglement in a superposition of macroscopically distinct
  states},}\ }\href {\doibase 10.1103/PhysRevLett.65.1838} {\bibfield
  {journal} {\bibinfo  {journal} {Phys. Rev. Lett.}\ }\textbf {\bibinfo
  {volume} {65}},\ \bibinfo {pages} {1838--1840} (\bibinfo {year}
  {1990}{\natexlab{b}})}\BibitemShut {NoStop}%
\bibitem [{\citenamefont {Barrett}\ \emph {et~al.}(2005)\citenamefont
  {Barrett}, \citenamefont {Linden}, \citenamefont {Massar}, \citenamefont
  {Pironio}, \citenamefont {Popescu},\ and\ \citenamefont
  {Roberts}}]{barrett2005nonlocal}%
  \BibitemOpen
  \bibfield  {author} {\bibinfo {author} {\bibfnamefont {Jonathan}\
  \bibnamefont {Barrett}}, \bibinfo {author} {\bibfnamefont {Noah}\
  \bibnamefont {Linden}}, \bibinfo {author} {\bibfnamefont {Serge}\
  \bibnamefont {Massar}}, \bibinfo {author} {\bibfnamefont {Stefano}\
  \bibnamefont {Pironio}}, \bibinfo {author} {\bibfnamefont {Sandu}\
  \bibnamefont {Popescu}}, \ and\ \bibinfo {author} {\bibfnamefont {David}\
  \bibnamefont {Roberts}},\ }\bibfield  {title} {\enquote {\bibinfo {title}
  {Nonlocal correlations as an information-theoretic resource},}\ }\href@noop
  {} {\bibfield  {journal} {\bibinfo  {journal} {Physical Review A—Atomic,
  Molecular, and Optical Physics}\ }\textbf {\bibinfo {volume} {71}},\ \bibinfo
  {pages} {022101} (\bibinfo {year} {2005})}\BibitemShut {NoStop}%
\bibitem [{\citenamefont {Pironio}\ \emph {et~al.}(2011)\citenamefont
  {Pironio}, \citenamefont {Bancal},\ and\ \citenamefont
  {Scarani}}]{Extremal_points_Pironio_2011}%
  \BibitemOpen
  \bibfield  {author} {\bibinfo {author} {\bibfnamefont {Stefano}\ \bibnamefont
  {Pironio}}, \bibinfo {author} {\bibfnamefont {Jean-Daniel}\ \bibnamefont
  {Bancal}}, \ and\ \bibinfo {author} {\bibfnamefont {Valerio}\ \bibnamefont
  {Scarani}},\ }\bibfield  {title} {\enquote {\bibinfo {title} {Extremal
  correlations of the tripartite no-signaling polytope},}\ }\href {\doibase
  10.1088/1751-8113/44/6/065303} {\bibfield  {journal} {\bibinfo  {journal}
  {Journal of Physics A: Mathematical and Theoretical}\ }\textbf {\bibinfo
  {volume} {44}},\ \bibinfo {pages} {065303} (\bibinfo {year}
  {2011})}\BibitemShut {NoStop}%
\bibitem [{\citenamefont {Abramsky}\ \emph {et~al.}(2018)\citenamefont
  {Abramsky}, \citenamefont {Barbosa}, \citenamefont {Car\`{u}}, \citenamefont
  {de~Silva}, \citenamefont {Kishida},\ and\ \citenamefont
  {Mansfield}}]{min_quntun_resources_abramsky2017}%
  \BibitemOpen
  \bibfield  {author} {\bibinfo {author} {\bibfnamefont {Samson}\ \bibnamefont
  {Abramsky}}, \bibinfo {author} {\bibfnamefont {Rui~Soares}\ \bibnamefont
  {Barbosa}}, \bibinfo {author} {\bibfnamefont {Giovanni}\ \bibnamefont
  {Car\`{u}}}, \bibinfo {author} {\bibfnamefont {Nadish}\ \bibnamefont
  {de~Silva}}, \bibinfo {author} {\bibfnamefont {Kohei}\ \bibnamefont
  {Kishida}}, \ and\ \bibinfo {author} {\bibfnamefont {Shane}\ \bibnamefont
  {Mansfield}},\ }\bibfield  {title} {\enquote {\bibinfo {title} {Minimum
  quantum resources for strong non-locality},}\ }\href {\doibase
  10.4230/LIPIcs.TQC.2017.9} {\bibfield  {journal} {\bibinfo  {journal}
  {Leibniz International Proceedings in Informatics (LIPIcs)}\ }\textbf
  {\bibinfo {volume} {73}},\ \bibinfo {pages} {9:1--9:20} (\bibinfo {year}
  {2018})}\BibitemShut {NoStop}%
\bibitem [{Note2()}]{Note2}%
  \BibitemOpen
  \bibinfo {note} {The correlation with the contextual fraction $\protect
  \mathrm {CF}=1$ but not the maximal marginal}\BibitemShut {NoStop}%
\bibitem [{\citenamefont {Rather}\ \emph {et~al.}(2022)\citenamefont {Rather},
  \citenamefont {Burchardt}, \citenamefont {Bruzda}, \citenamefont
  {Rajchel-Mieldzio\ifmmode~\acute{c}\else \'{c}\fi{}}, \citenamefont
  {Lakshminarayan},\ and\ \citenamefont {\ifmmode~\dot{Z}\else
  \.{Z}\fi{}yczkowski}}]{AME_construction}%
  \BibitemOpen
  \bibfield  {author} {\bibinfo {author} {\bibfnamefont {Suhail~Ahmad}\
  \bibnamefont {Rather}}, \bibinfo {author} {\bibfnamefont {Adam}\ \bibnamefont
  {Burchardt}}, \bibinfo {author} {\bibfnamefont {Wojciech}\ \bibnamefont
  {Bruzda}}, \bibinfo {author} {\bibfnamefont {Grzegorz}\ \bibnamefont
  {Rajchel-Mieldzio\ifmmode~\acute{c}\else \'{c}\fi{}}}, \bibinfo {author}
  {\bibfnamefont {Arul}\ \bibnamefont {Lakshminarayan}}, \ and\ \bibinfo
  {author} {\bibfnamefont {Karol}\ \bibnamefont {\ifmmode~\dot{Z}\else
  \.{Z}\fi{}yczkowski}},\ }\bibfield  {title} {\enquote {\bibinfo {title}
  {Thirty-six entangled officers of euler: Quantum solution to a classically
  impossible problem},}\ }\href {\doibase 10.1103/PhysRevLett.128.080507}
  {\bibfield  {journal} {\bibinfo  {journal} {Phys. Rev. Lett.}\ }\textbf
  {\bibinfo {volume} {128}},\ \bibinfo {pages} {080507} (\bibinfo {year}
  {2022})}\BibitemShut {NoStop}%
\bibitem [{\citenamefont {Huber}\ \emph {et~al.}(2018)\citenamefont {Huber},
  \citenamefont {Lami}, \citenamefont {Lancien},\ and\ \citenamefont
  {M\"uller-Hermes}}]{AME_construction_PPT}%
  \BibitemOpen
  \bibfield  {author} {\bibinfo {author} {\bibfnamefont {Marcus}\ \bibnamefont
  {Huber}}, \bibinfo {author} {\bibfnamefont {Ludovico}\ \bibnamefont {Lami}},
  \bibinfo {author} {\bibfnamefont {C\'ecilia}\ \bibnamefont {Lancien}}, \ and\
  \bibinfo {author} {\bibfnamefont {Alexander}\ \bibnamefont
  {M\"uller-Hermes}},\ }\bibfield  {title} {\enquote {\bibinfo {title}
  {High-dimensional entanglement in states with positive partial
  transposition},}\ }\href {\doibase 10.1103/PhysRevLett.121.200503} {\bibfield
   {journal} {\bibinfo  {journal} {Phys. Rev. Lett.}\ }\textbf {\bibinfo
  {volume} {121}},\ \bibinfo {pages} {200503} (\bibinfo {year}
  {2018})}\BibitemShut {NoStop}%
\bibitem [{\citenamefont {Goyeneche}\ \emph {et~al.}(2018)\citenamefont
  {Goyeneche}, \citenamefont {Raissi}, \citenamefont {Di~Martino},\ and\
  \citenamefont {\ifmmode~\dot{Z}\else
  \.{Z}\fi{}yczkowski}}]{AME_construction_combi}%
  \BibitemOpen
  \bibfield  {author} {\bibinfo {author} {\bibfnamefont {Dardo}\ \bibnamefont
  {Goyeneche}}, \bibinfo {author} {\bibfnamefont {Zahra}\ \bibnamefont
  {Raissi}}, \bibinfo {author} {\bibfnamefont {Sara}\ \bibnamefont
  {Di~Martino}}, \ and\ \bibinfo {author} {\bibfnamefont {Karol}\ \bibnamefont
  {\ifmmode~\dot{Z}\else \.{Z}\fi{}yczkowski}},\ }\bibfield  {title} {\enquote
  {\bibinfo {title} {Entanglement and quantum combinatorial designs},}\ }\href
  {\doibase 10.1103/PhysRevA.97.062326} {\bibfield  {journal} {\bibinfo
  {journal} {Phys. Rev. A}\ }\textbf {\bibinfo {volume} {97}},\ \bibinfo
  {pages} {062326} (\bibinfo {year} {2018})}\BibitemShut {NoStop}%
\bibitem [{\citenamefont {Santos}\ and\ \citenamefont
  {Amaral}(2021)}]{santos2021conditions}%
  \BibitemOpen
  \bibfield  {author} {\bibinfo {author} {\bibfnamefont {Leonardo}\
  \bibnamefont {Santos}}\ and\ \bibinfo {author} {\bibfnamefont {Barbara}\
  \bibnamefont {Amaral}},\ }\bibfield  {title} {\enquote {\bibinfo {title}
  {Conditions for logical contextuality and nonlocality},}\ }\href@noop {}
  {\bibfield  {journal} {\bibinfo  {journal} {Physical Review A}\ }\textbf
  {\bibinfo {volume} {104}},\ \bibinfo {pages} {022201} (\bibinfo {year}
  {2021})}\BibitemShut {NoStop}%
\bibitem [{\citenamefont {Gogioso}\ and\ \citenamefont
  {Zeng}(2019)}]{Secretsharingmaxcontext}%
  \BibitemOpen
  \bibfield  {author} {\bibinfo {author} {\bibfnamefont {Stefano}\ \bibnamefont
  {Gogioso}}\ and\ \bibinfo {author} {\bibfnamefont {William}\ \bibnamefont
  {Zeng}},\ }\bibfield  {title} {\enquote {\bibinfo {title} {Generalised
  mermin-type non-locality arguments},}\ }\href {\doibase
  10.23638/LMCS-15(2:3)2019} {\bibfield  {journal} {\bibinfo  {journal}
  {Logical Methods in Computer Science}\ }\textbf {\bibinfo {volume} {Volume
  15, Issue 2}},\ \bibinfo {eid} {3} (\bibinfo {year} {2019}),\
  10.23638/LMCS-15(2:3)2019}\BibitemShut {NoStop}%
\bibitem [{\citenamefont {Hillery}\ \emph {et~al.}(1999)\citenamefont
  {Hillery}, \citenamefont {Bu\ifmmode~\check{z}\else \v{z}\fi{}ek},\ and\
  \citenamefont {Berthiaume}}]{HBBscheme}%
  \BibitemOpen
  \bibfield  {author} {\bibinfo {author} {\bibfnamefont {Mark}\ \bibnamefont
  {Hillery}}, \bibinfo {author} {\bibfnamefont {Vladim\'{\i}r}\ \bibnamefont
  {Bu\ifmmode~\check{z}\else \v{z}\fi{}ek}}, \ and\ \bibinfo {author}
  {\bibfnamefont {Andr\'e}\ \bibnamefont {Berthiaume}},\ }\bibfield  {title}
  {\enquote {\bibinfo {title} {Quantum secret sharing},}\ }\href {\doibase
  10.1103/PhysRevA.59.1829} {\bibfield  {journal} {\bibinfo  {journal} {Phys.
  Rev. A}\ }\textbf {\bibinfo {volume} {59}},\ \bibinfo {pages} {1829--1834}
  (\bibinfo {year} {1999})}\BibitemShut {NoStop}%
\bibitem [{\citenamefont {Majumdar}(2026)}]{four_party_AMCC}%
  \BibitemOpen
  \bibfield  {author} {\bibinfo {author} {\bibfnamefont {Nripendra}\
  \bibnamefont {Majumdar}},\ }\bibfield  {title} {\enquote {\bibinfo {title}
  {Four party absolutely maximal contextual correlations},}\ }\href@noop {}
  {\bibfield  {journal} {\bibinfo  {journal} {arXiv preprint arXiv:2602.23883}\
  } (\bibinfo {year} {2026})}\BibitemShut {NoStop}%
\bibitem [{\citenamefont {Srinivasamurthy}\ and\ \citenamefont
  {Radhakrishna}(2015)}]{aravinda2013complementarity}%
  \BibitemOpen
  \bibfield  {author} {\bibinfo {author} {\bibfnamefont {Aravinda}\
  \bibnamefont {Srinivasamurthy}}\ and\ \bibinfo {author} {\bibfnamefont
  {Srikanth}\ \bibnamefont {Radhakrishna}},\ }\bibfield  {title} {\enquote
  {\bibinfo {title} {Complementarity between signaling and local indeterminacy
  in quantum nonlocal correlations},}\ }\href {\doibase 10.26421/QIC15.3-4-6}
  {\bibfield  {journal} {\bibinfo  {journal} {Quantum Information and
  Computation}\ }\textbf {\bibinfo {volume} {15}},\ \bibinfo {pages} {310--317}
  (\bibinfo {year} {2015})}\BibitemShut {NoStop}%
\end{thebibliography}
%

\end{document}